\documentclass[prx,twocolumn,superscriptaddress,citeautoscript,amsart,longbibliography]{revtex4}

\usepackage{graphicx}
\usepackage{multirow}
\usepackage{xcolor}
\usepackage{bm}
\usepackage{times}
\usepackage{amssymb}
\usepackage{amsmath,bm,amsfonts}
\usepackage{dcolumn}
\usepackage{graphicx}
\usepackage{latexsym}
\usepackage{bbm,amsthm,verbatim,tikz,tikz-cd,physics}
\usepackage[colorlinks=true,linkcolor=magenta,citecolor=magenta]{hyperref}

\usepackage{ulem} 
\usepackage{braket}
\usepackage{bbold}

\def\ket#1{|#1\rangle }
\def\bra#1{\langle #1 |}
\def\n{\nonumber \\ }

\def\bb{\mathbb}

\def\d{\partial}
\def\dd{\mathrm{d}}
\def\Tr{\mathrm{Tr}}
\def\Re{\mathrm{Re}}

\begin{document}
\title{Real Hopf insulator}

\author{Hyeongmuk \surname{Lim}$^\dagger$}
\affiliation{Department of Physics and Astronomy, Seoul National University, Seoul 08826, Korea}
\affiliation{Center for Correlated Electron Systems, Institute for Basic Science (IBS), Seoul 08826, Korea}
\affiliation{Center for Theoretical Physics (CTP), Seoul National University, Seoul 08826, Korea}

\author{Sunje \surname{Kim}$^\dagger$}
\affiliation{Department of Physics and Astronomy, Seoul National University, Seoul 08826, Korea}
\affiliation{Center for Correlated Electron Systems, Institute for Basic Science (IBS), Seoul 08826, Korea}
\affiliation{Center for Theoretical Physics (CTP), Seoul National University, Seoul 08826, Korea}

\author{Bohm-Jung \surname{Yang}}
\email{bjyang@snu.ac.kr}
\affiliation{Department of Physics and Astronomy, Seoul National University, Seoul 08826, Korea}
\affiliation{Center for Correlated Electron Systems, Institute for Basic Science (IBS), Seoul 08826, Korea}
\affiliation{Center for Theoretical Physics (CTP), Seoul National University, Seoul 08826, Korea}

\date{\today}

\begin{abstract}
Establishing the fundamental relation between the homotopy invariants and the band topology of Hamiltonians has played a critical role in the recent development of topological phase research.
In this work, we establish the homotopy invariant and the related band topology of three-dimensional (3D) real-valued Hamiltonians with two occupied and two unoccupied bands. Such a real Hamiltonian generally appears in $\mathcal{PT}$ symmetric spinless fermion systems where $\mathcal{P}$ and $\mathcal{T}$ indicate the inversion and time-reversal symmetries, respectively.
We show that the 3D band topology of the system is characterized by two independent Hopf invariants when the lower-dimensional band topology is trivial. 
Thus, the corresponding 3D band insulator with nonzero Hopf invariants can be called a real Hopf insulator (RHI).
In sharp contrast to all the other topological insulators discovered up to now, the topological invariants of RHI can be defined only when the fixed number of both the occupied and unoccupied states are simultaneously considered. Thus, the RHI belongs to the category of delicate topological insulators proposed recently.
We show that finite-size systems with slab geometry support surface states with nonzero Chern numbers in a $\mathcal{PT}$-symmetric manner independent of the Fermi level position, and establish the bulk-boundary correspondence.
We also discuss the bulk-boundary correspondence of rotation symmetric RHIs using the returning Thouless pump.
\end{abstract}

\maketitle
\def\thefootnote{$\dagger$}\footnotetext{These authors contributed equally to this work}

\section{Introduction}
Establishing the relation between mathematical homotopy invariants and physical properties of topological insulators is the most critical step in the study of topological phases.
Although possible topological invariants have been systematically categorized~\cite{ryu2010tenfold,altland1997AZclass,
chiu2016classificationTI,kitaev2009periodictableofTI,
bernevig2017TQC,bernevig2018TQC2,
bernevig2021magneticTQC,vishwanath2017SI,slager2017classAclassification,
morimoto2013TIclassification}, there are only few topological invariants whose integral form is explicitly known.
One representative example is the two-dimensional (2D) Chern number $C$ whose integral form is given by
\begin{gather}
    C=\frac{1}{2\pi}\int_{BZ} d^{2}k \mathrm{Tr}\mathrm{F}(\mathbf{k})\label{Cn},
\end{gather}
where $\mathrm{F}=\nabla\times\mathbf{A}+\left[\mathrm{A}_{x},\mathrm{A}_{y}\right]$ is a non-abelian Berry curvature of the occupied states, $\left[\mathrm{A}_{i}\right]_{nm}=i\langle u_{n}|\partial_{k_{i}} u_{m}\rangle$ is a non-abelian Berry connection for the Bloch states $|u_{n}\rangle$, and BZ denotes the Brillouin zone (BZ).
A Chern insulator with nonzero $C$ breaks time-reversal $\mathcal{T}$ symmetry, and carries chiral edge modes as long as the energy gap remains finite~\cite{kane2010TI}. Since the topological property is unaffected by adding trivial bands below the Fermi energy $E_F$, the Chern insulator belongs to the category of stable topological insulators~\cite{kitaev2009periodictableofTI,
chiu2016classificationTI}.


Interestingly, by superposing two Chern insulators with opposite Chern numbers, a new type of topological insulators can be generated when additional symmetries are supplemented. For example, 
in 2D spinless fermion systems with the combined $\mathcal{PT}$ symmetry where $\mathcal{P}$ indicates the inversion symmetry, such a superposition can give an Euler insulator when the number of the occupied bands is two~\cite{ahn2019failure,slager2020geometric,
slager2020braidingWeylwithEI,slager2021experimentalEI,
slager2022EImodels,bernevig2021TBGeulerclass} or a Stiefel-Whitney insulator when more than two bands are occupied~\cite{Ahn2019SWclass,ahn2018linking,
bzdusek2020monopolebynonabelian,
zhao2017realdiracfermions,zhao2020boundaryofmonopoleNL}.
For a given Euler insulator with two occupied bands, its integer $\bb{Z}$ valued Euler invariant $e$ can be computed by integrating the off-diagonal component of the $2\times 2$ Berry curvature matrix,
\begin{gather}
e=\frac{1}{2\pi}\int_{BZ} d^{2}k \left[\mathrm{F}(\mathbf{k})\right]_{12},\label{euler}
\end{gather}
where $\left[\mathrm{F}\right]_{12}=\nabla\times[\mathbf{A}]_{12}$.
An Euler insulator can support in-gap corner states in the presence of additional mirror symmetry and chiral symmetry~\cite{ahn2019failure}. As the Wannier obstruction of an Euler insulator can be relieved by adding additional trivial bands below $E_F$, it belongs to the category of fragile topological insulators~\cite{po2018fragile,hwang2019fragile,
slager2019fragilebyWilson,bernevig2019SIoffragileTI,
Song2020monoid,Kooi2019}.

In the case of three-dimensional (3D) systems, one representative topological invariant whose integral form is known is the Hopf invariant, and the relevant topological insulator, called the Hopf insulator, was first proposed by Moore-Ran-Wen (MRW)~\cite{moore2008Hopfinsulator}.
The Hopf insulator is a two-band insulator with one occupied band and one unoccupied band, characterized by the integer $\bb{Z}$-valued Hopf invariant $\chi$ defined as
\begin{gather}
\chi=-\frac{1}{4\pi^2}\int_{BZ}d^{3}k \mathbf{A}\cdot \mathbf{F},\label{Hopf}
\end{gather}
where $\mathbf{A}$ and $\mathbf{F}$ are the abelian Berry connection and the Berry curvature of the occupied state, respectively~\cite{moore2008Hopfinsulator}.
Unlike the Chern insulator or the Euler insulator,
the Hopf insulator with nonzero Hopf number can be trivialized by adding trivial bands either above or below the Fermi energy $E_F$.
Namely, it belongs to the category of delicate topological insulators~\cite{bzdusek2021multicellularity}, which is distinct from fragile topological insulators~\cite{po2018fragile,hwang2019fragile,
bernevig2019SIoffragileTI,slager2019fragilebyWilson,
Song2020monoid,Kooi2019} that can be trivialized by adding trivial bands below the Fermi level, or stable topological insulators (TIs) with robust band topology.
Generally, a Hopf insulator does not have non-trivial in-gap states at the boundary without additional symmetries~\cite{moore2021Hopfrealize,
moore2021realizingHI}.
However, it always has surface states with nonzero Chern number that is equal to the Hopf invariant of the bulk~\cite{alexandradinata2021HIsurface,
Alex2021quanizedsurfacemagHI}.
There have been various intriguing recent developments aiming at understanding the delicate band topology and the bulk-boundary correspondence of the Hopf insulator, taking into account additional symmetries~\cite{bzdusek2021multicellularity, hughes2022spinHopf,liu2017symmetryHI}.

In this work, we study the band topology of 3D $\mathcal{PT}$ symmetric spinless fermion systems with two occupied bands and two unoccupied bands. We show that the band topology of such a Hamiltonian can be characterized by two integer invariants, each of which is equivalent to the Hopf invariant of a two-band system. The corresponding topological insulator with nonzero Hopf invariants can be called a real Hopf insulator (RHI), because the $\mathcal{PT}$ symmetry imposes the reality of the Hamiltonian and the wave function. Thus, a RHI can be considered as a superposition of two conventional Hopf insulators in a $\mathcal{PT}$ symmetric manner, similar to the case of an Euler insulator constructed by superposing two Chern insulators preserving the $\mathcal{PT}$ symmetry.

In particular, we derive the integral form of the RHI invariants, thus provides an additional rare example of the topological invariants with explicit integral form.
Interestingly, we find that the Berry curvature and the Berry connection of both the occupied and unoccupied bands are simultaneously required to define the RHI invariant, directly manifesting the delicate band topology of RHIs. This is in sharp contrast to all the other known topological invariants, including those in Eq.~(\ref{Cn}),~(\ref{euler}),~(\ref{Hopf}) that depend only on the occupied states. 
In the case of RHIs, any combination of the Berry curvature or Berry connection of the occupied bands cannot give quantized topological invariants. Both the occupied and unoccupied states are equally important to characterize the RHI.

We also establish the bulk-boundary correspondence of RHIs. Especially, we show that RHIs generally support surface Chern bands in a $\mathcal{PT}$ symmetric manner.
In the presence of additional rotational symmetry, the band topology of RHI can be described by using the idea of returning Thouless pump (RTP)~\cite{bzdusek2021multicellularity}. 
Especially, we illustrate the physical meaning of two Hopf invariants of RHIs in the context of RTP and discuss the role of multicellularity on the crossing between neighboring Wannier sheets.

The rest of the paper is organized as follows. In Sec.II, we reformulate the conventional two-band Hopf insulator in terms of quaternions. In Sec. III, the classifying space for spinless $\mathcal{PT}$ symmetric 4-band insulators is discussed. Augmenting the conventional classifying space with ``modified'' classifying spaces, we provide a general construction for such insulators in 3D. It is followed by Sec. IV where we define the Real Hopf Insulator (RHI) as a subset of the aforementioned class of insulators and explicitly compute the bulk topological invariants. Sec. V reveals the bulk-boundary correspondence of RHI arising in the slab geometry. Finally, Sec. VI provides an interpretation of the RHI invariants in terms of Returning Thouless Pump (RTP). Mathematical details and supplementary discussion are compilated in the appendix.

\section{Quaternion formulation of Hopf insulator}
\subsection{Review of the Hopf insulator}
Let us first briefly review the formulation of the Hopf insulator proposed by Moore-Ran-Wen (MRW) in Ref.~\cite{moore2008Hopfinsulator}.
The Hopf insulator indicates a magnetic topological insulator in three-dimensions with one occupied band and one unoccupied band
whose band topology is characterized by a nonzero Hopf invariant.
In momentum $\mathbf{k}$ space, such a band insulator can be described by a $2\times2$ matrix Hamiltonian $H(\mathbf{k})$
which has one positive and one negative eigenvalues for any $\mathbf{k}$ assuming that the Fermi energy $E_F$ is zero.
By deforming all the positive eigenvalues to +1 and all the negative eigenvalues to -1,
$H(\mathbf{k})$ can be written as
\begin{align}
	H(\mathbf{k})=W(\mathbf{k})\sigma_zW^{\dag}(\mathbf{k}),
\end{align}
where $\sigma_{x,y,z}$ are Pauli matrices for two bands,
and $W(\mathbf{k})\in U(2)$, the vector space of unitary $2\times2$ matrices.
As the overall phase change within either the occupied subspace or the unoccupied subspace does not affect the $\sigma_z$ part of the Hamiltonian, for any $\mathbf{k}$, $H(\mathbf{k})$ is a point on the manifold $U(2)/[U(1)\times U(1)]$ that is isomorphic to a 2-sphere $S^2$, i.e.,
\begin{align}
	H(\mathbf{k})\in \frac{U(2)}{U(1)\times U(1)}=S^2.
\end{align}
So a two-band insulator in three-dimensions is described by the mapping from the 3D Brillouin zone torus $\bb{T}^3$ to $S^2$. When the Chern number of the system is zero, different classes of topological two-band insulators correspond to different classes of the mapping from $S^3$ to $S^2$ characterized by the integer Hopf index valued in the third homotopy group $\pi_3(S^2)=\bb{Z}$.

The Hopf map can be described as follows.
We first consider $z(\mathbf{k})\in\bb{S}^3$ defined as $z(\mathbf{k})^{\top}=(z_{\uparrow}(\mathbf{k}),z_{\downarrow}(\mathbf{k}))^{\top}\in \mathbf{C}^2$ 
where two complex numbers $z_{\uparrow}, z_{\downarrow}\in \mathbf{C}$ can be written as $z_{\uparrow}=a + b \text{i}$, $z_{\downarrow}=c + d \text{i}$ with real numbers $a, b, c, d$, which satisfy $|z_{\uparrow}|^2+|z_{\downarrow}|^2=a^2+b^2+c^2+d^2=1$.
Then a map $p:\bb{S}^3 \longrightarrow \bb{S}^2$ can be defined by 
\begin{align}
	p:z(\mathbf{k})\in\bb{S}^3 \longrightarrow z^{\dagger}(\mathbf{k}) \bm{\sigma} z(\mathbf{k})\in \bb{S}^2,
\end{align}
where
\begin{align}
\label{eq.Hopf_v}
    z^{\dagger}(\mathbf{k}) \bm{\sigma} z(\mathbf{k})\equiv \mathbf{v}(\mathbf{k})=[v^1(\mathbf{k}),v^2(\mathbf{k}),v^3(\mathbf{k})], 
\end{align}
satisfying $\sum_{i=1,2,3} [v^{i}(\mathbf{k})]^2=1$.

In the real coordinate system $(a,b,c,d)$, the Hopf map $p$ is given by
\begin{gather}
    p:\bb{S}^3\subseteq\bb{R}^4 \longrightarrow \bb{S}^2\subseteq\bb{R}^3 \n (a,b,c,d)\mapsto(v^1,v^2,v^3),
\label{ftn.Hopf_1}
\end{gather}
\begin{align}
    & v^1=2(ac+bd), \n 
    & v^2=2(ad-bc), \n 
    & v^3=a^2+b^2-c^2-d^2,
\end{align}
in which $a^2+b^2+c^2+d^2=1$. 

The Hopf map projects a circle embedded in $\bb{S}^3$ onto a single point, thus mapping onto the lower dimensional sphere $\bb{S}^2$. This is often rephrased by saying that the Hopf map $p$ is a fibration over $\bb{S}^2$ with fiber $\bb{S}^1$,
\begin{gather}
\label{Hopf fibration_1}
    \bb{S}^1 \hookrightarrow \bb{S}^3 \xrightarrow[]{\;p\;}\bb{S}^2,
\end{gather}
where the hook arrow from $\bb{S}^1$ in Eq.~(\ref{Hopf fibration_1}) specifies that the preimage of every point in $\bb{S}^2$ by $p$ has the shape of a circle. The circular fibers of Hopf map will be precisely described in Sec.~\ref{Sec.modified}. 

In general, the maps from $\bb{S}^3$ to $\bb{S}^2$ are classified (up to homotopy) by an integer invariant called Hopf index. One can conceive of this index as an analogue of winding numbers, although the former is defined between spheres with different dimensions while the latter for spaces with the same dimension. In particular, the Hopf map given by Eq.~(\ref{ftn.Hopf_1}) has Hopf index $h[p]=1$ and serves as a representative in the class $h=1$. There are versions of Hopf maps representing those classes with different values of $h$.

Given a version of Hopf map $p$, the Hamiltonian $H(\mathbf{k}) = (p\circ q)(\mathbf{k})$ is the composition of $p$ and a deformation mapping 
\begin{gather}
    q: BZ = \bb{T}^3 \longrightarrow S^3
\end{gather}
that shrinks the surface of the BZ into a point, which is possible since the Chern number of the system vanishes. This implies the BZ can effectively be regarded as a 3-sphere to which the Hopf map can be applied. A prominent example of such deformation map appears in the Moore-Ran-Wen model $H(\mathbf{k}) = p\circ q_{MRW}^{t,h}$ with adjustable parameters $t, h$, where
\begin{gather}
    q_{MRW}^{t,h} : \bb{T}^3 \longrightarrow S^3\subset\bb{R}^4
    \n 
    \begin{pmatrix} k_x \\ k_y \\ k_z \end{pmatrix} \mapsto \frac{1}{\mathcal{N}_{\mathbf{k}}} \begin{pmatrix} \sin k_x \\ t \sin k_y \\ \sin k_z \\ h+\sum\limits_{i=x,y,z}\cos k_i\end{pmatrix} .
    \label{ftn.MRW_deform}
\end{gather}
Here $\mathcal{N}_{\mathbf{k}}$ is a normalization factor. 

Since $q$ maps between two spaces of same dimension, it is classified by the usual winding number~\cite{deng2013HImodels}
\begin{gather}
    w_3[q] = \frac{1}{12\pi^2}\int_{\bb{T}^3}d^3k\epsilon^{abcd}\epsilon^{ijk}q_a\partial_{i}q_b\partial_{j}q_c\partial_{k}q_d.
\end{gather}
For instance, $w_3[q_{MRW}^{t,h}] = 1$ for $t=1,h=\pm3/2$~\cite{moore2008Hopfinsulator,deng2013HImodels}. The $\bb{Z}$-valued Hopf invariant classifying the Hopf insulator is given by the product of the Hopf index $h[p]$ and the winding number $w_3[q]$.
\begin{gather}
    \chi = h[p] \cdot w_3[q].
\label{eq.Hopf-winding}
\end{gather}
Since every integer $n$ can be realized as the winding number of an appropriate deformation map, e.g. by substituting $k_z\to nk_z$ to the Moore-Ran Wen model with $w_3=1$, without loss of generality one can fix a version of Hopf fibration having index 1 and call it ``the" Hopf fibration. Then Eq.~(\ref{eq.Hopf-winding}) reduces to 
\begin{gather}
    \chi = w_3[q],
\label{eq.reduced Hopf index}
\end{gather}
which corresponds to the models with the fixed Hopf fibration map. These exhaust all homotopy classes of the Hopf insulator. In the remaining part of this paper, we will take this route and fix the Hopf fibration structure as in either Eq.~(\ref{ftn.Hopf_1}) or Eq.~(\ref{eq.Hopf_quaternion}). [These two choices are equivalent; see below Eq.~(\ref{eq.Hopf_quaternion}).] The integer in Eq.~(\ref{eq.reduced Hopf index}) will be referred to as the Hopf invariant. The corresponding Hamiltonian for the Hopf insulator is 
\begin{gather}
    \label{eq.H_Hopf_CP1}
    H_{C} = - [z^{\dagger}(\mathbf{k}) \bm{\sigma} z(\mathbf{k})]\cdot \bm{\sigma}\equiv - \mathbf{v}(\mathbf{k})\cdot \bm{\sigma}.
\end{gather}
The Moore-Ran-Wen model of Hopf insulator refers to the family of Hamiltonians in two parameters $(t,h)$
\begin{gather}
    H^{t,h}_{C,MRW} = - [z_{t,h}^{\dagger}(\mathbf{k}) \bm{\sigma} z_{t,h}(\mathbf{k})]\cdot \bm{\sigma} \n
    \mbox{} \n 
\label{def.MRW_Hopf}
    z_{t,h}(\mathbf{k}) = \frac{1}{\mathcal{N}_{\mathbf{k}}}
    \begin{pmatrix}
        \sin k_x + \text{i}t \sin k_y \\ 
        \sin k_z + \text{i} \left[ h+\sum\limits_{i=x,y,z}\cos k_i \right]
    \end{pmatrix} .
\end{gather} 
The normalization factor $\mathcal{N}_{\mathbf{k}}$ is such that $|z(\mathbf{k})|=1$. As a special case, $(t,h)=(1,\pm3/2)$ realizes the class $\chi=1$. 

The topological property of the Hopf insulator can be described by using $\mathbf{F}=(F^1,F^2,F^3)$ and $\mathbf{A}=(A^1,A^2,A^3)$ which are defined as
\begin{gather}
\label{eq.Gauss curvature}
    F^a
    =\epsilon^{abc} \mathbf{v} \cdot \frac{\d}{\d k_b}\mathbf{v} \times \frac{\d}{\d k_c}\mathbf{v},
\end{gather}
satisfying the magneto-static equations
\begin{gather}
    \label{eq.magnetostat 1}
    \bm{\nabla}_\mathbf{k}\times\mathbf{A}=\mathbf{F},
	\\
    \label{eq.magnetostat 2}
	\bm{\nabla}_\mathbf{k}\cdot\mathbf{A}=0.
\end{gather}

In fact, in terms of $z(\mathbf{k})=(z_{\uparrow}(\mathbf{k}),z_{\downarrow}(\mathbf{k}))$, $\mathbf{A}$ can be explicitly written as
\begin{gather}
	\label{eq.electrostat sol}
	\mathbf{A} = \text{i} (z_{\uparrow}^{*},z_{\downarrow}^{*})\nabla (z_{\uparrow},z_{\downarrow})^{\top},
\end{gather}
which is nothing but the familiar Berry connection.
Finally, the integer-valued Hopf index is given by
\begin{gather}
	\label{def.Hopf Inv vec}
	\chi = -\frac{1}{4\pi^2} \int_{BZ}d^3k \; \mathbf{F}\cdot\mathbf{A}.
\end{gather}

\subsection{Quaternion formulation of the Hopf insulator}
The Hopf insulator can also be formulated in terms of the quaternion as described below.

\subsubsection{Quaternions}
The quaternion algebra $\mathbf{H}$ is the 4-dimensional real vector space with basis vectors $\{1, \text{i} , \text{j}, \text{k}\}$,
\begin{gather}
	\mathbf{H}
	= \{ a+b\text{i}+c\text{j}+d\text{k}: a,b,c,d\in\bb{R} \},
\end{gather}
equipped with the multiplication rules
\begin{gather}
	\text{i}^2 = \text{j}^2 = \text{k}^2 = \text{i}\text{j}\text{k} = -1.
\end{gather}
We regard the complex vector space $\mathbf{C}$ as the subset
\begin{gather}
	\mathbf{C} = \{a+b\text{i}+c\text{j}+d\text{k}:c=d=0\}\subset\mathbf{H}.
\end{gather}
$\mathbf{H}$ has the conjugation operation 
\begin{align}
	& z=a+b\text{i}+c\text{j}+d\text{k}
	\nonumber
	\\
	\mapsto\;
	& \overline{z}=a-b\text{i}-c\text{j}-d\text{k},
\end{align}
which extends the complex conjugation defined on $\mathbf{C}$, and the norm function
\begin{gather}
	|z|^2 = z\overline{z}=a^2+b^2+c^2+d^2.
\end{gather}
These operations satisfy the equations
\begin{gather}
	\label{eq.conjugate of product}
	\overline{zw}=\overline{w}\cdot\overline{z},
	\\
	\label{eq.norm of conjugate}
	|\overline{z}|=|z|, 
	\\
	\label{eq.norm of product}
	|zw|=|z|\cdot|w|,
\end{gather}
where $z, w\in \mathbf{H}$. The order of $z, w$ in Eq.~(\ref{eq.conjugate of product}) is important since the quaternion multiplication is not commutative.

$\mathbf{H}$ has an obvious isomorphism to $\bb{R}^4$:
\begin{gather}
	\ket{\cdot}: \mathbf{H} \longrightarrow \bb{R}^4,
	\n
	q=p_0 + p_1 \text{i} + p_2 \text{j} + p_3 \text{k} \mapsto \ket{q} \equiv (p_0,p_1,p_2,p_3)^{\top}.
	\label{ftn.H=R4}
\end{gather}
Under the identification in Eq.~(\ref{ftn.H=R4}), we can model $\bb{S}^3$ as the set of unit quaternions:
\begin{gather}
	\nonumber
	\bb{S}^3
	= \{ q\in\mathbf{H}: |q|^2=1\}\subseteq \mathbf{H}.
\end{gather}

Also, the Euclidean inner product of two vectors $\ket{p}=(p_0,p_1,p_2,p_3)^{\top}$ and $\ket{q}=(q_0,q_1,q_2,q_3)^{\top}$,
\begin{gather}
	\braket{p|q}  = \sum_{i=0}^{3} p_i q_i,
\end{gather}
can be expressed using quaternions as
\begin{gather}
	\label{R4 inner product quaternion form}
	\braket{p|q} = \text{Re}[\;p\overline{q}\;] = \text{Re}[\;\overline{p}q\;],
\end{gather}
using the projection onto the real part
\begin{gather}
	\Re : \mathbf{H} \longrightarrow \bb{R} 
	\n 
	a+b\text{i}+c\text{j}+d\text{k} \mapsto a.
\end{gather}
$\Re$ also satisfies the following useful relation
\begin{gather}
	\Re[\;wq\overline{w}\;]=|w|^2 \Re[\;q\;].
	\label{eq.Re conj}
\end{gather}
There is another identification of vector spaces
\begin{gather}
	\mathbf{H} \longrightarrow \mathbf{C}^2 \n
	a+b\text{i}+c\text{j}+d\text{k} \mapsto (a+b\text{i},c+d\text{i})^{\top}.
\end{gather}
Since
\begin{gather}
	a+b\text{i}+c\text{j}+d\text{k} = a+b\text{i}+(c+d\text{i})\text{j},
\end{gather}
we can interchangeably use a complex vector $(\alpha,\beta)^{\top}\in\mathbf{C}^2$ in place of $\alpha+\beta \text{j} \in \mathbf{H}$ and vice versa.
To avoid confusion, we denote the quaternion conjugate by $\overline{(\cdot)}$ and the complex conjugate by $(\cdot)^{*}$ in the following.

\subsubsection{The quaternion form of the Hopf map}

The Hopf map can be compactly formulated in terms of the quaternion as pointed out in Ref.~\cite{slager2020dynamicEulerHopf}. Under the identification $\bb{R}^4 \approx \bb{H}$ of real 4-vectors and quaternions, the sphere $\bb{S}^3$ in $\bb{R}^4$ can also be seen as the unit quaternions.
For any $z\in \bb{S}^3$, observe that
\begin{align}
	\overline{z}\text{i}z
	& \in \{a \text{i} + b \text{j} + c \text{k} \in \mathbf{H}: a^2+b^2+c^2=1\}
	\n 
	& = \{\text{purely imaginary unit quaternions}\}
	\n 
	& = \bb{S}^2.
\end{align}
This can be checked applying Eq.~(\ref{eq.conjugate of product})-(\ref{eq.norm of product}):
\begin{gather}
	\overline{\overline{z}\text{i}z} =  \overline{z}(-\text{i})z
	= - \overline{z}\text{i}z,
\end{gather}
which implies $\overline{z}\text{i}z$ is purely imaginary, and
\begin{gather}
	\label{eq.|ziz'|=1_1}
	|\overline{z}\text{i}z|=|\overline{z}| \cdot |\text{i}| \cdot |z| = |z|^2 = 1,
\end{gather}
indicating that it has unit length. The last equality in Eq.~(\ref{eq.|ziz'|=1_1}) follows from $z\in \bb{S}^3$.

Now the Hopf map in Eq.~(\ref{ftn.Hopf_1}) can be alternatively defined as
\begin{gather}
	p: \bb{S}^3 \subseteq \mathbf{H} \longrightarrow \bb{S}^2 \subseteq \mathbf{H}
	\n 
	z \mapsto \overline{z}\text{i}z.
	\label{eq.Hopfmap_Q}
\end{gather}
Indeed, using the real coordinates $z=a+b\text{i}+c\text{j}+d\text{k}$, we can check
\begin{align}
    \label{eq.Hopf_quaternion}
    \overline{z}\text{i}z 
    = & (a-b\text{i}-c\text{j}-d\text{k})\cdot \text{i} \cdot (a+b\text{i}+c\text{j}+d\text{k}) \n 
    = & (a^2+b^2-c^2-d^2) \; \text{i} \n 
    & + 2(-ad+bc) \; \text{j} \n 
    & + 2(ac+bd) \; \text{k},
\end{align}
in agreement (up to a rotation) with Eq.~(\ref{ftn.Hopf_1}). Since Hopf index of a map $f:\bb{S}^3\to\bb{S}^2$ is invariant under continuous deformation of $f$ such as rotation, Eq.~(\ref{eq.Hopf_quaternion}) can be taken as a Hopf map with index unity.

\subsubsection{The quaternion form of the Hopf insulator}
\label{Subsec.quaternion_Hopf}
Using the quaternion $z=a+b\text{i}+c\text{j}+d\text{k}\equiv z_{\uparrow}+z_{\downarrow}\text{j}\in \mathbf{H}$, 
the Hamiltonian for the Hopf insulator can be written as
\begin{gather}
\label{eq.Hopf_Q}
	H_{Q} = \Re[\overline{z}\text{i}z\cdot\sigma_{Q}]=-u^1\sigma_x-u^2\sigma_y-u^3\sigma_z,
\end{gather}
where $\sigma_{Q}=\sigma_x\text{i}+\sigma_y\text{j}+\sigma_z\text{k}$, and
\begin{gather}
\overline{z}\text{i}z=u^1 \text{i} + u^2 \text{j} + u^3 \text{k},
\end{gather}
in which
\begin{align}
\label{eq.Hopf_u}
    u^1&=a^2+b^2-c^2-d^2=v^3,
    \n
    u^2&=2(bc-ad)=-v^2,
    \n
    u^3&=2(ac+bd)=v^1.	
\end{align}
Using a unitary matrix
\begin{gather}
	U=\frac{1}{\sqrt{2}}\begin{pmatrix} 1 & 1 \\ 1 & -1
	\end{pmatrix},
\end{gather}
it is straightforward to show that
\begin{align}
    \label{eq.H_identification}
    U^{-1}H_{Q}U=\Re[\overline{z}\text{i}z\cdot U^{-1}\sigma_{Q}U]=-v^1\sigma_x-v^2\sigma_y-v^3\sigma_z,	
\end{align}
which is nothing but the Hopf insulator Hamiltonian $H_C$ in Eq.~(\ref{eq.H_Hopf_CP1}).

By defining
\begin{gather}
	\mathbf{a} = \text{Re}[\;-\text{i} \cdot z \nabla \overline{z}\;], 
	\quad
	\mathbf{f} = \nabla\times\mathbf{a},
\end{gather}
one can show that Eq.~(\ref{def.Hopf Inv vec}) is equivalent to 
\begin{gather}
	\label{def.HI inv vec-q}
	\chi = - \frac{1}{4\pi^2} \int_{BZ}d^3k \; \mathbf{f}\cdot\mathbf{a}. 
\end{gather}
Moreover, we find the relation between $\mathbf{A}$ for $H^C$ and $\mathbf{a}$ for $H_{Q}$ such that
\begin{gather}
	\mathbf{A} = - a\nabla b + b\nabla a - c\nabla d + d\nabla c = \mathbf{a}. 
\end{gather}

\section{$\mathcal{PT}$ symmetric 4-band insulators in three-dimensions} 

In this work, we study the topological properties of the 4-band real Hamiltonian $H_{R}^{2,2}(\mathbf{k})$ with two occupied and two unoccupied bands in $\mathcal{PT}$ symmetric spinless fermion systems in three-dimensions.
In the absence of spin-orbit coupling, the combined symmetry $\mathcal{PT}$ can be represented by $\mathcal{PT=K}$ where $\mathcal{K}$ denotes the complex conjugation operator. As $\mathcal{PT}$ is local in $\mathbf{k}$ space, the $\mathcal{PT}$ invariance of the $\mathbf{k}$-space Hamiltonian $H_{R}^{2,2}(\mathbf{k})$ imposes the reality condition on $H_{R}^{2,2}(\mathbf{k})$ given by
\begin{align}
    (PT)H_{R}^{2,2}(\mathbf{k})(PT)^{-1}=[H_{R}^{2,2}(\mathbf{k})]^*=H_{R}^{2,2}(\mathbf{k}).
\end{align}
The Hamiltonian, being both hermitian and real, allows a diagonalization with a real orthogonal basis. To be precise, the spectral theorem for real symmetric matrices states that
\begin{gather}
    H = H^\dagger \text{ and } H^* = H
    \n
    \Rightarrow H = O\cdot diag(E_1,E_2,\ldots E_n) \cdot O^\top
\end{gather}
where $n$ is the dimension. Hence, the wave function corresponding to $\mathcal{PT=K}$ can be chosen to be real.
Therefore, the $\mathcal{PT}$ symmetric spinless fermion systems are an ideal avenue to study the band topology of real wave functions~\cite{ahn2019failure,ahn2018linking,
Ahn2019SWclass}.

\subsection{Classifying space}
To examine the topological property of $H_{R}^{2,2}(\mathbf{k})$, one can deform all the positive eigenvalues to +1 and all the negative eigenvalues to -1 assuming the Fermi energy at 0 ($E_F=0$).
The resulting flattened Hamiltonian $\overline{H}_{R}^{2,2}(\mathbf{k})$ can generally be written as
\begin{gather}
	\label{4band Hamiltonian}
	\overline{H}_{R}^{2,2}(\mathbf{k})
	= R
	\begin{pmatrix}
		1&&&\\&1&&\\&&-1&\\&&&-1
	\end{pmatrix}
	R^{\top},
\end{gather}
where $R\in \mathrm{O}(4)$, the space of real orthogonal $4\times4$ matrices.
As the energy levels are fixed, the Hamiltonian is completely determined by the projection operator onto the occupied subspace
\begin{gather}
\label{def.projector}
    {P} : BZ=\bb{T}^3\longrightarrow \mathrm{Gr}(2,4),
\end{gather}
where the classifying space, or Grassmannian, $\mathrm{Gr}(2,4)$ is the set of 2-dimensional subspaces of $\bb{R}^4$. The unoccupied subspace is automatically determined as the complement of ${P}(\mathbf{k})$.
Since $\mathrm{O}(2)$ rotation within the occupied (unoccupied) band subspace generated by the first (last) two columns of $R$ does not affect the projector in Eq.~(\ref{def.projector}), $\mathrm{Gr}(2,4)\approx \mathrm{O}(4)/[\mathrm{O}(2)\times \mathrm{O}(2)]$. These two spaces can be used interchangeably. In the rest of this paper, we completely identify these spaces. That is, we identify the projector ${P}$ with the equivalence class of the matrix $R$.

Replacing the orthogonal groups in the above consideration with special orthogonal groups, one obtains the ``oriented" Grassmannian
\begin{gather}
\label{def.Gr+_1}
    \mathrm{Gr}^{+}(2,4) = \mathrm{SO}(4)/  [\mathrm{SO}(2)\times\mathrm{SO}(2)].
\end{gather}
Eq.~(\ref{def.Gr+_1}) can be geometrically interpreted as the projectors onto the occupied subspace, provided we distinguish the two projectors of the same subspace with different order of the occupied states. Once the occupied states are ordered, we can fix the order of unoccupied states so that the matrix $R$ has determinant 1. Hence, appears $\mathrm{SO}(4)$ in Eq.~(\ref{def.Gr+_1}). Note that the oriented projector is invariant under the $\mathrm{SO}(2)$ rotation of the occupied (unoccupied) band subspace.

In fact, $\mathrm{Gr}^{+}(2,4)$ is a covering space of $\mathrm{Gr}(2,4)$ with the covering map
\begin{gather}
    q: \mathrm{Gr}^{+}(2,4) \longrightarrow \mathrm{Gr}(2,4) \n 
    [A]^{+} \mapsto [A],\label{ftn.cover 1}
\end{gather}
which maps the oriented subspace represented by $A\in \mathrm{SO}(4)$ to the subspace represented by $A\in \mathrm{O}(4)$. The plus sign in the superscript indicates that the equivalence class is taken in the oriented Grassmannian. The map in Eq.~(\ref{ftn.cover 1}) is ``two-to-one" in the sense that the preimage $q^{-1}([A])$ of a point $[A]\in\mathrm{Gr}(2,4)$ is always a two-point set $\{[A]^{+}, [A']^{+}\}$. Here $A'\in\mathrm{SO}(4)$ is a special orthogonal matrix whose first two columns span the same subspace as that of $A$ but are oriented in the opposite order. Though there are infinitely many different choices for such $A'$, we shall canonically define $A'$ as the matrix obtained from $A$ by multiplying $-1$ to the second and fourth columns. That is,
\begin{gather}
	A'=\begin{pmatrix}
		\mathbf{a}_1 & -\mathbf{a}_2 & \mathbf{a}_3 & -\mathbf{a}_4
	\end{pmatrix},
	\label{def.A'}
\end{gather}
where $\mathbf{a}_{j}$ is the $j$-th column vector of $A$. 

\begin{figure}[t!]
	\includegraphics[width=8.5cm]{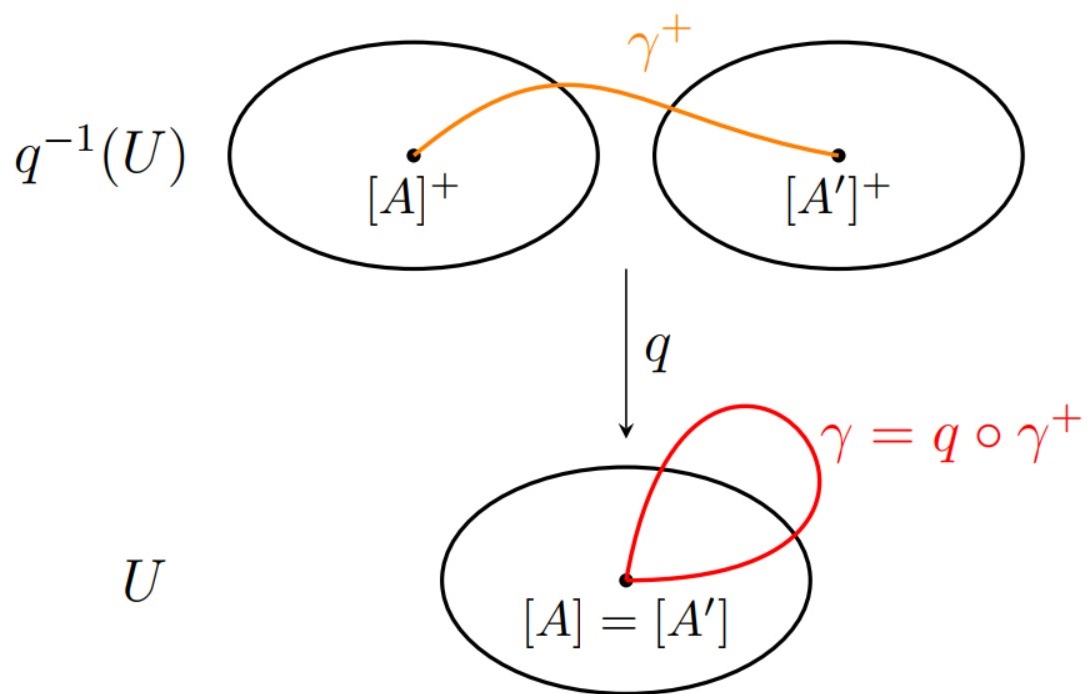}
	\caption{
		The structure of a two to one universal cover. To each point in the base space, denoted by $x=[A]=[A']$, correspond exactly two points $y_1=[A]^{+}$ and $y_2=[A']^{+}$ in the covering space. Every continuous path that connects these two points project onto a non-contractible loop in the base space, and the homotopy class of the projected loop does not depend on the original path.
	}
	\label{fig.cover}
\end{figure}

As will be shown in the next subsection, $\mathrm{Gr}^{+}(2,4)$ is homeomorphic to $\bb{S}^2\times\bb{S}^2$ with trivial fundamental group, thus being the universal cover of $\mathrm{Gr}(2,4)$. Having a two-to-one universal cover, a standard theorem of topology guarantees that $\mathrm{Gr}(2,4)$ has a non-contractible loop, unique up to homotopy, which becomes nullhomotopic when multiplied by itself: $\pi_1[\mathrm{Gr}(2,4)]=\bb{Z}_2$. Since we will rely on the relatively simple space $\mathrm{Gr}^{+}(2,4)\approx\bb{S}^2\times\bb{S}^2$ when classifying the topological phases of real 4-band insulators, it is worth noting the relation between the non-contractible loop of $\mathrm{Gr}(2,4)$ and the cover $\mathrm{Gr}^{+}(2,4)$ [see Fig.~\ref{fig.cover}]. For any point $[A]\in\mathrm{Gr}(2,4)$, there is a continuous path
\begin{gather}
    \gamma^{+}: [0,1] \longrightarrow \mathrm{Gr}^{+}(2,4)
\end{gather}
in $\mathrm{Gr}^{+}(2,4)$ connecting the two points in its preimage $q^{-1}([A])=\{[A]^{+}, [A']^{+}\}$. Then its projection $\gamma=q\circ\gamma^{+}$ onto $\mathrm{Gr}(2,4)$ represents the nonzero element $[\gamma]\in\pi_1[\mathrm{Gr}(2,4)]$.

In the viewpoint of homotopy groups, the difference between the base space and the covering space are all contained in the fundamental group: The homotopy groups of higher dimension coincide. 
In particular, $\pi_2[\mathrm{Gr}(2,4)]=\pi_2[\mathrm{Gr}^{+}(2,4)]=\pi_2[\bb{S}^2\times \bb{S}^2]=\bb{Z}\oplus\bb{Z}$ and $\pi_3[\mathrm{Gr}(2,4)]=\pi_3[\mathrm{Gr}^{+}(2,4)]=\pi_3[\bb{S}^2\times \bb{S}^2]=\bb{Z}\oplus\bb{Z}$. This fact will be crucial in the homotopy classification of 3D $\mathcal{PT}$-symmetric 4-band insulators as shown below.

\subsection{Modified classifying space and real Hopf invariant}
\label{Sec.modified}

As shown above, the classifying space $\mathrm{Gr}(2,4)$ is usually constructed as the quotient space $\mathrm{O}(4)/\mathrm{O}(2) \times \mathrm{O}(2)$ and interpreted as the set of projectors onto the occupied subspace. Though this construction applies to every non-interacting insulator with finite numbers of occupied and unoccupied bands, it is hard to understand geometrically how two integer invariants of $\pi_{2,3}[\mathrm{Gr}(2,4)]$ arise and how to compute them.

To improve on this, we used the double cover as a ``modified" classifying space $\mathrm{Gr}^{+}(2,4)$ and found $\pi_{n>1}[\mathrm{Gr}(2,4)]=\pi_{n>1}[\bb{S}^2\times\bb{S}^2]$. In the rest of this section, we will further this strategy and construct another ``classifying space" $\bb{S}^3\times\bb{S}^3$ enveloping both $\mathrm{Gr}(2,4)$ and $\mathrm{Gr}^{+}(2,4)$. The promised isomorphism $\mathrm{Gr}^{+}(2,4)\approx \bb{S}^2\times \bb{S}^2$ will be established en route.

What we will observe is that the projector ${P}$ can be determined from a pair of unit quaternions. This suggests the possibility of employing the set of pairs of unit quaternions to express arbitrary 4-band $\mathcal{PT}$-symmetric Hamiltonian. In turn, the set of unit quaternions can also be seen as the sphere $\bb{S}^3$ inside $\bb{R}^4$ under the identification $\bb{R}^4 \approx \bb{H}$ of real 4-vectors and quaternions. Thus using $\bb{S}^3\times\bb{S}^3$ as the enveloping classifying space, one can examine the relation between the Hopf insulator and the general 3D 4-band Hamiltonians with spinless $\mathcal{PT}$ symmetry at half filling. The latter, which will be shown to be the $\mathcal{PT}$-symmetric superposition of two Hopf insulators and its generalizations, will be examined in the framework of homotopy theory. When the lower-dimensional contribution to the homotopy (``weak topological invariants") is trivial, it will be called the Real Hopf Insulator (RHI).

To be precise, $\bb{S}^3 \times \bb{S}^3$ forms a Serre fibration~\cite{hatcher2005algebraic} over the Grassmannian and can be seen as the composition of two familiar fibrations
\begin{gather}
	\label{Serre}
	\bb{S}^3 \times \bb{S}^3 \xrightarrow[]{\;p_2\;} \mathrm{Gr}^{+}(2,4) \xrightarrow[]{\;p_1\;} \mathrm{Gr}(2,4).
\end{gather}
The first map is the product of two copies of Hopf fibration, which is commonly used to construct Hopf insulators, while the second is the 2-1 covering space over the Grassmannian. 

Being a Serre fibration, Eq.~(\ref{Serre}) enjoys the homotopy lifting property with respect to the 3-dimensional cube $ \mathrm{I}^3$: For each continuous map $ f: \mathrm{I}^3 \longrightarrow \mathrm{Gr}(2,4)$, there exists its (not necessarily unique) lifts such as
\begin{gather}
    f^{+}:  \mathrm{I}^3 \longrightarrow \mathrm{Gr}^{+}(2,4), \n
 \widetilde{f}:  \mathrm{I}^3 \longrightarrow \bb{S}^3 \times \bb{S}^3,
\end{gather}
where the continuous maps $f$, $f^{+}$, $\widetilde{f}$ satisfy $p_2\circ\widetilde{f} = f^{+}$ and $p_1\circ f^{+} = f$.

This lifting property is crucial for using the modified classifying spaces. One might try to lift the projector of a flattened Hamiltonian
\begin{gather}
    {P}: BZ=\bb{T}^3 \longrightarrow \mathrm{Gr}(2,4)
\end{gather}
up along the fibration, however, one cannot define such maps on BZ since the fibration in Eq.~(\ref{Serre}) does not have the lifting property with respect to the 3-torus. To bypass this difficulty, we interpret the projector as a map on the cube with periodic boundary conditions:
\begin{gather}
    f :  \mathrm{I}^3 \longrightarrow \mathrm{Gr}(2,4) \n 
    \mathbf{k} \mapsto {P}(\mathbf{k}),
\end{gather}
for which the fibration admits lifting. The lifted maps $f^{+},\widetilde{f}$ are then maps on the cube with certain boundary conditions as will be described [see Eq.~(\ref{eq.BC})].

To reach the aforementioned goal, we begin with the following homomorphism of groups
\begin{gather}
	g: \bb{S}^3\times \bb{S}^3 \longrightarrow \mathrm{SO}(4)
	\n 
	(z,w) \mapsto R_{z,w},
	\label{SO(4) qform}
\end{gather}
where $\mathrm{SO}(N)$ is the group of the $N\times N$ special orthogonal matrices. For a fixed pair of unit quaternions $z,w\in\bb{S}^3$, the matrix $R_{z,w}$ is defined by
\begin{gather}
	\label{def.Rzw}
	R_{z,w} \ket{x} = \ket{ \overline{z}xw }
	\qquad (x\in\mathbf{H}).
\end{gather}
We will use without proof the fact that $g$ is surjective, i.e. every $R\in\mathrm{SO}(4)$ arises as $R=R_{z,w}$ for some $z,w\in\bb{S}^3$, and that
\begin{gather}
    \text{ker}g = \{\pm(1,1)\} \approx \bb{Z}_2.
\end{gather}
It follows that by factoring out the kernel, we get the group isomorphism
\begin{gather}
	\overline{g}: \bb{S}^3\times \bb{S}^3/\{\pm(1,1)\} \longrightarrow \mathrm{SO}(4)
	\n 
	[(z,w)] \mapsto R_{z,w}.
	\label{ftn.iso0.5_1}
\end{gather}
Before proceeding, we shall describe the matrix $R_{z,w}$ by directly computing its column vectors. Observe that for any $4\times4$ matrix $R$, the four columns of $R$ are precisely the images of the elementary basis vectors
\begin{gather}
	R\begin{pmatrix} 1\\0\\0\\0\end{pmatrix},
	R\begin{pmatrix} 0\\1\\0\\0\end{pmatrix},
	R\begin{pmatrix} 0\\0\\1\\0\end{pmatrix},
	R\begin{pmatrix} 0\\0\\0\\1\end{pmatrix}.
\end{gather}
Accordingly, the column vectors of $R_{z,w}$ are
\begin{gather}
	\label{eq.eigenstates qform}
	\ket{\overline{z}w},\; \ket{\overline{z}\text{i}w},\; \ket{\overline{z}\text{j}w},\; \ket{\overline{z}\text{k}w}.
\end{gather}
Now we want to pass to the classifying space by factoring the right-hand side of Eq.~(\ref{ftn.iso0.5_1}) by the $\mathrm{SO}(2)\times\mathrm{SO}(2)$-action. We can transfer this action using the isomorphism (\ref{ftn.iso0.5_1}) and factor the left-hand side of Eq.~(\ref{ftn.iso0.5_1}) by this induced action. The resulting space, being isomorphic to $\mathrm{SO}(4)/\mathrm{SO}(2)\times\mathrm{SO}(2)$, provides an alternative description to the oriented Grassmannian.

To see things more vividly, here we write down explicitly how $\mathrm{SO}(2)$ acts on the occupied (unoccupied) space. Acting
\begin{gather}
	\begin{pmatrix} \cos\theta_1 & -\sin\theta_1 \\ \sin\theta_1 & \cos\theta_1
	\end{pmatrix}
	\oplus
	\begin{pmatrix} \cos\theta_2 & -\sin\theta_2 \\ \sin\theta_2 & \cos\theta_2
	\end{pmatrix}
	\in \mathrm{SO}(2)\times \mathrm{SO}(2)
	\label{SO(2)xSO(2) el}
\end{gather}
to a matrix $R\in\mathrm{SO}(4)$, $R$ moves to a new special orthogonal matrix whose column vectors are
\begin{gather}
	R\begin{pmatrix} \cos\theta_1\\\sin\theta_1\\0\\0\end{pmatrix},
	R\begin{pmatrix} -\sin\theta_1\\\cos\theta_1\\0\\0\end{pmatrix},
	R\begin{pmatrix} 0\\0\\\cos\theta_2\\\sin\theta_2\end{pmatrix},
	R\begin{pmatrix} 0\\0\\-\sin\theta_2\\\cos\theta_2\end{pmatrix}.
	\label{S1xS1.SO(4)-col}
\end{gather}
Now let us assume $R=R_{z,w}$. Then the vectors in Eq.~(\ref{S1xS1.SO(4)-col}) can be recast as
\begin{gather}
	\nonumber
	\cos\theta_1 \; \ket{\overline{z}w} + \sin\theta_1 \; \ket{\overline{z}\text{i}w} = \ket{ \overline{z}e^{\text{i}\theta_{1}}w },
	\\
	\nonumber
	-\sin\theta_1 \; \ket{\overline{z}w} + \cos\theta_1 \; \ket{\overline{z}\text{i}w} = \ket{\overline{z}e^{\text{i}\theta_{1}}\text{i}w},
	\\
	\nonumber
	\cos\theta_2 \; \ket{\overline{z}\text{j}w} + \sin\theta_2 \; \ket{\overline{z}\text{k}w} = \ket{ \overline{z} e^{\text{i}\theta_2} \text{j} w },
	\\
	-\sin\theta_2 \; \ket{\overline{z}\text{j}w} + \cos\theta_2 \; \ket{\overline{z}\text{k}w} = \ket{\overline{z} e^{\text{i}\theta_2} \text{k} w}.
	\label{S1xS1.SO(4)-q}
\end{gather}
The form of vectors in Eq.~(\ref{S1xS1.SO(4)-q}) indicates that the action could be understood as changing the quaternions $z,w\in\bb{S}^3$ instead of rotating the columns of $R_{z,w}$. From now on, let us identify the $\mathrm{SO}(2)\times \mathrm{SO}(2)$ matrix in Eq.~(\ref{SO(2)xSO(2) el}) with a pair of complex numbers $(e^{\text{i}\theta_1}, e^{\text{i}\theta_2})\in \bb{S}^1\times\bb{S}^1 $. We define a new action
\begin{gather}
	\big( \bb{S}^3\times\bb{S}^3 \big) \times
	\big( \bb{S}^1\times\bb{S}^1  ) \longrightarrow 
	\big( \bb{S}^3\times\bb{S}^3 \big)
	\n 
	\big( (z,w), (e^{\text{i}\phi_1},e^{\text{i}\phi_2}) \big) \mapsto (e^{\text{i}\phi_1}z , e^{\text{i}\phi_2}w ).
	\label{def.S1xS1.S3xS3}
\end{gather}
Notice that Eq.~(\ref{def.S1xS1.S3xS3}) maps $(-z,-w)$ to $(-e^{\text{i}\phi_1}z , -e^{\text{i}\phi_2} w )$, that is, it moves a pair of elements congruent under the $\bb{Z}_2$-action to another congruent pair. Hence, it induces an action on the quotient space $\bb{S}^3\times\bb{S}^3/\{\pm(1,1)\}$:
\begin{gather}
	\left( \frac{ \bb{S}^3\times\bb{S}^3 }{ \{\pm(1,1)\} } \right) \times
	\big(  \bb{S}^1\times\bb{S}^1  ) \longrightarrow 
	\left( \frac{ \bb{S}^3\times\bb{S}^3 }{ \{\pm(1,1)\} } \right)
	\n 
	\big( [(z,w)], (e^{\text{i}\phi_1},e^{\text{i}\phi_2}) \big) \mapsto [(e^{\text{i}\phi_1}z , e^{\text{i}\phi_2}w )].
	\label{def.S1xS1/pm.S3xS3}
\end{gather}
Direct computation shows that acting $(e^{\text{i}\theta_1}, e^{\text{i}\theta_2})$ on $R_{z,w}$ corresponds, via the isomorphism Eq.~(\ref{ftn.iso0.5_1}), to acting $(e^{\text{i}\phi_1},e^{\text{i}\phi_2})$ on $[(z,w)]$ if $\theta_{1,2},\phi_{1,2}$ are related by
\begin{gather}
	\theta_1(\phi_1,\phi_2) = \phi_2 - \phi_1,
	\nonumber
	\\
	\theta_2(\phi_1,\phi_2) = -(\phi_1 + \phi_2).
	\label{eq.action rel}
\end{gather}
Since any pair $(\theta_1, \theta_2)$ can be expressed as the difference and sum of a pair $(\phi_1, \phi_2)$, the orbit of Eq.~(\ref{def.S1xS1/pm.S3xS3}) is isomorphic to the orbit of Eq.~(\ref{SO(2)xSO(2) el}). [See SM for the brief review of some group theory including the definition of the orbit.]

It implies that after modding out the $ \bb{S}^1\times \bb{S}^1 $ actions on both sides of Eq.~(\ref{ftn.iso0.5_1}), it induces an isomorphism between orbit spaces:
\begin{gather}
    g^{+}: \frac{\bb{S}^3 \times \bb{S}^3} { \bb{S}^1 \times \bb{S}^1 } \longrightarrow \frac{\mathrm{SO}(4)}{ \mathrm{SO}(2)\times\mathrm{SO}(2)} \n
    [(z,w)] \mapsto [R_{z,w}].
 \label{ftn.iso1_1}
\end{gather}
In Eq.~(\ref{ftn.iso1_1}), we do not write $\big((\bb{S}^3\times\bb{S}^3)/\{\pm(1,1)\}\big) / (\bb{S}^1\times\bb{S}^1)$ since the orbits of the $\bb{S}^1\times\bb{S}^1$-action always contain that of the $\{\pm(1,1)\}$-action. To be precise, given $(z,w)\in\bb{S}^3\times\bb{S}^3$,
\begin{gather}
	\{\pm(z,w)\} \subset
	\{( e^{\text{i}\phi_1}z, e^{\text{i}\phi_2}w ) : \phi_1,\phi_2\in\bb{R} \}.
\end{gather}
It follows that
\begin{gather}
	\big((\bb{S}^3\times\bb{S}^3)/\{\pm(1,1)\}\big) / (\bb{S}^1\times\bb{S}^1)
	= (\bb{S}^3\times\bb{S}^3) / (\bb{S}^1\times\bb{S}^1).
\end{gather}
Since each multiplicand of $ \bb{S}^1\times\bb{S}^1 $ acts independently [see Eq.~(\ref{def.S1xS1.S3xS3})] on the respective component of $\bb{S}^3 \times \bb{S}^3$, the domain of the isomorphism Eq.~(\ref{ftn.iso1_1}) can also be written as the product of two independent orbit spaces:
\begin{gather}
    g^{+}: \frac{\bb{S}^3}{ \bb{S}^1 } \times \frac{\bb{S}^3}{ \bb{S}^1 } \longrightarrow \frac{\mathrm{SO}(4)}{\mathrm{SO}(2)\times \mathrm{SO}(2)} \n
    ([z], [w]) \mapsto [R_{z,w}].
\label{ftn.iso2}
\end{gather}
In Eq.~(\ref{ftn.iso2}), $\bb{S}^3/\bb{S}^1$ is the orbit space of the $\bb{S}^1$-action
\begin{gather}
	\bb{S}^3 \times \bb{S}^1 \longrightarrow \bb{S}^3
	\n 
	(z, e^{\text{i}\theta }) \mapsto e^{\text{i} \theta }z.
\end{gather}

Now, let us show that the orbit space $\bb{S}^3/ \bb{S}^1 $ can be identified with the sphere $\bb{S}^2$ precisely as in the Hopf fibration sequence in Eq.~(\ref{Hopf fibration_1}) by using the alternative form of the Hopf map $p:z \mapsto \overline{z}\text{i}z$ in Eq.~(\ref{eq.Hopfmap_Q}).
In Eq.~(\ref{ftn.iso1_1}), we formed the quotient space $\bb{S}^3/\bb{S}^1$ by factoring the orbits of the $\bb{S}^1-$action on $\bb{S}^3$ given by
\begin{gather}
	e^{\text{i}\theta}\cdot z = e^{\text{i}\theta} z.
	\label{def.S1.S3}
\end{gather}
It is easy to see that
\begin{gather}
    \overline{ e^{\text{i}\theta}z}\cdot \text{i} \cdot e^{\text{i}\theta}z
    = \overline{z} \underbrace{e^{-\text{i}\theta}\text{i}e^{i\theta}}_{=\text{i}}z
	= \overline{z}\text{i}z,
\end{gather} 
i.e. $e^{\text{i}\theta} z\in p^{-1}(\overline{z}\text{i}z)$. Conversely, suppose $w\in p^{-1}(\overline{z} \text{i} z)$, i.e. $\overline{z}\text{i}z=\overline{w}\text{i}w$ for some $z,w\in \bb{S}^3$. We want to show $z\overline{w}=e^{\text{i}\theta}$ for some $\theta$, since then $w=e^{-\text{i}\theta}z$ and $w$ belongs to the orbit of $z$ under the action Eq.~(\ref{def.S1.S3}). Note that
\begin{alignat}{2}
	& \overline{z}\text{i}z &&= \overline{w}\text{i}w \\
	\Rightarrow\quad
	& z(\overline{z}\text{i}z)\overline{z} &&= z(\overline{w}\text{i}w)\overline{z} \\
	\Rightarrow\quad
	& \text{i} &&= (z\overline{w})\text{i}(w\overline{z}) \\
	\Rightarrow\quad
	& \text{i}(z\overline{w}) &&= (z\overline{w})\text{i}.
\end{alignat} 
Since $z\overline{w}$ has unit length and commutes with $\text{i}$, it must be of the form
\begin{gather}
    z\overline{w} = a+b\text{i} \quad(a^2+b^2=1),
\end{gather} 
i.e. $z\overline{w}=e^{\text{i}\theta} \in \bb{S}^1$ for some $\theta\in\bb{R}$.
\begin{gather}
	p^{-1}(\overline{z}\text{i}z) = \{e^{\text{i}\theta}z:\theta\in\bb{R}\} \approx \bb{S}^1.
\end{gather}
In conclusion, we can represent the equivalence class $[z]$ in Eq.~(\ref{ftn.iso2}) by using $\overline{z}\text{i}z$. This establishes an isomorphism between $\bb{S}^2\times\bb{S}^2$ and the oriented classifying space given by
\begin{gather}
    g^{+}: \bb{S}^2\times\bb{S}^2 \longrightarrow 
	\frac{\mathrm{SO}(4)}{ \mathrm{SO}(2)\times \mathrm{SO}(2) }
	\n 
	(\overline{z}\text{i}z, \overline{w}\text{i}w) \mapsto [R_{z,w}].
\label{ftn.iso3}
\end{gather}

\section{Real Hopf insulator}

The conclusion we have obtained so far is twofold. First, we have established the isomorphism
\begin{gather}
	\label{isomorphism: conclusion}
	\bb{S}^2\times \bb{S}^2 \xrightarrow[]{\;\approx\;} \mathrm{Gr}^{+}(2,4),
\end{gather}
which enables us to classify the band topology of any $4\times4$ real Hamiltonian $H_{R}^{2,2}(\mathbf{k})$ with two occupied and two unoccupied bands by $\pi_3(\bb{S}^2\times \bb{S}^2)=\bb{Z}\oplus\bb{Z}$ indices. 
Second, we have shown that each $\bb{S}^2$ factor in Eq.~(\ref{isomorphism: conclusion}) can be regarded as the base space of the Hopf fibration in Eq.~(\ref{Hopf fibration_1}). 
This indicates the topological classes of $H_{R}^{2,2}(\mathbf{k})$ can be classified by two integer Hopf invariants $(\chi_z,~\chi_w)$. 
The topological insulator described by $H_{R}^{2,2}(\mathbf{k})$ carrying nontrivial Hopf invariant pairs can be called a real Hopf insulator (RHI).

\subsection{Definition and Construction of Real Hopf Insulator}
\label{subsec.construction}
What have been shown so far is that for any flattened Hamiltonian $\overline{H}_{R}^{2,2}(\mathbf{k})$, there are two lifts of the projector $f^{+}$ and $\widetilde{f}$ that map into the modified classifying spaces. From now on, we denote the lifted projectors by
\begin{gather}
    \underline{P}^{+} :  \mathrm{I}^3 \longrightarrow \mathrm{Gr}^{+}(2,4), \n 
    \widetilde{\underline{P}} : \mathrm{I}^3 \longrightarrow \bb{S}^3\times\bb{S}^3,
\end{gather}
and call them `oriented' and `enveloping' projectors, respectively.
The underlines in $\underline{P}^{+},\widetilde{\underline{P}}$ signify that the domain of projectors changed from torus to cube, while the symbols $P^{+},\widetilde{P}$ are reserved for when the domain remains to be a torus; see Eqs.~(\ref{ftn.oriented_proj}), (\ref{ftn.enveloping_proj}).

Let us comment on the boundary conditions of the projectors. Since the genuine projector ${P}(\mathbf{k})$ is a function on $\bb{T}^3$~(equivalently, a periodic function on $ \mathrm{I}^3$), the lifted projectors must satisfy certain boundary conditions. In view of the fibration structure between genuine and modified classifying spaces,
\begin{gather}
    \bb{S}^3\times\bb{S}^3 \xrightarrow[]{p_2} \bb{S}^2\times\bb{S}^2 \xrightarrow[]{p_1} \mathrm{Gr}(2,4) \n
    (z,w) \mapsto (\overline{z}\text{i}z, \overline{w}\text{i}w) \mapsto [R_{z,w}],
\label{ftn.RHI classifying}
\end{gather}
the boundary condition for $\widetilde{\underline{P}}=(z(\mathbf{k}),w(\mathbf{k}))\equiv (z,w)(\mathbf{k})$ can be determined as follows. Since the (unoriented) classifying space is $\mathrm{Gr}(2,4)=\mathrm{Gr}^{+}(2,4)/\bb{Z}_2 \approx \bb{S}^2\times\bb{S}^2/\{\pm(1,1)\}$, $(z,w)(\mathbf{k})$ can be any $\bb{S}^3\times\bb{S}^3$-valued functions on $ \mathrm{I}^3$ subject to the boundary condition
\begin{gather}
    (\overline{z}\text{i}z, \overline{w}\text{i}w) (\mathbf{k}) = \pm (\overline{z}\text{i}z, \overline{w}\text{i}w) (\mathbf{k}+2\pi\mathbf{e}_j) \n
	(\mathbf{e}_j \text{ is the unit vector along $k_j$ axis, } j=1,2,3),
\label{eq.BC}
\end{gather}
The plus (minus) sign in Eq.~(\ref{eq.BC}) corresponds to the contractible (non-contractible) loop along $k_j$ in $\pi_1[\mathrm{Gr}(2,4)]$. It is easy to see that the periodic boundary condition
\begin{gather}
	(\overline{z}\text{i}z, \overline{w}\text{i}w)(\mathbf{k})
	= (\overline{z}\text{i}z, \overline{w}\text{i}w) (\mathbf{k}+2\pi\mathbf{e}_j) \n
	(\mathbf{e}_j \text{ are unit vectors }, j=1,2,3).
\label{eq.BC+}
\end{gather}
is equivalent to saying that the oriented projector can be written as a function on the torus,
\begin{gather}
    {P}^{+} : \bb{T}^3\longrightarrow \bb{S}^2\times\bb{S}^2 \n 
    \mathbf{k} \mapsto (\overline{z}\text{i}z, \overline{w}\text{i}w) (\mathbf{k}),
    \label{ftn.oriented_proj}
\end{gather}
instead of just a function on $ \mathrm{I}^3$. It is explained in Appendix~\ref{Sec.BC} that this is possible if and only if the Berry phases along three axes all vanish:
\begin{gather}
    \int_{\bb{T}^1} \Tr{A}(\mathbf{k}) dk_j = 0 \quad\text{mod }2\pi\quad(j=1,2,3).
\label{eq.triviality 1}    
\end{gather}
On the other hand, the anti-periodic boundary condition (minus sign in Eq.~(\ref{eq.BC})) along any direction $k_j$ is equivalent to the nontrivial Berry phase 
\begin{gather}
    \int_{\bb{T}^1} \Tr{A}(\mathbf{k}) dk_j = \pi \quad\text{mod }2\pi.
\end{gather}
We focus our attention to those Hamiltonians with periodic oriented projectors, i.e. all loops along the three independent directions in $BZ=\bb{T}^3$ being contractible in the classifying space $\mathrm{Gr}(2,4)$; In this case, Eq.~(\ref{eq.BC+}) implies
\begin{gather}
    z(k_1,k_2,2\pi)= e^{\text{i}\theta_3(k_1,k_2)}  z(k_1,k_2,0),\n 
    w(k_1,k_2,2\pi)= e^{\text{i}\phi_3(k_1,k_2)}  w(k_1,k_2,0), \n 
    \text{for some } \theta_3, \phi_3 : \mathrm{I}^2\rightarrow\bb{R}, \n 
    \text{similarly for cyclic permutations of }1,2,3.
\end{gather}
The functions $\theta_{j},\phi_{j}\; (j=1,2,3)$ can be set to constant zero, that is, the enveloping projector can be set (up to homotopy) as a periodic function
\begin{gather}
    \widetilde{{P}} : \bb{T}^3 \longrightarrow \bb{S}^3\times\bb{S}^3 \n 
    \mathbf{k} \mapsto (z,w)(\mathbf{k})
    \label{ftn.enveloping_proj}
\end{gather}
if and only if the Euler class associated to the occupied (unoccupied) bands vanishes on all three faces [see Appendix~\ref{Sec.BC} and also~\ref{Subsec.2d}]:
\begin{gather}
    \int_{\bb{T}^2}\text{Eu}^v(\mathbf{k}) dk_1dk_2 = \int_{\bb{T}^2}\text{Eu}^c(\mathbf{k}) dk_1dk_2 = 0, \n 
    \text{as well as for cyclic permutations of }1,2,3.
    \label{eq.triviality 2}
\end{gather}
The Euler form $\text{Eu}^{v,c}(\mathbf{k})$, which can be defined for $\mathcal{PT}$-symmetric insulators in spatial dimension two, will be defined in Eqs.~(\ref{Euler connection})-(\ref{Euler form}). 

It is easy to observe that the condition in Eq.~(\ref{eq.triviality 2}) implies Eq.~(\ref{eq.triviality 1}), since triviality on faces guarantees that on edges. This is consistent with the fact that the Euler invariant is well-defined when the Berry phase is vanishing.

We define the Real Hopf Insulators (RHIs) to be the class of 4-band spinless $\mathcal{PT}$-symmetric Hamiltonians at half filling, with additional constraint in Eq.~(\ref{eq.triviality 2}). A RHI can be specified (up to homotopy) by either one of the two ways: One (the level of genuine classifying space) is by the periodic $\bb{R}^4$-valued real wave functions $\{\ket{u^v_1(\mathbf{k})}, \ket{u^v_2(\mathbf{k})}\}$ and $\{\ket{u^c_1(\mathbf{k})}, \ket{u^c_2(\mathbf{k})}\}$ on $BZ=\bb{T}^3$ that span the occupied and unoccupied subspaces, respectively; the other (the level of enveloping classifying space) is by two unit quaternion-valued periodic functions $z(\mathbf{k}), w(\mathbf{k})$ on $BZ=\bb{T}^3$. In the second case, using the notation in Eq.~(\ref{ftn.H=R4}), the wave functions are determined by 
\begin{gather}
\label{valence states qform_1}
    \ket{u^{v}_1} = \ket{\overline{z}\text{j}w}, \quad
    \ket{u^{v}_2} = \ket{\overline{z}\text{k}w},
\end{gather}
for the occupied bands, and
\begin{gather}
	\label{conduction states qform_1}
	\ket{u^{c}_1} = \ket{\overline{z}w},\quad
	\ket{u^{c}_2} = \ket{\overline{z}\text{i}w},
\end{gather}
for the unoccupied bands. In both ways of defining RHI, analyticity of the wave functions guarantees the triviality condition in Eq.~(\ref{eq.triviality 2}).

A concrete model of RHI with two integer invariants $(\chi_z,\chi_w)$ can be constructed as follows. 
First, we can choose $z(\mathbf{k}), w(\mathbf{k})$ as the unit quaternions corresponding to the occupied states of the conventional Hopf insulator, described by the Moore-Ran-Wen model in Eq.~(\ref{def.MRW_Hopf}), with the Hopf indices $\chi_z, \chi_w$, respectively. We then assign $\{\ket{\overline{z}\text{j}w}, \ket{\overline{z}\text{k}w}\}$ as the occupied eigenstates of $\overline{H}_{R}^{2,2}(\mathbf{k})$ with the energy eigenvalue $-1$ while $\{\ket{\overline{z}w},\ket{\overline{z}\text{i}w}\}$ as the unoccupied states with the eigenvalue +1. 
The occupied (unoccupied) space is invariant under the action
\begin{gather}
    (z,w)\mapsto (e^{\text{i}\theta_1}z, e^{\text{i}\theta_2}w),
\end{gather}
which can be understood from Eq.~(\ref{S1xS1.SO(4)-q}).
Then the matrix form of the flattened RHI Hamiltonian is given by
\begin{gather}
\label{eq.RHI_spectral}
    \overline{H}_{R}^{2,2}(\mathbf{k})
    = R_{z,w}
    \begin{pmatrix}
        1& & & \\
        &1& & \\
        & &-1& \\
        & & &-1
    \end{pmatrix}
    R_{z,w}^{\top}
\end{gather}
where $R_{z,w}$ is defined in Eq.~(\ref{def.Rzw}).
A more general unflattened Hamiltonian $H_{R}^{2,2}(\mathbf{k})$ can be obtained from $\overline{H}_{R}^{2,2}(\mathbf{k})$ by allowing the momentum-dependent variation of the energy eigenvalues while keeping the finite energy gap between the occupied and unoccupied bands.

The Moore-Ran-Wen RHI, defined in Eq.~(\ref{eq.RHI_spectral}), is especially useful when constructing RHI models with additional symmetries. For instance, Appendix~\ref{subsec.P} describes the models having the inversion ($\mathcal{P}$) and time reversal ($\mathcal{T}$) symmetries separately.

At this point, let us clarify that 2-band Hopf insulator and 4-band RHI have relations stronger than merely sharing a mathematical construction using the Hopf map. In fact, a $\mathcal{PT}$-symmetric superposition of a pair of Hopf insulators realizes a subset of RHIs. To be precise, let us consider the RHI with $(\chi_z,\chi_w)=(0,n)$ for integer $n$. Setting $z\equiv 1$, the corresponding Hamiltonian reads
\begin{gather}
    \overline{H}_{R}^{2,2}(\mathbf{k}) = 
    \begin{pmatrix}
    u^1&0&u^3&-u^2\\
    0&u^1&u^2&u^3\\
    u^3&u^2&-u^1&0\\
    -u^2&u^3&0&-u^1
    \end{pmatrix},
\end{gather}
where $(u^1,u^2,u^3)(\mathbf{k})$ is determined by
\begin{gather}
    u^1\text{i} + u^2\text{j} + u^3\text{k} = (\overline{w}\text{i}w)(\mathbf{k}).
\end{gather}
Under a basis change,
\begin{gather}
\label{eq.RHI_diag}
    U \overline{H}_{R}^{2,2}(\mathbf{k}) U^{\dagger} = 
    \begin{pmatrix}
    H_C^{*} & O \\
    O & H_C
    \end{pmatrix}, \\ 
    H_C = 
    \begin{pmatrix}
    u^1 & u^3+iu^2 \\
    u^3-iu^2 & -u^1
    \end{pmatrix},
\end{gather} 
where $H_C$ is the Hamiltonian of a Hopf insulator associated to the function $(z^{\dagger}\bm{\sigma}z)(\mathbf{k}) = [v^1(\mathbf{k}),v^2(\mathbf{k}),v^3(\mathbf{k})] = [u^3(\mathbf{k}),-u^2(\mathbf{k}),u^1(\mathbf{k})]$ [compare Eqs.(\ref{eq.Hopf_v}) and (\ref{eq.Hopf_u})], $H_C^{*}$ its complex conjugate, and $O$ the zero matrix. The relevant unitary transformation is given by
\begin{gather}
    U = 
    \begin{pmatrix}
    c&-c^{*}&0&0\\
    0&0&c&-c^{*}\\
    c&c^{*}&0&0\\
    0&0&c&c^{*}
    \end{pmatrix}, 
    c = -\frac{1}{2}+\frac{1}{2}i.
\end{gather}
It turns out that $H_C^{*}$ is also a Hamiltonian of a Hopf insulator with the same Hopf invariant $\chi$ as $H_C$.

Hence, every RHI with the first invariant trivial, $\chi_z=0$, is the $\mathcal{PT}$-symmetric superposition of two independent copies of Hopf insulator. As one turns on the interaction between the two, without breaking $\mathcal{PT}$, nontrivial terms appear in the off-diagonal entries in Eq.~(\ref{eq.RHI_diag}). The energy gap should be closed during the process, passing to another homotopy class $(\chi_z,\chi_w)=(m,n)$ of RHI.

\subsection{The bulk invariants of RHI}
\label{Subsec.bulk invariants}
Let us explain how one can compute the two integer topological invariants of $H_{R}^{2,2}(\mathbf{k})$ related to the third homotopy group
\begin{gather}
\pi_3[\mathrm{Gr}(2,4)]= 
	\pi_3[\bb{S}^2\times\bb{S}^2]=\bb{Z}\oplus\bb{Z}.
\end{gather}
Here we assume that the 1D and 2D topological invariant of $H_{R}^{2,2}(\mathbf{k})$, that is, the Berry phase and the Euler invariant, respectively, are trivial because the two Hopf invariants are well-defined only under such a condition. The influence of the 1D and 2D invariants is further discussed in Appendix~\ref{Sec.Class}.

For a given flattened Hamiltonian $\overline{H}_{R}^{2,2}(\mathbf{k})$, one can find a map,
\begin{gather}
    \bb{T}^3 \longrightarrow \bb{S}^3\times\bb{S}^3 \n 
    \mathbf{k} \mapsto (z,w)(\mathbf{k}).
\end{gather}
The two integer invariants are the degree of the maps 
\begin{gather}
	z(\mathbf{k}), w(\mathbf{k}) : \bb{T}^3 \longrightarrow \bb{S}^3 .
\end{gather}

These degrees can be quantified by the following manner: First, we define the connection 1-forms on the BZ
\begin{gather}
    \label{def.zw connection}
    a_z = \Re[ -\text{i} \cdot z~\dd \overline{z} ], \quad
    a_w = \Re[ -\text{i} \cdot w~\dd \overline{w} ],
\end{gather}
where $\dd$ is the differential operator defined by
\begin{gather}
    \dd f
    = \frac{\d f}{\d k_x} dk_x +
    \frac{\d f}{\d k_y} dk_y +
    \frac{\d f}{\d k_z} dk_z.
\end{gather}
Differentiating $a_z$ and $a_w$, we get
\begin{gather}
	\label{def.zw curvature}
	f_z = \dd a_z, \quad
	f_w = \dd a_w.
\end{gather}
Then the integers in $\pi_3[\mathrm{Gr}(2,4)]=\bb{Z}\oplus\bb{Z}$ are computed in the same way as the Hopf invariant:
\begin{align}
	& (\chi_z, \chi_w) \n
	& =  \left(
	\frac{-1}{4\pi^2} \int_{BZ} a_z \wedge f_z,~
	\frac{-1}{4\pi^2} \int_{BZ} a_w \wedge f_w \right) 
	\in \bb{Z} \oplus \bb{Z}.
\end{align}
In the following two subsections, we provide alternative definitions, equivalent to the one above, of RHI invariants using the eigenstates of the Hamiltonian, instead of the abstract functions $z,~w$. 

\subsection{ RHI invariants of unflattened Hamiltonian }
\label{Subsec.RHI inv eig}
To develop methods to compute the 3D invariants of a general unflattened Hamiltonian $H_{R}^{2,2}(\mathbf{k})$, one needs the relation between $\overline{z}\text{i}z, \overline{w}\text{i}w$ and the occupied/unoccupied wave functions.  
For example, one can see from Eq.~(\ref{valence states qform_1}) that
\begin{gather}
    u^{v}_2 \overline{u^{v}_1} = \overline{z}\text{i}z, \n
    - \overline{u^{v}_1} u^{v}_2 = \overline{w}\text{i}w.
\label{RHI invariant ingredient 1}
\end{gather}
Then, by combining two orthogonal occupied eigenstates $u^{v}_1(\mathbf{k})$ and $u^{v}_2(\mathbf{k})$,
we can define two functions 
\begin{gather}
    \mathbf{v}_z : \bb{T}^3 \longrightarrow \bb{S}^2 \n
    \mathbf{k} \mapsto \mathbf{v}_z(\mathbf{k}) = u^{v}_2(\mathbf{k}) \overline{u^{v}_1}(\mathbf{k}),
 \label{ftn.Gauss map 1} \\
 \mbox{} \n
    \mathbf{v}_w : \bb{T}^3 \longrightarrow \bb{S}^2 \n
    \mathbf{k} \mapsto \mathbf{v}_w(\mathbf{k}) = -\overline{u^{v}_1}(\mathbf{k}) u^{v}_2(\mathbf{k}).
 \label{ftn.Gauss map 2}
\end{gather}
Then by applying the procedures described in Eq.~(\ref{eq.Gauss curvature})-(\ref{def.Hopf Inv vec}) to $\mathbf{v}_z(\mathbf{k})$
and $\mathbf{v}_w(\mathbf{k})$, one can determine the corresponding connection and curvature tensors, from which the bulk invariants of RHI can be computed. The resulting integer invariants $(\chi_z,\chi_w)$ are independent of the choice of orthogonal basis $u^{v}_1(\mathbf{k})$ and $u^{v}_2(\mathbf{k})$. In Appendix~\ref{Sec.invariance}, we prove the invariance of $(\chi_z,\chi_w)$ under the rotation
\begin{gather}
	\nonumber
	u^{v}_1 \mapsto \cos\theta\; u^{v}_1 - \sin\theta\; u^{v}_2
	\\
	\label{RHI eigenstates rotation}
	u^{v}_2 \mapsto \sin\theta\; u^{v}_1 + \cos\theta\; u^{v}_2
\end{gather}
and the reflection
\begin{gather}
	\nonumber
	u^{v}_1 \mapsto u^{v}_2
	\\
	\label{RHI eigenstates reflection}
	u^{v}_2 \mapsto u^{v}_1.
\end{gather}
Similar relations can also be found by using the unoccupied wave functions such as
\begin{gather}
	\nonumber
	u^{c}_2 \overline{u^{c}_1} = \overline{z}\text{i}z,
	\\
	\label{RHI index from unoccupied band}
	\overline{u^{c}_1} u^{c}_2 =\overline{w}\text{i}w.
\end{gather}

\subsection{RHI invariants in terms of Euler connections}
\label{Subsec.RHI inv Eu}
More importantly, we can define the bulk invariants from the conduction (valence) band Euler connections. This will reveal a characteristic feature of RHI: The integral formula for the bulk invariants necessarily include the Euler connections of both conduction and valence bands,
\begin{gather}
    \label{Euler connection}
    \text{a}^c = \bra{u^c_2} \dd \ket{u^c_1},
    \quad
    \text{a}^v = \bra{u^v_2} \dd \ket{u^v_1},
\end{gather}
and the corresponding curvatures
\begin{gather}
    \label{Euler form}
    \text{Eu}^c = \dd \text{a}^c,
    \quad
    \text{Eu}^v = \dd \text{a}^v.
\end{gather}
In Eq.~(\ref{Euler connection}), the wave functions $\ket{u}_{1,2}^{v,c}$ are set to be real vectors, which is enabled by the $\mathcal{PT}$ symmetry. For such a real basis, the 2 by 2 connection matrix for the occupied (unoccupied) subspace $a_{ij}^{v,c}\equiv\bra{u_{j}^{v,c}}\dd\ket{u_{i}^{v,c}}$ becomes real antisymmetric. Hence, the connection matrix is determined by its off-diagonal component $a_{12}^{v,c}=\text{a}^{v,c}$, which is the definition of Euler connection~\cite{ahn2019failure,ahn2018linking}. 

Using Eq.~(\ref{valence states qform_1}) and Eq.~(\ref{conduction states qform_1}), we can compute Eq.~(\ref{Euler connection}) in terms of $z, w$:
\begin{align}
    \text{a}^c
    & = \bra{ \overline{z}\text{i}w} \dd \ket{\overline{z}w}
    \n
    & = \text{Re}\bigg[ \overline{w}(-\text{i})z \big( \dd  \overline{z} \cdot w + \overline{z} \dd w \big) \bigg]
    \n
	& = \text{Re}\bigg[ -\text{i}\cdot z \dd  \overline{z} + \overline{w}(-\text{i}) \dd w \bigg]
    \n
	& = \text{Re}\bigg[ -\text{i} \cdot z  \dd  \overline{z} + \text{i} \cdot w \dd \overline{w} \bigg]
    \n
    & = a_z - a_w,
    \label{aczw}
\end{align}
where the first equality is the definition. In the second and third equalities, we use Eq.~(\ref{R4 inner product quaternion form}) and Eq.~(\ref{eq.Re conj}), respectively. The fourth again follows from Eq.~(\ref{R4 inner product quaternion form}). Analogously, we have
\begin{gather}
	\text{a}^v
	= a_z + a_w.
	\label{avzw}
\end{gather}
Now it directly follows that
\begin{align}
	&  -\frac{1}{8\pi^2}  \int_{BZ} \text{a}^c \wedge \text{Eu}^c + \text{a}^v \wedge \text{Eu}^v
	= \chi_z + \chi_w,
	\n
	&  -\frac{1}{8\pi^2}  \int_{BZ} \text{a}^c \wedge \text{Eu}^v + \text{a}^v \wedge \text{Eu}^c
	= \chi_z - \chi_w.
	\label{eq.forms & invariants 1}
\end{align}   
Or equivalently,
\begin{align}
	&  -\frac{1}{16\pi^2}  \int_{BZ} \left( \text{a}^c + \text{a}^v \right)\wedge \left(\text{Eu}^c + \text{Eu}^v \right)
	= \chi_z ,
	\n
	&  -\frac{1}{16\pi^2}  \int_{BZ} \left( \text{a}^c - \text{a}^v \right) \wedge \left( \text{Eu}^c - \text{Eu}^v \right)
	= \chi_w .
\label{eq.forms & invariants 2}
\end{align}
It is worth noting that the 3D bulk invariants ($\chi_z$, $\chi_w$) of RHI are defined in terms of both occupied and unoccupied bands. One may try to find other invariants, similar to above ones, using only the Euler connection and curvature of occupied bands. However, as shown in Appendix~\ref{Sec.uniqueness}, any attempt to build such new invariants ends up with an expression which is either gauge non-invariant or non-quantized. This is in sharp contrast with all the other topological invariants including the Chern number, Euler number, Hopf invariant, which can be computed by using only the occupied subspace.

This fact has a consequence on the bulk-boundary correspondence of RHI. Analogous to the other integral formulae such as Berry phase in 1D and Chern class in 2D, which contain physical information of the boundary charge and edge current, the integral formulae for the RHI bulk invariants allow interpretation as the Berry curvature polarization, which would leave a mark on the boundary of a finite-size sample. More precisely, we will show later that the first equation of Eq.~(\ref{eq.forms & invariants 1}) is directly proportional to the surface Chern number~\cite{vanderbilt2017BBCofCSA} of the total bands, including both occupied and unoccupied bands. This fact is intimately connected to the appearance of the connection and curvature of both the occupied and unoccupied states in the definition of quantized topological invariants, which is a direct manifestation of the delicate nature of the RHI band topology. 

As the final remark for this section, we discuss the apparent contradiction between Eq.~(\ref{RHI invariant ingredient 1}) and the uniqueness statement of Eq.~(\ref{eq.forms & invariants 2}). The fact that Eq.~(\ref{RHI invariant ingredient 1}) provides a way to express $\chi_z$, say, in terms of occupied states only may render the uniqueness of Eq.~(\ref{eq.forms & invariants 2}) invalid. However, the definition of $\chi_z$ from Eq.~(\ref{RHI invariant ingredient 1}) requires one to solve the magneto-static equations in Eqs.~(\ref{eq.magnetostat 1})-(\ref{eq.magnetostat 2}), whose solution cannot be expressed in terms of occupied state Euler connection. In fact, the solution is
\begin{gather}
    a_z = \frac{\text{a}^{c} + \text{a}^{v}}{2}.
\end{gather}
This inevitably enforces the use of both occupied and unoccupied bands when computing the bulk invariants of RHI in terms of Euler connection.

\section{Bulk-boundary correspondence}
\label{sec.bulk-boundary}
\subsection{Review on the bulk-boundary correspondence of the Hopf insulator}
Let us first review the bulk-boundary correspondence of the Hopf insulator which has non-trivial surface Chern numbers at the boundary~\cite{alexandradinata2021HIsurface}.
Generally, the surface state of a topological insulator is described by using its occupied states only.
On the other hand, in the Hopf insulator, the unoccupied states as well as the occupied states are equally important to understand its surface states as discussed below.

More specifically, let us consider a slab geometry of a Hopf insulator with $N_z$ layers stacked in the $z$-direction while keeping the periodic boundary condition in the $x,~y$-directions with the corresponding momentum denoted by $\boldsymbol\kappa=(\kappa_{x},\kappa_{y})$.
When the system is periodic along all three directions, the Wannier states of the occupied and unoccupied bands are well-defined because there is a finite bulk energy gap $E_{g,\mathrm{bulk}}$ and the Chern number of the system is zero. On the other hand, in a slab geometry, possible band crossings between the surface occupied and unoccupied bands can cause some difficulty in defining the corresponding Wannier states. However, in the case of two-band Hopf insulator, it was shown that the band crossing between surface states can be removed by applying appropriate surface perturbations~\cite{fidkowski2011WilsonBBC,alexandradinata2021HIsurface}.

For an eigenstate $|u_{\mathrm{slab}}\rangle$ of the slab Hamiltonian, one can define the expectation value of the $z$-coordinate as  $\langle u_{\mathrm{slab}}|Z_{\mathrm{slab}}|u_{\mathrm{slab}}\rangle$ where $Z_{\mathrm{slab}}$ is the $z$-directional position operator.
In general, there is a unitary transformation $U\in \mathrm{U}(N_{z})$ for the occupied states $\{|u_{n,\mathrm{slab}}^{v}\rangle:~n=1,\cdots ,N_{z}\}$ such that the transformed occupied states $|\tilde{u}_{n,\mathrm{slab}}^{v}\rangle=\sum_{m}U_{nm}|u_{m,\mathrm{slab}}^{v}\rangle$ are localized in the $z$-direction.
The transformed occupied state $|\tilde{u}_{n,\mathrm{slab}}^{v}\rangle$ and its $z$-directional position $z_{n,\mathrm{slab}}^{v}$ are equivalent to the eigenstate and the eigenvalue of $P_{\mathrm{slab}}Z_{\mathrm{slab}}P_{\mathrm{slab}}$ where $P_{\mathrm{slab}}$ is the projection operator onto the occupied states $\{|u_{n,\mathrm{slab}}^{v}\rangle\}$~\cite{Vanderbilt1997maximalWan}.
Similarly, using the projection operator $Q_{\mathrm{slab}}$ onto the unoccupied states, one can define the $z$-directionally localized unoccupied states $|\tilde{u}_{n,\mathrm{slab}}^{c}\rangle$ and their $z$-positions $z_{n,\mathrm{slab}}^{c}$
as the eigenstates and eigenvalues of $Q_{\mathrm{slab}}Z_{\mathrm{slab}}Q_{\mathrm{slab}}$.
Without loss of generality, the indices of $|\tilde{u}_{n,\mathrm{slab}}^{v}\rangle$ and $z_{n,\mathrm{slab}}^{v}$ (resp. $|\tilde{u}_{n,\mathrm{slab}}^{c}\rangle$ and $z_{n,\mathrm{slab}}^{c}$) can be chosen as $z_{1,\mathrm{slab}}^{v}\leq\cdots \leq z_{N_{z},\mathrm{slab}}^{v}$ (resp. $z_{1,\mathrm{slab}}^{c}\leq\cdots \leq z_{N_{z},\mathrm{slab}}^{c}$).
By plotting $\{z_{n,\mathrm{slab}}^{v(c)}:n=1,\cdots ,N_{z}\}$ as a function of $\boldsymbol\kappa$, we obtain the Wannier sheets stacked along the $z$-direction.
Note that $z_{1,\mathrm{slab}}^{v(c)}$ corresponds to the bottom sheet and $z_{N_{z},\mathrm{slab}}^{v(c)}$ corresponds to the top sheet.
Also, $|\tilde{u}^{v}_{n,\mathrm{slab}}\rangle , |\tilde{u}^{c}_{n,\mathrm{slab}}\rangle$ are called by the slab hybrid Wannier functions(HWF) of the occupied and unoccupied states~\cite{vanderbilt2017BBCofCSA}.

Let us first consider the occupied states.
Suppose that, for some $n_{0}$, $z_{n_{0},\mathrm{slab}}^{v}$ is decoupled from adjacent Wannier sheets, i.e., $z_{n_{0}-1,\mathrm{slab}}^{v}(\boldsymbol\kappa)<z_{n_{0},\mathrm{slab}}^{v}(\boldsymbol\kappa)<z_{n_{0}+1,\mathrm{slab}}^{v}(\boldsymbol\kappa)$ for all $\boldsymbol\kappa\in rBZ$ where $rBZ$ indicates the reduced Brillouin zone for in-plane momenta $\boldsymbol\kappa$.
In this case, the Chern number of $|\tilde{u}_{n_{0},\mathrm{slab}}^{v}\rangle$ can be computed by integrating the abelian Berry curvature $\frac{i}{2\pi}\left(\langle \partial_{\kappa_{x}} \tilde{u}_{n_{0},\mathrm{slab}}^{v}|\partial_{\kappa_{y}} \tilde{u}_{n_{0},\mathrm{slab}}^{v} \rangle - \langle \partial_{\kappa_{y}} \tilde{u}_{n_{0},\mathrm{slab}}^{v}|\partial_{\kappa_{x}} \tilde{u}_{n_{0},\mathrm{slab}}^{v} \rangle\right)$ of the slab HWF $|\tilde{u}_{n_{0},\mathrm{slab}}^{v}\rangle$ over the $rBZ$.
One can also define the Chern number $C_{n_{1},n_{2}}$ for a group of adjacent Wannier sheets $\{z_{n,\mathrm{slab}}^{v}: n_{1}\leq n\leq n_{2}\}$ given by
\begin{gather}
C_{n_{1},n_{2}}^{v} = \frac{1}{2\pi}\int_{rBZ} d\boldsymbol\kappa ~\mathrm{Tr}{\mathcal{F}},
\label{surfCocc}
\end{gather}
where $\mathrm{Tr}$ is a trace over $\{|\tilde{u}_{n,\mathrm{slab}}^{v}\rangle : n_{1}\leq n\leq n_{2}\}$ and $\mathcal{F}$ is the corresponding non-abelian Berry curvature~\cite{alexandradinata2021HIsurface}.
We note that $C_{n_{1},n_{2}}$ is well-defined regardless of possible crossing among $\{z_{n,\mathrm{slab}}^{v}: n_{1}\leq n\leq n_{2}\}$ as long as $z_{n_{1}-1,\mathrm{slab}}^{v}(\boldsymbol\kappa)<z_{n_{1},\mathrm{slab}}^{v}(\boldsymbol\kappa)$ and $z_{n_{2},\mathrm{slab}}^{v}(\boldsymbol\kappa)<z_{n_{2}+1,\mathrm{slab}}^{v}(\boldsymbol\kappa)$ for all $\boldsymbol\kappa\in rBZ$.
Similarly, one can define the Chern number $C_{n_{1},n_{2}}^{c}$ for the unoccupied states by using $\{|\tilde{u}_{n,\mathrm{slab}}^{c}\rangle : n_{1}\leq n\leq n_{2}\}$.

Now we define the bottom surface Chern number  $C_{\mathrm{bottom}}^{v}$ of the occupied states as
\begin{gather}
C_{\mathrm{bottom}}^{v}=C_{1,n_{\mathrm{bulk}}}^{v},
\label{Csufocc}
\end{gather}
where $n_{\mathrm{bulk}}$ is a sufficiently large integer such that $|\tilde{u}_{n_{\mathrm{bulk}},\mathrm{slab}}^{v}\rangle$ is \textit{bulk-like}.
Namely, we assume that $|\tilde{u}_{n_{\mathrm{bulk}},\mathrm{slab}}^{v}\rangle$ is similar to the bulk HWF $|w_{l}^{v}(k_{x},k_{y})\rangle$ obtained from the bulk state $|u^{v}(\mathbf{k})\rangle$ by taking the Fourier transformation in the $z$-direction as
\begin{gather}
|w_{l}^{v}(k_{x},k_{y})\rangle = \frac{c}{2\pi}\int dk_{z} e^{ik_{z}(z-lc)}|u^{v}(\mathbf{k})\rangle,
\label{defofHWF}
\end{gather}
where $l$ is an integer, $c$ is the lattice constant along the $z$-direction, and $\mathbf{k}=(k_x,k_y,k_z)$ is a 3D momentum~\cite{vanderbilt2015CSApumping}.

It is known that there is a unitary transformation $U\in\mathrm{U}(1)$ for the occupied state $|u^{v}\rangle$ such that the bulk HWF $|\tilde{w}_{l}^{v}\rangle$ obtained from the unitary-transformed occupied states are localized in the $z$-direction~\cite{Vanderbilt1997maximalWan}.
Then the \textit{bulk-like} state $|\tilde{u}_{n_{\mathrm{bulk}},\mathrm{slab}}^{v}\rangle$ should have the following properties similar to those of $|\tilde{w}_{l}^{v}\rangle$.
First, the distance in the $z$-direction between adjacent Wannier sheets should be $c$, i.e., $|z_{n_{bulk}+1,\mathrm{slab}}^{v}-z_{n_{bulk},\mathrm{slab}}^{v}|\approxeq c$.
This follows from the fact that in the case of $|\tilde{w}_{l}^{v}\rangle$, its translational invariance along the $z$-direction
\begin{gather}
|\tilde{w}_{l}^{v}(z+c)\rangle=|\tilde{w}_{l-1}^{v}(z)\rangle,
\end{gather}
gives
\begin{gather}
z_{l}^{v}=z_{l-1}^{v}+c,
\label{Wandist}
\end{gather}
where $z_{l}^{v}$ is the Wannier center of $|\tilde{w}_{l}^{v}\rangle$ in the $z$-direction.
Therefore, the adjacent Wannier sheets are always apart from each other by the distance $c$.

Second, the Chern number of each Wannier sheet should be zero.
This comes from the fact that the Chern number of an arbitrary 2D submanifold of the 3D Brillouin zone is zero for the Hopf insulator~\cite{moore2008Hopfinsulator,
vanderbilt2015CSApumping}.
From these two properties of the \textit{bulk-like} states, we can deduce that the bottom surface Chern number $C_{\mathrm{bottom}}^{v}$ of the occupied states in Eq.~(\ref{Csufocc}) is invariant with respect to the choice of $n_{\mathrm{bulk}}$:
when $n_{\mathrm{bulk}}$ is changed, $C_{\mathrm{bottom}}^{v}$ will also be changed as much as the Chern number of the added or subtracted \textit{bulk-like} states, which is equal to zero.
Therefore, $C_{\mathrm{bottom}}^{v}$ does not vary under the change of $n_{\mathrm{bulk}}$.

We can also define the top surface Chern number $C_{\mathrm{top}}^{v}$ of the occupied states as
\begin{gather}
C_{\mathrm{top}}^{v}=C^{v}_{n_{\mathrm{bulk}},N_{z}}.
\label{Ctopocc}
\end{gather}
Similarly, the top and bottom surface Chern numbers $C_{\mathrm{top}}^{c},~C_{\mathrm{bottom}}^{c}$ of the unoccupied states can be defined as those of the occupied states.
Then we define the bottom surface Chern number $C_{\mathrm{bottom}}$ as
\begin{gather}
C_{\mathrm{bottom}}=C_{\mathrm{bottom}}^{v}+C_{\mathrm{bottom}}^{c}.
\end{gather}
Likewise, the top surface Chern number $C_{\mathrm{top}}$ is given by $C_{\mathrm{top}}^{v}+C_{\mathrm{top}}^{c}$.

The surface Chern numbers of the Hopf insulator can be related to its Hopf invariant through the surface theorem for the Chern-Simons 3-form $\theta_{3}$~\cite{vanderbilt2017BBCofCSA},
which is defined as
\begin{gather}
\theta_{3}^{v}=-\frac{1}{4\pi}\int_{BZ} d\mathbf{k} \epsilon^{ijk}\mathrm{Tr}\left[A_{i}\partial_{j}A_{k}-\frac{2}{3}i A_{i}A_{j}A_{k}\right],
\label{th3occ}
\end{gather}
where $\epsilon^{ijk}$ is the Levi-Civita symbol and $A_{i}$ is the non-abelian Berry connection of the occupied bulk states with components $\left[A_{i}\right]_{nm}=i\langle u_{n}|\partial_{i}|u_{m}\rangle$, and the trace is over the occupied states.
Note that, for the Hopf insulator with a Hopf invariant $\chi$, we have $\theta_{3}^{v}=\pi \chi$.

$\theta_{3}^{v}$ can be related to the quantity $\theta_{\mathrm{slab}}^{v}$ defined for the slab geometry of the same Hamiltonian.
For a slab composed of $N_{z}$ layers stacked in the $z$-direction, $\theta_{\mathrm{slab}}^{v}$ is given by
\begin{gather}
\theta_{\mathrm{slab}}^{v}=\frac{1}{N_{z}}\int_{rBZ} d\boldsymbol\kappa \mathrm{Tr}\left[Z_{\mathrm{slab}}^{v}\Omega_{xy,\mathrm{slab}}^{v}\right],
\label{th2occ}
\end{gather}
where $Z_{\mathrm{slab}}^{v}$ and $\Omega_{xy,\mathrm{slab}}^{v}$ are the $z$-directional position operator and the non-abelian Berry curvature of the occupied states of the slab Hamiltonian, and the trace is over the occupied states of the slab Hamiltonian.
In Ref.~\cite{vanderbilt2017BBCofCSA}, it is shown that
\begin{gather}
\theta_{\mathrm{slab}}^{v}\rightarrow \theta_{3}^{v}-2\pi C_{\mathrm{top}}^{v}~~\mathrm{for}~N_{z}\rightarrow \infty ,
\label{surfthmocc}
\end{gather}
where $C_{\mathrm{top}}^{v}$ is defined similar to Eq.~(\ref{Ctopocc}).
This theorem holds when the Chern number of the occupied states of the entire slab is trivial, which is satisfied for the Hopf insulator.
We can deduce a similar equation for the unoccupied states given by
\begin{gather}
\theta_{\mathrm{slab}}^{c}\rightarrow \theta_{3}^{c}-2\pi C_{\mathrm{top}}^{c}~~\mathrm{for}~N_{z}\rightarrow \infty.
\label{surfthmunocc}
\end{gather}
Interestingly, we also find that
\begin{gather}
\theta_{\mathrm{slab}}^{v}+\theta_{\mathrm{slab}}^{c}=0,
\label{tslabtot}
\end{gather}
as shown in Appendix~\ref{sec:prooftslabtot}.
After adding Eq.~(\ref{surfthmocc}) and (\ref{surfthmunocc}), we obtain $\theta_{3}^{v}+\theta_{3}^{c}=2\pi (C_{\mathrm{top}}^{v}+C_{\mathrm{top}}^{c})=2\pi C_{\mathrm{top}}$.
For a Hopf insulator with the Hopf invariant $\chi$, it was shown that $\theta_{3}^{v}$ and $\theta_{3}^{c}$ are the same and equal to $\pi \chi$~\cite{trifunovic2021nbandHI,slager2019hopffloquet},
therefore, we obtain
\begin{gather}
\chi=C_{\mathrm{top}}=-C_{\mathrm{bottom}},
\label{HIBBC}
\end{gather}
where $C_{\mathrm{top}}+C_{\mathrm{bottom}}=0$ since not only the total Chern number of the conduction band and the valence band of the slab but also the total Chern number of the bulk region are zero.

\subsection{Bulk-boundary correspondence of real Hopf insulators}
Similar to the Hopf insulator, topologically protected surface Chern number can also exist in the real Hopf insulator.
Let us consider a real Hopf insulator slab with $N_{z}$ layers stacked in the $z$-direction.
As in the case of the Hopf insulator, an energy gap between the surface occupied and unoccupied states is needed to define the surface Chern numbers of the occupied and unoccupied states.
When there are band crossings between surface occupied and unoccupied bands, one can open a gap by adding suitable perturbations preserving $\mathcal{PT}$ symmetry.
We note that although gapless nodes can be symmetry-protected in 2D $\mathcal{PT}$ symmetric systems, as it relates the top and bottom surfaces in the case of a slab, gapless nodes can always be removed by applying $\mathcal{PT}$ symmetric perturbations.

When the energy gap is open, we can define a projection operator on the occupied states, $P_{\mathrm{slab}}$, and the unoccupied states, $Q_{\mathrm{slab}}$, in the $rBZ$.
And the eigenstates and eigenvalues of $P_{\mathrm{slab}}Z_{\mathrm{slab}}P_{\mathrm{slab}}$ are slab HWFs $|\tilde{u}_{n,\mathrm{slab}}^{v}\rangle$ and their Wannier sheets $z_{n,\mathrm{slab}}^{v}$.
Since there are two occupied states in the bulk unit cell of the real Hopf insulator, the index $n$ can be from $1$ to $2N_{z}$.
For the unoccupied states, we can define slab HWFs  $|\tilde{u}_{n,\mathrm{slab}}^{c}\rangle$ and their Wannier sheets $z_{n,\mathrm{slab}}^{c}$ similarly.
Without loss of generality, the order of the index $n$ is chosen to satisfy $z_{1,\mathrm{slab}}^{v}\leq\cdots \leq z_{2N_{z},\mathrm{slab}}^{v}$ and  $z_{1,\mathrm{slab}}^{c}\leq\cdots \leq z_{2N_{z},\mathrm{slab}}^{c}$.

To define the surface Chern number of the RHI, one needs to be more careful than the case of the Hopf insulator, because of the possible band crossings between adjacent Wannier sheets.
For example, the surface Chern number in Eq.~(\ref{Csufocc}) is well-defined when $n_{\mathrm{bulk}}$ is an index belonging to the bulk region.
This is because, in the Hopf insulator, the Wannier sheets $z_{n_{\mathrm{bulk}},\mathrm{slab}}^{v}$ in the bulk region are not touching with their adjacent Wannier sheets and have Chern number $0$.
In the case of the RHI, as there are two occupied states per unit cell,
Eq.~(\ref{Wandist}) might not be satisfied, especially for the adjacent Wannier sheets belonging to the same unit cell.
However, as long as adjacent Wannier sheets belonging to different unit cells do not touch each other, we can define the Chern number of the two Wannier sheets in one unit cell, which is zero because the Chern number of the RHI on any 2D submanifold is trivial.

Then let us divide the Wannier sheets $\{z_{n,\mathrm{slab}}^{v}:n=1,\cdots ,2N_{z}\}$ into three regions: the bottom surface region, the bulk region, and the top surface region.
The states with indices $1\leq n\leq n_{\mathrm{bottom}}$ and $n_{\mathrm{top}}\leq n\leq 2N_{z}$ belong to the bottom surface region and top surface region, respectively.
And the other states belong to the bulk region.
Note that $n_{\mathrm{bottom}}<n_{\mathrm{top}}$ and top surface region is automatically determined after setting the bottom surface region due to the $\mathcal{PT}$ symmetry.
Since there are integer numbers of unit cells in the bulk region, $n_{\mathrm{bottom}}$ should be set not to cut the interior of a unit cell in the bulk region.
If we change $n_{\mathrm{bottom}}$ obeying this condition, Wannier sheets are added to or subtracted from the bottom surface region in units of unit cells.
Then we define the bottom surface Chern number $C_{\mathrm{bottom}}^{v}$ of the occupied states by
\begin{gather}
C_{\mathrm{bottom}}^{v}=C^{v}_{1,n_{\mathrm{bottom}}},
\label{rCbotocc}
\end{gather}
where $C^{v}_{1,n_{\mathrm{bottom}}}$ is given by Eq.~(\ref{surfCocc}).
We can confirm that $C_{\mathrm{bottom}}^{v}$ is well-defined since the Chern number of a unit cell is zero.
Also, if we define the top surface Chern number $C_{\mathrm{top}}^{v}$ of the occupied states by
\begin{gather}
C_{\mathrm{top}}^{v}=C^{v}_{n_{\mathrm{top}},2N_{z}},
\label{rCtopocc}
\end{gather}
$C_{\mathrm{top}}^{v}$ is well-defined.

Note that the surface Chern number is not well-defined when \textit{bulk-like} Wannier sheets in different unit cells are touching each other.
We observed this phenomenon in several RHI models and explain why these models have such property in the context of multicellularity of delicate states in Appendix~\ref{sec.crossing}.
The Wannier sheets of the unoccupied states can also be divided into the bottom surface region, the bulk region, and the top surface region, and the corresponding surface Chern numbers  $C_{\mathrm{bottom}}^{c}$ and $C_{\mathrm{top}}^{c}$ can also be defined as in Eq.~(\ref{rCbotocc}) and (\ref{rCtopocc}), respectively, using localized unoccupied states.

Applying the surface theorem for the Chern-Simons 3-form~\cite{vanderbilt2017BBCofCSA}, we can get Eq.~(\ref{surfthmocc}) for the occupied states and Eq.~(\ref{surfthmunocc}) for the unoccupied states.
The definitions of $\theta_{3}$ and $\theta_{\mathrm{slab}}$ are given by Eq.~(\ref{th3occ}) and (\ref{th2occ}), respectively.
After adding Eq.~(\ref{surfthmocc}) and (\ref{surfthmunocc}), we can get
\begin{equation}
\label{rth3Csurf}
\begin{split}
\theta_{3}^{v}+\theta_{3}^{c}
& =2\pi C_{\mathrm{top}}\\
& =-2\pi C_{\mathrm{bottom}},
\end{split}
\end{equation}
where $\theta_{\mathrm{slab}}^{v}+\theta_{\mathrm{slab}}^{c}=0$, which is proved in Appendix \ref{sec:prooftslabtot}, $C_{\mathrm{top}}=C_{\mathrm{top}}^{v}+C_{\mathrm{top}}^{c}$ and $C_{\mathrm{bottom}}=C_{\mathrm{bottom}}^{v}+C_{\mathrm{bottom}}^{c}$.
We used $C_{\mathrm{top}}=-C_{\mathrm{bottom}}$ in the second line, which is due to the $\mathcal{PT}$ symmetry.

To find a relation between $\theta_{3}^{v}+\theta_{3}^{c}$ and the real Hopf invariants $\chi_{z}$ and $\chi_{w}$, let us express $\theta_{3}^{v}$ and $\theta_{3}^{c}$ using the Euler connections and the Euler forms of the occupied and unoccupied states, which are in Eq.~(\ref{Euler connection}),~(\ref{Euler form}), respectively.
The non-abelian Berry connection $\mathbf{A}^{v}$ of the occupied states is given by
\begin{gather}
\mathbf{A}^{v}=\begin{pmatrix}
0 && -i\langle u_{2}^{v}|\nabla |u_{1}^{v}\rangle \\ i\langle u_{2}^{v}|\nabla |u_{1}^{v}\rangle && 0
\end{pmatrix}
=\textbf{a}^{v}\sigma_{y},
\label{Aocc}
\end{gather}
where $|u_{1,2}^{v}\rangle$ are the occupied states, $\textbf{a}^{v}=\langle u_{2}^{v}|\nabla |u_{1}^{v}\rangle$, and $\sigma_{y}$ is the Pauli matrix.
Note that the diagonal components of $\mathbf{A}^{v}$ are zero since $|u_{1,2}^{v}\rangle$ are real-valued.
Substituting $\mathbf{A}^{v}$ into the definition of $\theta_{3}^{v}$ in Eq.~(\ref{th3occ}), we can get
\begin{equation}
\label{rth3occ}
\begin{split}
\theta_{3}^{v}
 & =-\frac{1}{2\pi}\int_{BZ}d\mathbf{k}~\textbf{a}^{v}\cdot \nabla\times\textbf{a}^{v},\\
 & =-\frac{1}{2\pi}\int_{BZ}\text{a}^{v}\wedge \text{Eu}^{v},
\end{split}
\end{equation}
where we used the definition of $\text{a}^{v}$ and $\text{Eu}^{v}$ in the second line, which are in Eq.~(\ref{Euler connection}),~(\ref{Euler form}).
Similarly, one can easily show that $\theta_{3}^{c}$ is given by
\begin{gather}
\label{rth3unocc}
\theta_{3}^{c}=-\frac{1}{2\pi}\int_{BZ}\text{a}^{c}\wedge \text{Eu}^{c}.
\end{gather}
Adding Eq.~(\ref{rth3occ}) and (\ref{rth3unocc}), we can find that
\begin{equation}
\label{rth3chi}
\begin{split}
\theta_{3}^{v}+\theta_{3}^{c}
 & = -\frac{1}{2\pi}\int_{BZ}(\text{a}^{v}\wedge \text{Eu}^{v}+\text{a}^{c}\wedge \text{Eu}^{c})\\
 & =4\pi (\chi_{z}+\chi_{w}),
\end{split}
\end{equation}
where we used the first equation of Eq.~(\ref{eq.forms & invariants 1}) in the second line.
Using Eq.~(\ref{rth3Csurf}) and (\ref{rth3chi}), we find that
\begin{gather}
    \label{RHIBBC}
    C_{\mathrm{bottom}} = -2(\chi_{z}+\chi_{w}) = -C_{\mathrm{top}},
\end{gather}
where we use $C_{\mathrm{top}}+C_{\mathrm{bottom}}=0$.

\begin{table}[t]
\includegraphics[width=\linewidth]{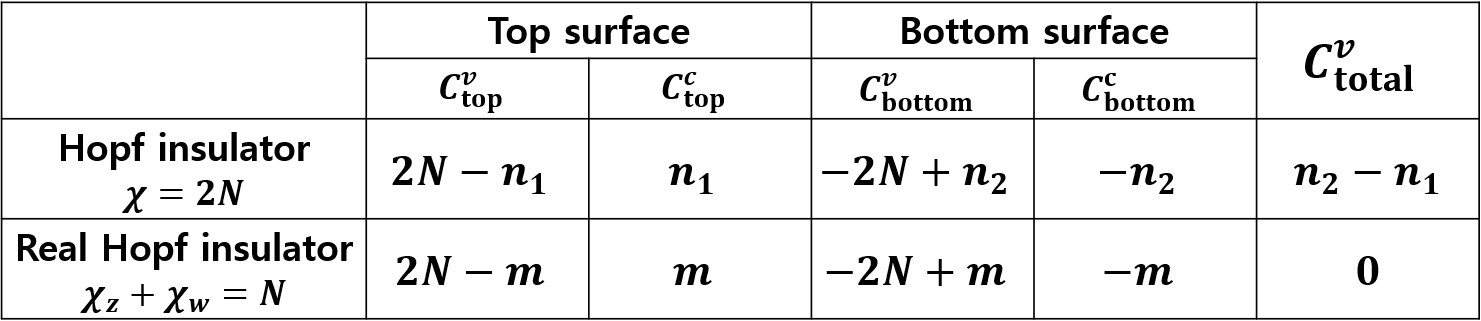}
\caption{Possible surface Chern numbers of the Hopf insulator and the real Hopf insulator. $n_{1},\,n_{2},$ and $m$ are arbitrary integers and $C_{\mathrm{total}}^{v}$ means the total Chern number of the occupied states.}
\label{fig:Diff}
\end{table}

Let us compare the surface Chern numbers of a Hopf insulator with $\chi$ and a real Hopf insulator with $(\chi_{z}, \chi_{w})$  that satisfy $\chi=2(\chi_{z}+\chi_{w})=2N$.
According to Eq.~(\ref{HIBBC}),~(\ref{RHIBBC}), both the Hopf insulator and the real Hopf insulator have topologically protected surface Chern number $C_{\mathrm{top}}=2N$.
However, the Chern number distribution between the occupied and unoccupied states is quite different as shown below.

Let us separately consider the surface Chern numbers of the occupied states and the unoccupied states.
$C_{\mathrm{top}}^{v}$ and $C_{\mathrm{top}}^{c}$ can exchange their values by closing the energy gap between the occupied and unoccupied states in the top surface region.
This is the same for the bottom region.
We write possible surface Chern numbers of the Hopf insulator with $\chi=2N$ in the first row of Table.~\ref{fig:Diff}.
Here, $n_{1}$ (resp. $n_{2}$) is an integer that can be changed by closing the energy gap between the occupied and unoccupied states in the top (resp. bottom) region.
Note that $C_{\mathrm{top}}^{v}+ C_{\mathrm{bottom}}^{v}$ is not necessarily zero for the Hopf insulator.
However, in the case of the real Hopf insulator, $\mathcal{PT}$ symmetry constraints $C_{\mathrm{top}}^{v}=- C_{\mathrm{bottom}}^{v}$.
This difference affects the total Chern numbers of their occupied states, $C_{\mathrm{total}}^{v}$.
Since the Chern numbers of the bulk states are zero for both Hopf insulator and real Hopf insulator, $C_{\mathrm{total}}^{v}$ of both cases are $C_{\mathrm{top}}^{v}+ C_{\mathrm{bottom}}^{v}$, which are written in the last column of Table.~\ref{fig:Diff}.
Therefore, the occupied states of the Hopf insulator slab can have non-trivial Chern number but those of the real Hopf insulator slab always have trivial Chern number.
 
At this point, let us recall that the bulk-boundary correspondence of conventional Hopf insulator also relates the total surface Chern number, including both occupied and unoccupied states, to the bulk invariant, as a result of delicate topology. However, in this case, the bulk invariant can be computed only using the occupied state. In contrast, RHI illustrates the feature of delicate topology in a more decisive way: The bulk invariant has integral formula that unavoidably contain both occupied and unoccupied bands. Furthermore, if one tries to construct another integral formula written solely in terms of occupied (unoccupied) bands, such as
\begin{gather}
    \int_{BZ} a^c\wedge\text{Eu}^c\text{ or } \int_{BZ} a^v\wedge\text{Eu}^v,
\end{gather}
one is bound to get gauge non-invariant or non-quantized results [see Appendix~\ref{Sec.uniqueness}]. Hence, RHI provides a clear evidence that the bulk-boundary correspondence of occupied and unoccupied bands as a whole is a signature of delicate topology. 
We note that the bulk invariant of Hopf insulator can also be defined as a property of the total bands including both occupied and unoccupied bands. However, due to the intrinsic particle-hole symmetry, Hopf insulator partially eluded the above observation regarding the nature of delicate TI. In fact, particle-hole symmetry makes possible the integral formula for bulk invariant only containing the occupied band.

Before turning to the numerical results, we point out an implication of Eq.~(\ref{RHIBBC}) on the time reversal ($\mathcal{T}$) and space inversion ($\mathcal{P}$) symmetries of RHI. Since Berry curvature is odd under $\mathcal{T}$ transformation, $\mathcal{T}$ symmetry requires vanishing surface Chern number, hence $\chi_{z}+\chi_{w}=0$. It turns out that every homotopy class $(\chi_z,\chi_w)=(m,-m)\;(m\in\bb{Z})$ of RHI contains at least one model equipped with both $\mathcal{T}$ and $\mathcal{P}$ symmetries [Note that all RHIs have the combined $\mathcal{PT}$ symmetry, but individual symmetries $\mathcal{P}$ and $\mathcal{T}$ are not warranted in general.] and such an example is given by the Moore-Ran-Wen RHIs. To be specific, converting the MRW model in Eq.~(\ref{ftn.MRW_deform}) into the quaternion form as in Sec.~\ref{Subsec.quaternion_Hopf} yields appropriate $z(\mathbf{k})$ such that $\chi_z=m$. Then $w(\mathbf{k})$ is defined by putting $k_{z}\to -k_{z}$ into $z(\mathbf{k})$. The resulting flattened RHI Hamiltonian $\overline{H}^{2,2}_{R}$ [see Eq.~(\ref{eq.RHI_spectral})] realizes the desired symmetries for every $m\in\bb{Z}$. A detailed derivation appears in Appendix~\ref{subsec.P}.

\subsection{Model calculation of bulk-boundary correspondence}

\begin{figure}[t]
\includegraphics[width=\linewidth]{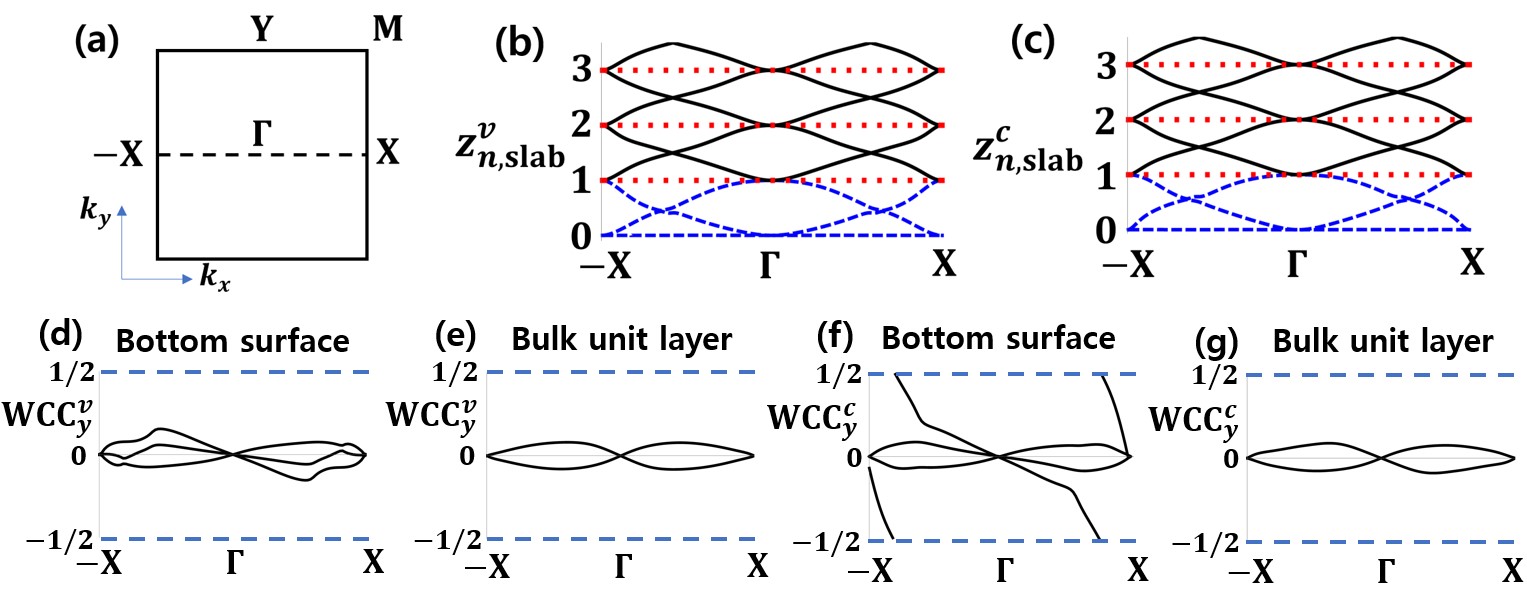}
\caption{
Wannier sheets and their Wilson loop spectra of the real Hopf insulator slab with $(\chi_{z},\chi_{w})=(1,0)$ described in Eq.~(\ref{RHI10}).
(a) Brillouin zone of the slab Hamiltonian.
(b, c) Cross-sections of valence (conduction) band Wannier sheets, along $k_{x}$ axis, of a surface-deformed slab Hamiltonian near the bottom surface. The blue dashed lines indicate the Wannier sheets inside bottom surface region while the black solid lines are those in bulk region. The red dotted lines divide the slab into unit layers whose length is set to be $1$.
(d, e) Wilson loop spectra of the valence band Wannier sheets (d) inside the bottom surface region, (e) inside a single bulk layer.
$\mathrm{WCC}_{y}$ stands for the Wannier charge center in spatial $y$ direction.
(f, g) Wilson loop spectra of the conduction band Wannier sheets (f) inside the bottom surface region, (g) inside a single bulk layer.
}
\label{fig:BBC}
\end{figure}

Let us confirm Eq.~(\ref{RHIBBC}) using the tight-binding models with the real Hopf invariants $(\chi_{z},\chi_{w})=(\pm 1,0),~(\pm 2,0),~(0,\pm 1),~(0,\pm 2)$, respectively.

In Section.~\ref{subsec.construction}, we show that, for given two quaternions $z,w$, the eigenstates of RHI with invariants $(\chi_{z},\chi_{w})$ are given by normalized states $\ket{\bar{z}w},|\bar{z}\text{i}w\rangle,|\bar{z}\text{j}w\rangle,|\bar{z}\text{k}w\rangle$, which are column vectors of the matrix $R_{z,w}$ defined in Eq.~(\ref{def.Rzw}).
Then we can define a real-valued Hamiltonian $\overline{H}_{R}^{2,2}(\mathbf{k})$ as in Eq.~(\ref{4band Hamiltonian})
\begin{gather}
\overline{H}_{R}^{2,2}(\mathbf{k})
=R_{z,w}\begin{pmatrix}
1 & & & \\
 &1 & & \\
 & &-1& \\
 & & &-1
\end{pmatrix}
R_{z,w}^{\top} .
\label{defHzw}
\end{gather}
We can easily show that $\overline{H}_{R}^{2,2}(\mathbf{k})$ has eigenstates $\ket{\bar{z}w}$, $|\bar{z}\text{i}w\rangle$, $|\bar{z}\text{j}w\rangle$, $|\bar{z}\text{k}w\rangle$.
Also, the energy of $\ket{\bar{z}w}$,$|\bar{z}\text{i}w\rangle$ ($|\bar{z}\text{j}w\rangle$, $|\bar{z}\text{k}w\rangle$) is $|z||w|$ ($-|z||w|$).
Therefore, for given quaternions $z(\mathbf{k}),w(\mathbf{k})$ which have non-trivial Hopf invariants, we can make a RHI model with invariants $(\chi_{z},\chi_{w})$ by using Eq.~(\ref{defHzw}).

For example, two quaternions $z,w$ of a real Hopf insulator model with $(\chi_{z},\chi_{w})=(1,0)$ is given by~\cite{deng2013HImodels}
\begin{gather}
    z=\sin{k_{x}}+\text{i} \sin{k_{y}}+\text{j} \sin{k_{z}}+\text{k}\left(\sum_{i=x,y,z}\cos{k_{i}}-\frac{3}{2}\right),\n
    w=1. 
    \label{RHI10}
\end{gather}
We study a slab Hamiltonian of this model with 16 layers stacked in the $z$-direction.
To calculate Wannier sheets $z_{n,\mathrm{slab}}^{v},~z_{n,\mathrm{slab}}^{c}$ of the occupied and unoccupied states, it is important to make an energy gap between the surface occupied and unoccupied states of the slab Hamiltonian, which achieved by adding a small $\mathcal{PT}$-symmetric perturbation
\begin{gather}
H_{pert}=
\begin{pmatrix}
0&&0&&0&&0\\
0&&1&&0&&0\\
0&&0&&1&&0\\
0&&0&&0&&0
\end{pmatrix},
\label{Hpert1}
\end{gather}
to the top and bottom layers.

\begin{figure}[t]
\includegraphics[width=\linewidth]{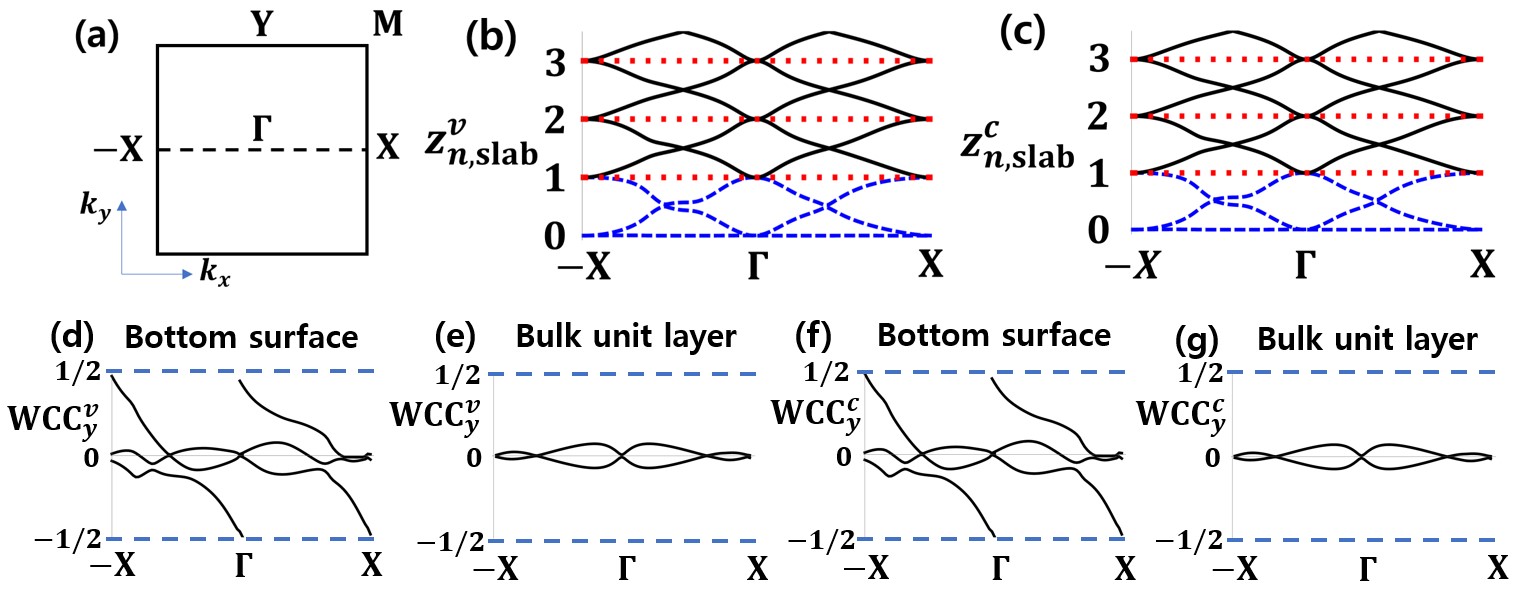}
\caption{
Wannier sheets and their Wilson loop spectra of the real Hopf insulator slab with $(\chi_{z},\chi_{w})=(2,0)$ described in Eq.~(\ref{RHI20}).
(a) Brillouin zone of the slab Hamiltonian.
(b, c) Cross-sections of valence (conduction) band Wannier sheets, along $k_{x}$ axis, of a surface-deformed slab Hamiltonian near the bottom surface. The blue dashed lines indicate the Wannier sheets inside bottom surface region while the black solid lines are those in bulk region. The red dotted lines divide the slab into unit layers whose length is set to be $1$.
(d, e) Wilson loop spectra of the valence band Wannier sheets (d) inside the bottom surface region, (e) inside a single bulk layer. 
$\mathrm{WCC}_{y}$ stands for the Wannier charge center in spatial $y$ direction.
(f, g) Wilson loop spectra of the conduction band Wannier sheets (f) inside the bottom surface region, (g) inside a single bulk layer.
}
\label{fig:BBC2}
\end{figure}

To check the surface Chern number, we divide the slab into the surface regions and the bulk region.
The bulk region is divided into bulk unit layers which nearly satisfy the $z$-directional translation symmetry.
We illustrate the Wannier sheets of the occupied and unoccupied states near the bottom surface in Fig.~\ref{fig:BBC} (a), (d).
The Wilson loop spectra of the bottom surface region are shown in Fig.~\ref{fig:BBC} (b), (e),
from which the bottom surface Chern numbers are given by $C_{\mathrm{bottom}}=C_{\mathrm{bottom}}^{v}+C_{\mathrm{bottom}}^{c}=-2$ consistent with Eq.~(\ref{RHIBBC}).
Note that $C_{\mathrm{top}}^{v(c)}=-C_{\mathrm{bottom}}^{v(c)}$ due to the $\mathcal{PT}$ symmetry.
Also, we calculate the Wilson loop spectra of all the bulk unit layers and find that the relevant Chern numbers are all zero.

To make a real Hopf insulator model with invariant $(2,0)$, we use two quaternions $z$, $w$ given by~\cite{deng2013HImodels}
\begin{gather}
z=\sin{k_{x}}-\text{i} \sin{k_{y}}+\text{j} \sin{k_{z}}+\text{k}\left(\sum_{i=x,y,z}\cos{k_{i}}-\frac{1}{2}\right),\n
w=1.\label{RHI20}
\end{gather}
For the corresponding slab Hamiltonian with 16 layers, we add
\begin{gather}
H'_{pert}=
\begin{pmatrix}
0&&0&&0&&0\\
0&&0&&1&&0\\
0&&1&&0&&0\\
0&&0&&0&&0
\end{pmatrix}
\label{Hpert2}
\end{gather}
to open the gap between the occupied and unoccupied states.
The relevant Wannier sheets for the occupied states near the bottom surface and the unoccupied states are shown in Fig.~\ref{fig:BBC2}~(a), (d), respectively.
We choose the lowest three sheets as a surface region, the blue dashed lines in Fig.~\ref{fig:BBC2}~(a), (d). 
The Wilson loop spectra near the bottom surface region for the occupied states and the unoccupied states are illustrated in Fig.~\ref{fig:BBC2}~(b), (e), respectively.
We find that $C_{\mathrm{bottom}}=C_{\mathrm{bottom}}^{v}+C_{\mathrm{bottom}}^{c}=-4$, which satisfies Eq.~(\ref{RHIBBC}).
We have also confirmed that the Chern number of each bulk unit layer is zero.

It is worth noting that we have chosen odd numbers of the surface Wannier sheets in Fig.~\ref{fig:BBC}~(a),~(d) and  Fig.~\ref{fig:BBC2}~(a),~(d) to avoid the crossing between adjacent Wannier sheets in neighboring unit cells. For validity of the surface theorem, the boundary of the surface and bulk regions must be put in a way that it does not intersect any of the Wannier sheets. Otherwise, the Wannier sheet would be partially included in the surface region so that the surface region is ill-defined.
To avoid the intersections, one regards the $(k_x,k_y)$-planes anchored at the integer points $z=N$ along the $z$ axis as virtual unit cell boundaries. After perturbing the Hamiltonian slightly, the Wannier sheets near the top (bottom) region shift away from the top (bottom) virtual unit cell so that the top (bottom) virtual unit cell can play the role of top (bottom) surface region. The rest of the unit cells are then regarded as the bulk region.

We can get a RHI model with invariants $(\chi_{z},\chi_{w})=(-1,0)$ by applying a simple transformation $k_{y}\rightarrow -k_{y}$ to Eq.~(\ref{RHI10})~\cite{deng2013HImodels}.
This transformation can be regarded as a $y$-directional mirror symmetry transformation, which inverts the sign of the surface Chern number. In accordance with this, one can confirm that
the bottom surface Chern number of this model is $C_{\mathrm{bottom}}=2$, being consistent with Eq.~(\ref{RHIBBC}).

Similarly, we can get a RHI model with invariants $(\chi_{z},\chi_{w})=(-2,0)$ by applying the transformation $k_{y}\rightarrow -k_{y}$ to Eq.~(\ref{RHI20})~\cite{deng2013HImodels}.
By the same reasoning as above, $C_{\mathrm{bottom}}=4$ for this model, which satisfies Eq.~(\ref{RHIBBC}).

We can also get RHI models with invariants $(\chi_{z},\chi_{w})=(0,\pm 1),~(0,\pm 2)$ by applying a simple transformation to Eq.~(\ref{RHI10}),~(\ref{RHI20}).
For given two quaternions $a(\mathbf{k}),b(\mathbf{k})$ which are functions on BZ, let us make two RHI Hamiltonians $H_{a,b},H_{b,a}$ by using Eq.~(\ref{defHzw}).
Then the RHI invariants of $H_{a,b}$ (resp. $H_{b,a}$) are $(\chi_{a},\chi_{b})$ (resp. $(\chi_{b},\chi_{a})$).
In this case, we find that 
\begin{gather}
H_{a,b}=OH_{b,a}O^{\top}
\label{HabHba}
\end{gather}
for an orthogonal matrix $O=\mathrm{diag}(-1,1,1,1)$.
For the slab Hamiltonians $H_{a,b,N_{z}}^{\mathrm{slab}}$ and $H_{b,a,N_{z}}^{\mathrm{slab}}$, made by stacking $N_{z}$ layers of $H_{a,b}$ and $H_{b,a}$ in $z$-direction, there is an orthogonal transformation $O_{\mathrm{slab}}\in \mathrm{O}(4N_{z})$ such that $H_{a,b,N_{z}}^{\mathrm{slab}}=O_{\mathrm{slab}}H_{b,a,N_{z}}^{\mathrm{slab}}O_{\mathrm{slab}}^{\top}$.
Since the surface Chern number defined in Eq.~(\ref{Csufocc}) is invariant under the transformation by $O_{\mathrm{slab}}$, the surface Chern number of $H_{a,b,N_{z}}^{\mathrm{slab}}$ and $H_{b,a,N_{z}}^{\mathrm{slab}}$ are the same.
Then the same bulk-boundary correspondence of RHI described in Eq.~(\ref{RHIBBC}) should be applied to these two Hamiltonians. 
Since we already verified the bulk-boundary correspondence of the RHI models with $(\chi_{z},\chi_{w})=(\pm 1,0),~(\pm 2,0)$,
by applying the orthogonal transformation $O$ in Eq.~(\ref{HabHba}) to these models, we can show that Eq.~(\ref{RHIBBC}) is also satisfied for RHI models with invariants $(\chi_{z},\chi_{w})=(0,\pm 1),~(0,\pm 2)$.

For other examples of real Hopf insulators with arbitrary $(\chi_{z},\chi_w)$, we observed that the well-definedness of the surface Chern number is not guaranteed due to the crossings between Wannier sheets, which are robust against small perturbations.
In certain models featuring $C_{4z}$ symmetry, such crossings can be shown to be unavoidable [see Appendix~\ref{sec.crossing}.] and originate from the multicellular nature of delicate topological insulators, which will be remarked at the end of the next section.

\section{Returning Thouless pump in rotational symmetric RHI}
\label{sec.RTP}
\subsection{Review of the returning Thouless pump in Hopf insulator}
\label{Subsec.RTP_review}
In a class of Hopf insulator with additional $n$-fold rotational symmetry about the $z$ axis $C_{nz}$, the rotation eigenvalues of the occupied (unoccupied) states behave coherently, resulting in the Hopf-returning Thouless pump (RTP) relation where the Hopf invariant $\chi$ controls the $z$-directional polarization of the occupied (unoccupied) state wave functions at $C_{nz}$-invariant momenta. There are numerous sufficient conditions for this to happen. Without digging into the general theory of RTP~\cite{bzdusek2021multicellularity,bzdusek2022rotationdelicateTI}, we illustrate the Hopf-RTP phenomenon with a class of $C_{4z}$-symmetric Hopf insulator. 
In this section, $\Gamma,\mathrm{M,X,Y}$ denote the $C_{2z}$-invariant lines in the 3D BZ. Namely,
\begin{gather}
    \Gamma = \{(0,0,k_z):k_z\in[0,2\pi]\} \n 
    \mathrm{M} = \{(\pi,\pi,k_z):k_z\in[0,2\pi]\} \n
    \mathrm{X} = \{(\pi,0,k_z):k_z\in[0,2\pi]\} \n
    \mathrm{Y} = \{(0,\pi,k_z):k_z\in[0,2\pi]\} 
    \nonumber
\end{gather} 
Defining the $z$-polarization
\begin{gather}
    \mathsf{p}_{z}(k_{x},k_{y})=\frac{1}{2\pi}\int_{0}^{2\pi} dk_z\mathrm{Tr}\left[A_{z}(k_{x},k_{y},k_{z})\right], 
    \label{def.polarization}
\end{gather}
where $A_{z,mn}=i\bra{ u_{m}}\partial_{k_z}\ket{u_{n}}$ is the $z$ component of the non-abelian Berry connection and the trace is over the occupied subspace~\cite{resta2007polarization}, we will find
\begin{gather}
    \mathsf{p}_{z}(\Pi_1) \equiv \mathsf{p}_{z}(\Pi_2) \mod 1
\end{gather}
between two $C_{4z}$-invariant momenta $\{\Pi_1,\Pi_2\}=\{\Gamma,\mathrm{M}\}$. Suppose $\mathsf{p}_{z}(\mathrm{M})-\mathsf{p}_{z}(\Gamma)=1$, say. While one sweeps the entire BZ starting from $\Gamma$, passing $\mathrm{M}$, and returning again to $\Gamma$, one will observe an increase of $\mathsf{p}_{z}(k_x,k_y)$ by 1 followed by a decrease by the same amount. This is the rationale behind the name ``returning Thouless pump". Note that the RTP behavior is consistent with the vanishing Chern class of Hopf insulator, since the total change in polarization is zero. 

 The Moore-Ran-Wen Hopf insulator, defined in Eq.~(\ref{def.MRW_Hopf}), instantly provides an example of Hopf-RTP physics. The $(t,h)=(1,3/2)$ model ($\chi=1$) obeys $C_{4z}$ symmetry given by
\begin{gather}
    \label{eq.C4_MRW_1}
    R_{C_{4z}} H_{C,MRW}^{1,3/2}(\mathbf{k}) R_{C_{4z}}^{-1} = H_{C,MRW}^{1,3/2}(C_{4z}\mathbf{k}), \\ 
    \label{eq.C4_MRW_2}
    R_{C_{4z}} = 
    \begin{pmatrix}
        i&0\\0&1
    \end{pmatrix}.
\end{gather}
Equivalently, [We drop the cumbersome subscripts of $z_{1,3/2}$.]
\begin{gather}
\label{eq.rotational_Hopf_C}
    R_{C_{4z}} (z^{\dagger}\bm{\sigma}z)(\mathbf{k}) R^{-1}_{C_{4z}} = (z^{\dagger}\bm{\sigma}z)(C_{4z}\mathbf{k}), \\ 
\label{eq.rotational_Hopf_Q}
    e^{-\text{i}\pi/4} (\overline{z}\text{i}z)(\mathbf{k}) e^{\text{i}\pi/4} = (\overline{z}\text{i}z)(C_{4z}\mathbf{k}).
\end{gather}
In Eq.~(\ref{eq.rotational_Hopf_C}), $z$ is a complex vector in $\mathbf{C}^2$. Eq.~(\ref{eq.rotational_Hopf_Q}) is a rewriting of Eq.~(\ref{eq.rotational_Hopf_C}) in terms of unit quaternion $z\in\bb{S}^3\subset\mathbf{H}$ employing the unitary transformation in Eq.~(\ref{eq.H_identification}). It will be useful in the next subsection.

Since the valence (conduction) band of Hopf insulator has rank 1, the occupied (unoccupied) subspace is spanned in real space by a $C_{4z}$-symmetric exponentially localized Wannier functions (WFs)~\cite{alexandradinata2018nogoTI}. These WFs control the rotation eigenvalues of the occupied (unoccupied) states at invariant momenta as follows. The symmetric WFs $W^{v,c}(\mathbf{r})$ are characterized by the properties
\begin{gather}
    \psi_{\mathbf{k}}^{v,c}(\mathbf{r}) = \frac{1}{\sqrt{N}} \sum_{\mathbf{R}} e^{\text{i}\mathbf{k}\cdot\mathbf{R}} W_{\mathbf{R}}^{v,c}(\mathbf{r}), \n 
    W_{C_{4z}^{-1} \mathbf{R}}^{v,c}(C_{4z}^{-1}\mathbf{r}) = \lambda^{v,c} W_{\mathbf{R}}^{v,c}(\mathbf{r}), \n 
    (\lambda^v, \lambda^c ) = (i,1) \text{ or } (1,i),
\label{eq.WF}
\end{gather}
where $\psi_{\mathbf{k}}^{v,c}$ is the Bloch wave functions and $N$ is the number of unit cells of the Bravais lattice whose points are the position vectors $\mathbf{R}$.
Without loss of generality, suppose $(\lambda^v, \lambda^c ) = (i,1)$. Operating the $C_{4z}$ rotation to the first line of Eq.~(\ref{eq.WF}),
\begin{align}
    \psi_{C_{4z}^{-1} \mathbf{k}}^{v,c}(C_{4z}^{-1} \mathbf{r}) & = \frac{1}{\sqrt{N}} \sum_{\mathbf{R}} e^{\text{i}\mathbf{k}\cdot\mathbf{R}}  W_{C_{4z}^{-1}\mathbf{R}}^{v,c}(C_{4z}^{-1} \mathbf{r}) \n 
    & = \frac{1}{\sqrt{N}} \sum_{\mathbf{R}} e^{\text{i}\mathbf{k}\cdot\mathbf{R}}  \lambda^{v,c} W_{\mathbf{R}}^{v,c}(\mathbf{r}) \n
    & = \lambda^{v,c} \psi_{\mathbf{k}}^{v,c}(\mathbf{r}).
    \label{eq.Bloch_transform}
\end{align}
Since there are two orbitals for a primitive unit cell of Hopf insulator, $\psi_{\mathbf{k}}^{v}$ is represented by a 2-component vector $\ket{\mathbf{k}}\in\mathbf{C}^2$ whose components correspond to the coefficients relevant to the orbital basis.
Also, $\ket{\mathbf{k}}$ serves as the occupied eigenstate of the Bloch Hamiltonian $H(\mathbf{k})$. Then Eqs.~(\ref{eq.C4_MRW_1})-(\ref{eq.C4_MRW_2}) imply that $\ket{\mathbf{k}}$ is multiplied by the matrix $R_{C_{4z}}$ when passing from $C_{4z}^{-1}\mathbf{k}$ to $\mathbf{k}$. Together with Eq.~(\ref{eq.Bloch_transform}), this consideration leads to the result
\begin{gather}
    R_{C_{4z}} \ket{\mathbf{k}} = \lambda^{v} \ket{C_{4z}^{-1}\mathbf{k}}.
\end{gather}  
Substituting the invariant momentum $\Pi$ into $\mathbf{k}$, we find that the rotation eigenvalue of the occupied state is identically $\lambda^{v}$ on all $\Pi$. Hence, the occupied (unoccupied) Bloch state on $\Pi$ is the eigenvector of $R_{C_{4z}}$ associated to the eigenvalue $\lambda^{v}$ ($\lambda^{c}$). To guarantee the uniqueness of such an eigenvector, it is important to take different rotation eigenvalues between the conduction and valence bands as in Eq.~(\ref{eq.C4_MRW_2}). This is known as the mutually exclusive condition to define the RTP~\cite{bzdusek2021multicellularity,bzdusek2022rotationdelicateTI}. 

$H^{1,3/2}_{C,MRW}$ also enjoys $C_{2z}$ symmetry with
\begin{gather}
    R_{C_{2z}} = (R_{C_{4z}})^2 = 
    \begin{pmatrix}
        -1&0\\0&1
    \end{pmatrix}.
\end{gather}
Repeating the previous discussion now for $R_{C_{2z}}$, we get a stronger result: $\ket{\Pi}$ is the same for all $C_{2z}$-invariant momenta $\Pi=\Gamma,\mathrm{M,X,Y}$. There, the polarization $\mathsf{p}_{z}(\Pi)$ has the same value modulo 1. That is, the Hopf insulator with rotational symmetry Eq.~(\ref{eq.rotational_Hopf_C}) harbors a rich set of RTP between the invariant momenta $\Pi=\Gamma,\mathrm{M,X,Y}$. 

Interestingly, the polarizations $\mathsf{p}_{z}(\Gamma,\mathrm{M,X,Y})$ are constrained by a relation, which is determined by the topological invariant $\chi$. Since the occupied subspace is constant on all $C_{2z}$-invariant momenta, so are the coefficients $(\overline{z}\text{i}z)(\mathbf{k})$ of $\bm{\sigma}$ in the Hamiltonian, as a function from BZ into $\bb{S}^2$. We can denote the constant image of $(\overline{z}\text{i}z)(\mathbf{k})$ at an invariant momentum $\Pi$ as $\text{im}(\Pi)$. Then the Whitehead formula~\cite{Whitehead1947Hopf} implies
 \begin{gather}
    \chi = 
    \int_{\Sigma} \mathbf{F}\cdot d\mathbf{\Sigma} = \int_{\partial\Sigma} \mathbf{A}\cdot d\mathbf{l} \n 
    = \mathsf{p}_{z}(\Gamma) + \mathsf{p}_{z}(\mathrm{M}) - \mathsf{p}_{z}(\mathrm{X}) - \mathsf{p}_{z}(\mathrm{Y}) \n 
    = \Delta \mathsf{p}_{\mathrm{X}\Gamma} +\Delta \mathsf{p}_{\mathrm{YM}},
    \label{HopfRTP}
\end{gather} 
where $\Delta \mathsf{p}_{\mathrm{AB}}=\mathsf{p}_{z}(\mathrm{B})-\mathsf{p}_{z}(\mathrm{A})$ and $\Sigma$ is any surface embedded in BZ whose boundary $\partial\Sigma=(\overline{z}\text{i}z)^{-1}(\text{im}(\Gamma))$ is precisely the pre-image of $\text{im}(\Gamma)$. 
In general, the pre-image $\partial\Sigma$ of the Hamiltonian may contain arbitrary closed curves, arranged in a $C_{4z}$-symmetric fashion, in addition to the $C_{2z}$-invariant lines $\Gamma,\mathrm{M,X,Y}$. These additional components contribute a multiple of 4 to the second integral in Eq.~(\ref{HopfRTP}) [For detailed derivation, consult Refs.~\cite{bzdusek2021multicellularity,bzdusek2022rotationdelicateTI}]. Hence, the Hopf-RTP relation can be stated as
\begin{gather}
    \chi \equiv 
    \int_{\Sigma} \mathbf{F}\cdot d\mathbf{\Sigma} = \int_{\partial\Sigma} \mathbf{A}\cdot d\mathbf{l} \mod 4, \n 
    \chi \equiv \Delta \mathsf{p}_{\mathrm{X}\Gamma} +\Delta \mathsf{p}_{\mathrm{YM}} \mod 4,
    \n 
    \partial\Sigma = \Gamma^{+} \cup \mathrm{M}^{+} \cup \mathrm{X}^{-} \cup \mathrm{Y}^{-}.
    \label{HopfRTP_mod4}
\end{gather} 
The superscript ``$+$ ($-$)" in the final line of Eq.~(\ref{HopfRTP_mod4}) indicates that the corresponding component is oriented in the $+\hat{k}_z$ ($-\hat{k}_z$) direction. For the general rule to assign orientation to $\partial\Sigma$, see Appendix~\ref{sec.orientation}. 

While RTP is a name for a general 3D insulator whose polarization difference is quantized between at least two lines in the BZ, the Hopf-RTP relation further constrains the quanta in terms of the bulk Hopf invariant. Were it not for the Hopf-RTP relation, in the model discussed here, $\{\Delta \mathsf{p}_{\mathrm{X}\Gamma}, \Delta \mathsf{p}_{\mathrm{YM}}\}=\{\Delta \mathsf{p}_{\mathrm{X}\Gamma}, \Delta \mathsf{p}_{\mathrm{XM}}\}$ would be a complete set of independent polarization differences since $\Delta \mathsf{p}_{\mathrm{YM}}=\Delta \mathsf{p}_{\mathrm{XM}}$ due to $C_{4z}$ symmetry. However, the bulk topology constrains them as in Eq.~(\ref{HopfRTP}). 

\subsection{Returning Thouless pump in the real Hopf insulator}
\label{sec:RTPforRHI}
A $\mathcal{PT}$-symmetric analogue of Hopf-RTP relation arises in rotational symmetric models. We illustrate this again with an example, which can be investigated with ease thanks to the discussion in the previous section. 

We investigate the RHI Hamiltonian with $C_{4z}$ symmetry
\begin{gather}
    R_{C_{4z}} H^{2,2}_{R}(\mathbf{k}) R_{C_{4z}}^{-1} = H^{2,2}_{R}(C_{4z}\mathbf{k}), 
\label{eq.RHI_C4} \\ 
    R_{C_{4z}} =
    \begin{pmatrix}
    1&&0&&0&&0\\
    0&&1&&0&&0\\
    0&&0&&0&&1\\
    0&&0&&-1&&0
    \end{pmatrix}.
\label{C4z1}
\end{gather}
The models satisfying Eqs.~(\ref{eq.RHI_C4})-(\ref{C4z1}) are easily realized using the Moore-Ran-Wen models [see Appendix~\ref{subsec.C4}], however, we do not specify a model here in order to demonstrate that the existence of RTP is determined by the symmetry operator (\ref{C4z1}) regardless of the functional form of the Hamiltonian. 

In the quaternion notation, the $C_{4z}$ operator in Eq.~(\ref{C4z1}) transforms an eigenstate $\ket{u}$ by
\begin{gather}
    R_{C_{4z}} \ket{u} = \ket{e^{-\text{i}\pi/4} u e^{\text{i}\pi/4}} ,
\end{gather}
where $u=u_{1}+u_{2}\text{i}+u_{3}\text{j}+u_{4}\text{k}$ when $|u\rangle=(u_{1},u_{2},u_{3},u_{4})^{\top}$.

According to the Eq.~(\ref{eq.RHI_C4}), the unoccupied subspace at two points $\mathbf{k}, R_{C_{4z}}\mathbf{k}$ are connected by $R_{C_{4z}}$: For some function valued in the orthogonal group $U(\mathbf{k})\in\mathrm{O}(2)$,
\begin{align}
    R_{C_{4z}} & \ket{\overline{z}w}(\mathbf{k}) \n 
    & = \ket{e^{-\text{i}\pi/4}\overline{z}we^{\text{i}\pi/4}}(\mathbf{k}) \n 
    & = \big(U_{11} \ket{\overline{z}w} + U_{12} \ket{\overline{z}\text{i}w}\big) \big(C_{4z}\mathbf{k}\big),
\label{eq.C4 relation 1} \\
\mbox{} \n 
    R_{C_{4z}} & \ket{\overline{z}\text{i}w}(\mathbf{k}) \n 
    & = \ket{e^{-\text{i}\pi/4} \overline{z}\text{i}w e^{\text{i}\pi/4}}(\mathbf{k}) \n 
    & = \big(U_{21} \ket{\overline{z}w} + U_{22} \ket{\overline{z}\text{i}w} \big) \big(C_{4z}\mathbf{k}\big).
\label{eq.C4 relation 2} 
\end{align}
Reading Eqs.~(\ref{eq.C4 relation 1})-(\ref{eq.C4 relation 2}) in quaternions instead of 4-vectors, we get
\begin{gather}
    (e^{-\text{i}\pi/4}\overline{z}we^{\text{i}\pi/4})(\mathbf{k}) = (U_{11}\overline{z}w + U_{12}\overline{z}\text{i}w)(C_{4z}\mathbf{k}),
\label{eq.C4 relation 3}  \\
    (e^{-\text{i}\pi/4}\overline{z}\text{i}we^{\text{i}\pi/4})(\mathbf{k}) = (U_{21}\overline{z}w + U_{22}\overline{z}\text{i}w)(C_{4z}\mathbf{k}).
\label{eq.C4 relation 4}
\end{gather}
These equations give us transformation rules of $\overline{z}\text{i}z,\overline{w}\text{i}w$ by $C_{4z}$, the 3-vectors that are two entries of the oriented projector. Namely,
\begin{align}
    (e^{-\text{i}\pi/4} \overline{z}\text{i}z e^{\text{i}\pi/4}) (\mathbf{k}) & = (\overline{z}\text{i}z\cdot\det U)(C_{4z}\mathbf{k}) \n 
    & = (\overline{z}\text{i}z)(C_{4z}\mathbf{k}), 
\label{eq.C4 relation 5} \\
    (e^{-\text{i}\pi/4} \overline{w}\text{i}w e^{\text{i}\pi/4})(\mathbf{k}) & = (\overline{w}\text{i}w\cdot\det U)(C_{4z}\mathbf{k}) \n 
    & = (\overline{w}\text{i}w)(C_{4z}\mathbf{k}).
\label{eq.C4 relation 6}
\end{align}
Eq.~(\ref{eq.C4 relation 5}) can be obtained by taking Eq.~(\ref{eq.C4 relation 4}) and multiplying by the conjugate of Eq.~(\ref{eq.C4 relation 3}) from the right. Eq.~(\ref{eq.C4 relation 6}) follows similarly if the multiplication is taken from the left. Finally, the value of $\det U$ can be determined by evaluating both equations at a $C_{4z}$-invariant momentum; to do this, observe that
\begin{gather}
    e^{-\text{i}\pi/4} (a_1\text{i} + a_2\text{j} + a_3\text{k}) e^{\text{i}\pi/4} = a_1\text{i} -a_3\text{j} + a_2\text{k}.
\end{gather}
Suppose $\mathbf{\Pi}=C_{4z}\mathbf{\Pi}$ and $(\overline{z}\text{i}z)(\mathbf{\Pi}) = a_1\text{i} + a_2\text{j} + a_3\text{k}$. Since $\det U = \pm 1$ and the left-hand side of Eq.~(\ref{eq.C4 relation 5}) is a $\mathrm{SO}(3)$ rotation of 3-vectors along the $x$-axis by $-\pi/2$, $\det U=1$ is necessary. 

Interpreting $(\overline{z}\text{i}z)(\mathbf{k})$ as the Hopf insulator Hamiltonian [see Eq.~(\ref{eq.Hopf_Q})]
\begin{gather}
    H_{Q}^{z} = \Re [ \overline{z}\text{i}z \cdot \sigma_Q ],
\end{gather}
Eq.~(\ref{eq.C4 relation 5}) implies the Hamiltonian $H_{Q}^{z}$ has the same symmetry as $H_{C,MRW}^{1,3/2}$ discussed in the previous subsection [see Eq.~(\ref{eq.rotational_Hopf_Q})]. This establishes via Whitehead formula
\begin{gather}
    \chi_{z} \equiv \int_{\Sigma}\mathbf{f}_{z}\cdot d\mathbf{\Sigma} = \int_{\partial\Sigma} \mathbf{a}_{z}\cdot d\mathbf{l} \mod 4, \n 
    \partial\Sigma = \Gamma^{+}\cup \mathrm{M}^{+}\cup \mathrm{X}^{-}\cup \mathrm{Y}^{-}.
    \label{eq.Whitehead_z}
\end{gather}
Analogously, Eq.~(\ref{eq.C4 relation 6}) produces
\begin{gather}
    \chi_{w} \equiv \int_{\Sigma}\mathbf{f}_{w}\cdot d\mathbf{\Sigma}  = \int_{\partial\Sigma} \mathbf{a}_{w}\cdot d\mathbf{l} \mod 4, \n 
    \partial\Sigma = \Gamma^{+}\cup \mathrm{M}^{+}\cup \mathrm{X}^{-}\cup \mathrm{Y}^{-}.
    \label{eq.Whitehead_w}
\end{gather} 
The functions $\mathbf{a_z,f_z,a_w,f_w}$ are the vector forms of Eqs.~(\ref{def.zw connection}), (\ref{def.zw curvature}).

Eqs.~(\ref{eq.Whitehead_z})-(\ref{eq.Whitehead_w}) have consequences on the Wilson loop spectra of RHIs as follows. Using the relation between valence (conduction) band Euler class and the curvature vectors $\mathbf{f}_{z,w}$, we arrive at
\begin{align}
    \chi_{z}+\chi_{w} &\equiv  \int_{\Sigma}\mathbf{Eu}^{v}\cdot d\mathbf{\Sigma} \mod 4 \n 
    & = \Delta \mathsf{p}^{v}_{\mathrm{X}\Gamma} +\Delta \mathsf{p}^{v}_{\mathrm{YM}} \mod 4, 
    \label{eq.Whitehead +} \\
    \chi_{z}-\chi_{w} &\equiv  \int_{\Sigma}\mathbf{Eu}^{c}\cdot d\mathbf{\Sigma} \mod 4 \n
    & = \Delta \mathsf{p}^{c}_{\mathrm{X}\Gamma} +\Delta \mathsf{p}^{c}_{\mathrm{YM}} \mod 4, 
    \label{eq.Whitehead -} \\
    \text{where } \partial\Sigma & = \Gamma^{+} \cup \mathrm{M}^{+} \cup \mathrm{X}^{-} \cup \mathrm{Y}^{-}.
\end{align}
In Eqs.~(\ref{eq.Whitehead +})-(\ref{eq.Whitehead -}), $\mathbf{Eu}^{v}, \mathbf{Eu}^{c}$ is the vector notation for the Euler forms defined in Eqs.~(\ref{Euler connection})-(\ref{Euler form}). Their relationship to $\mathbf{a_z,a_w}$ can be found using Eqs.~(\ref{aczw})-(\ref{avzw}). They can be regarded as the Berry curvatures of the complex-valued state vectors $|u_{+}^{v,c}\rangle=\frac{1}{\sqrt{2}}\left(|u_{1}^{v,c}\rangle+i |u_{2}^{v,c}\rangle\right)$ [see Appendix~\ref{Sec.Chern_basis}].
Passing to the boundary $\partial\Sigma$ with the help of Stokes theorem, Eqs.~(\ref{eq.Whitehead +})-(\ref{eq.Whitehead -}) are polarization differences between $C_{2z}$ invariant lines, where polarizations are calculated using $|u_{+}^{v,c}\rangle$.
These polarization differences leave a mark on the hybrid Wannier functions (HWFs) in the slab geometry finitely layered in the spatial $z$ direction. The result conforms to the multicellular obstruction of delicate topology, which is reported in the next subsection. 

We point out that the form of the Whitehead formula Eqs~(\ref{eq.Whitehead +})-(\ref{eq.Whitehead -}), obtained by assuming $C_{4z}$ rotation symmetry represented by Eq.~(\ref{C4z1}), may alter if the Hamiltonian $H^{2,2}_{R}(\mathbf{k})$ has a different symmetry condition. For instance, see the model in the next subsection, Eq.~(\ref{RTP2}), and the discussions below.

\subsection{Model calculation of the returning Thouless pump}
\label{subsec.RTP_numerical}

\begin{figure}[t]
\includegraphics[width=\linewidth]{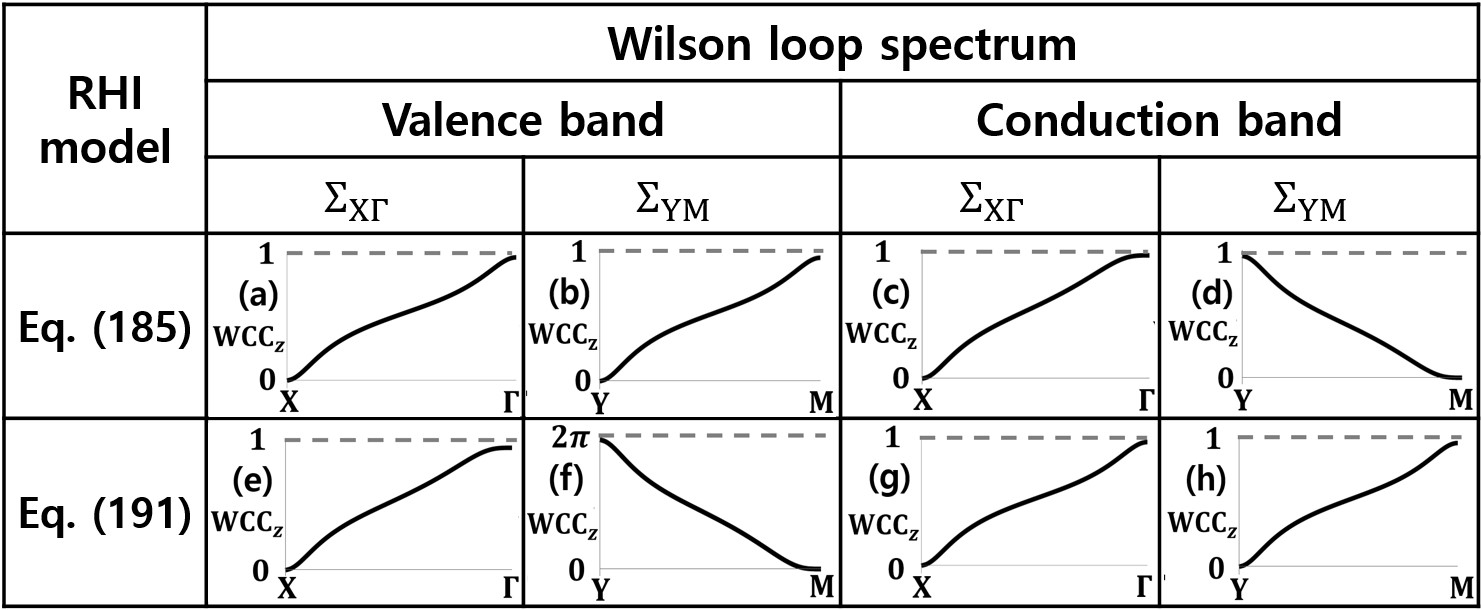}
\caption{
Comparing the flow of the Wannier charge centers (WCCs) between rotation invariant momenta for two RHI (1,1) models with different $C_{4z}$ symmetry.
(a, b) ((c, d)) Wilson loop spectra of $|u_{1}^{v}\rangle+i |u_{2}^{v}\rangle \; (|u_{1}^{c}\rangle+i |u_{2}^{c}\rangle)$ on $\Sigma_{\mathrm{X}\Gamma}$ and $\Sigma_{\mathrm{YM}}$, respectively, calculated using Eq.~(\ref{RTP1}).
$\mathrm{WCC}_{z}$ stands for the Wannier charge center in spatial $z$ direction.
The unit cell length is set to be $1$.
(e, f) ((g, h)) Wilson loop spectra of $|u_{1}^{v}\rangle+i |u_{2}^{v}\rangle \; (|u_{1}^{c}\rangle+i |u_{2}^{c}\rangle)$, respectively, computed for another model in Eq.~(\ref{RTP2}). Difference with the above model manifests the change of real Hopf-RTP relation from Eqs.~(\ref{eq.Whitehead +})-(\ref{eq.Whitehead -}) to Eqs.~(\ref{eq.Whitehead_swap +})-(\ref{eq.Whitehead_swap -}).
} 
\label{fig:RTP}
\end{figure}

We numerically confirm the RTP relations of two RHI tight binding models both having $(\chi_{z},\chi_{w})=(1,1)$.
The first model $H_{R,1}(\mathbf{k})$ is given by Eq.~(\ref{defHzw}), but with unnormalized quaternions (hence unflattened Hamiltonian) $z=z_{0}+z_{1}\text{i}+z_{2}\text{j}+z_{3}\text{k}$ and $w=w_{0}+w_{1}\text{i}+w_{2}\text{j}+w_{3}\text{k}$ where~\cite{deng2013HImodels}
\begin{gather}
    z_{0}=\sin k_{x},~z_{1}=\sin k_{y},~z_{2}=\sin k_{z}, \n 
    z_{3}=\cos k_{x}+\cos k_{y}+\cos k_{z} -\frac{3}{2}, \n 
    w_{0}=\sin k_{x},~w_{1}=\sin k_{y},~w_{2}=\sin k_{z}, \n 
    w_{3}=\cos k_{x}+\cos k_{y}+\cos k_{z} +\frac{3}{2}. 
    \label{RTP1}
\end{gather}
The associated Hopf insulators are defined by 
\begin{gather}
    H_{C,1}^{z}(\mathbf{k}) = - (z^{\dagger}\bm{\sigma}z)(\mathbf{k})\cdot\bm{\sigma}, \n 
    H_{C,1}^{w}(\mathbf{k}) = - (w^{\dagger}\bm{\sigma}w)(\mathbf{k})\cdot\bm{\sigma}.
\end{gather}
These Hamiltonians have $C_{4z}$ symmetry represented in the real gauge as in the previous subsection, which we rewrite here:
\begin{gather}
    R_{R,1} H_{R,1}(\mathbf{k}) R_{R,1}{}^{-1} = H_{R,1}(C_{4z}\mathbf{k}), \\ 
    R_{C,1} H_{C,1}^{z}(\mathbf{k}) R_{C,1}{}^{-1} = H_{C,1}^{z}(C_{4z}\mathbf{k}), \\ 
    R_{C,1} H_{C,1}^{w}(\mathbf{k}) R_{C,1}{}^{-1} = H_{C,1}^{w}(C_{4z}\mathbf{k}), \\ 
    R_{R,1} = \begin{pmatrix}
    1&0&0&0\\
    0&1&0&0\\
    0&0&0&1\\
    0&0&-1&0
    \end{pmatrix}, 
    R_{C,1} = \begin{pmatrix}
    i&0\\
    0&1
    \end{pmatrix}.
\end{gather}

\begin{figure}[t]
\includegraphics[width=\linewidth]{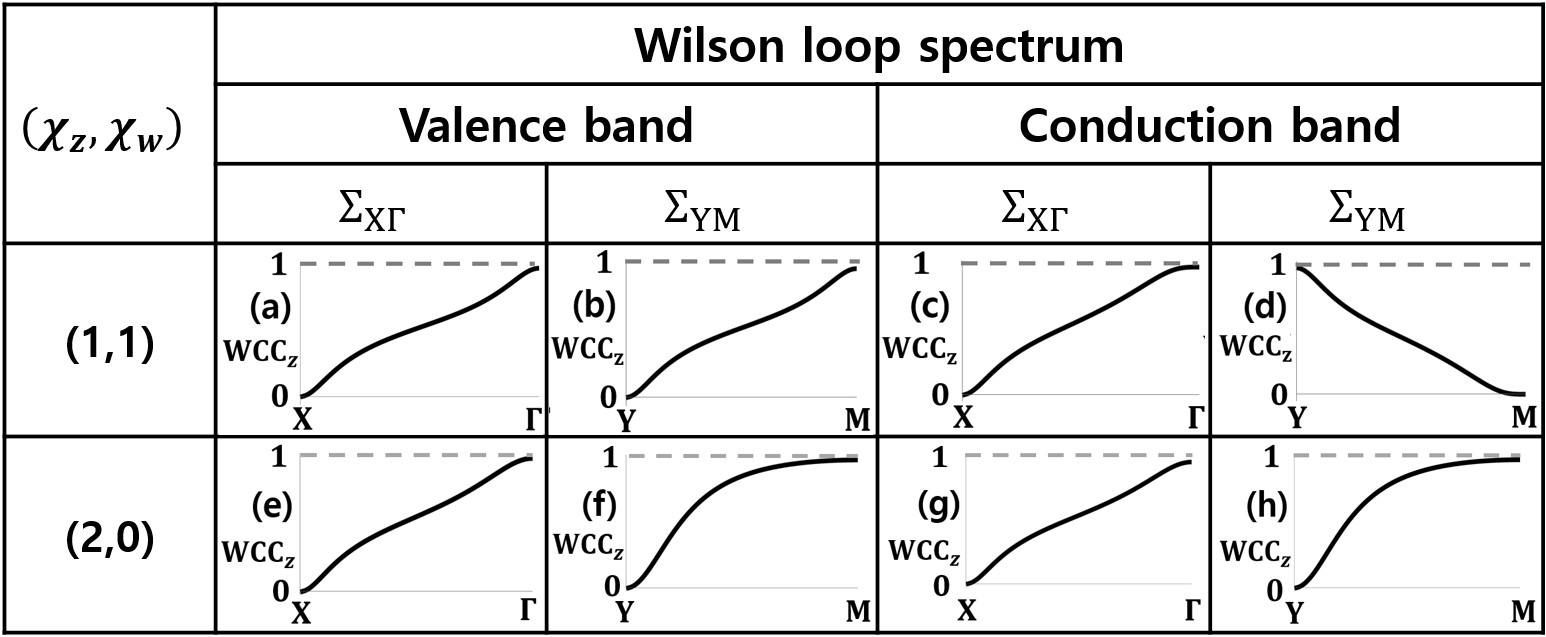}
\caption{
Comparing the flow of the Wannier charge centers between rotation invariant momenta for RHI $(1,1)$ and $(2,0)$ models.
(a, b) ((c, d)) Wilson loop spectra of $|u_{1}^{v}\rangle+i |u_{2}^{v}\rangle \; (|u_{1}^{c}\rangle+i |u_{2}^{c}\rangle)$ on $\Sigma_{\mathrm{X}\Gamma}$ and $\Sigma_{\mathrm{YM}}$, respectively, calculated using Eq.~(\ref{RTP1}) which has $(\chi_{z},\chi_{w})=(1,1)$.
$\mathrm{WCC}_{z}$ stands for the Wannier charge center in spatial $z$ direction.
The unit cell length is set to be $1$.
(e, f) ((g, h)) Wilson loop spectra of $|u_{1}^{v}\rangle+i |u_{2}^{v}\rangle \; (|u_{1}^{c}\rangle+i |u_{2}^{c}\rangle)$, computed for another model in Eq.~(\ref{RHI20}) which has $(\chi_{z},\chi_{w})=(2,0)$. Both models follow the real-Hopf RTP relation Eqs.~(\ref{eq.Whitehead +})-(\ref{eq.Whitehead -}).
}
\label{fig:chiz-chiw}
\end{figure}

Similarly, the second model $H_{R,2}(\mathbf{k})$ with associated Hopf insulators $H_{C,2}^{z}(\mathbf{k}), H_{C,2}^{w}(\mathbf{k})$ is given by
\begin{gather}
    z_{0}=\sin k_{x},~z_{1}=\sin k_{y},~z_{2}=\sin k_{z},\n
    z_{3}=\cos k_{x}+\cos k_{y}+\cos k_{z} -\frac{3}{2},\n
    w_{0}=\sin k_{x},~w_{1}=-\sin k_{y},~w_{2}=-\sin k_{z},\n
    w_{3}=\cos k_{x}+\cos k_{y}+\cos k_{z} +\frac{3}{2}. 
    \label{RTP2}
\end{gather}
Its $C_{4z}$ symmetry has a slightly different representation:
\begin{gather}
    R_{R,2} H_{R,2}(\mathbf{k}) R_{R,2}{}^{-1} = H_{R,2}(C_{4z}\mathbf{k}), \\ 
    R_{C,2}^{z} H_{C,2}^{z}(\mathbf{k}) R_{C,2}^{z}{}^{-1} = H_{C,2}^{z}(C_{4z}\mathbf{k}), \\ 
    R_{C,2}^{w} H_{C,2}^{w}(\mathbf{k}) R_{C,2}^{w}{}^{-1} = H_{C,2}^{w}(C_{4z}\mathbf{k}), \\ 
    R_{R,2} = \begin{pmatrix}
        0&1&0&0\\
        -1&0&0&0\\
        0&0&1&0\\
        0&0&0&1
    \end{pmatrix}, 
    R_{C,2}^{z} = \begin{pmatrix}
        i&0\\
        0&1
    \end{pmatrix},
    R_{C,2}^{w} = \begin{pmatrix}
        -i&0\\
        0&1
    \end{pmatrix}.
\end{gather}
The differing symmetry condition results in a different type of Whitehead formula. Adapting the method in Appendix~\ref{sec.orientation}, one gets an extra minus sign in Eq.~(\ref{eq.Whitehead_w}). The upshot is that while the first model Eq.~(\ref{RTP1}) follows Eqs.~(\ref{eq.Whitehead +})-(\ref{eq.Whitehead -}), the second satisfies a slightly different formulae 
\begin{gather}
    \chi_{z}+\chi_{w} \equiv  \int_{\Sigma}\mathbf{Eu}^{c}\cdot d\mathbf{\Sigma} \mod 4, 
\label{eq.Whitehead_swap +} \\
    \chi_{z}-\chi_{w} \equiv  \int_{\Sigma}\mathbf{Eu}^{v}\cdot d\mathbf{\Sigma} \mod 4.
\label{eq.Whitehead_swap -}
\end{gather} 
Note that the place of valence and conduction bands switched.

For both models, the corresponding ``real Hopf-RTP" relations can be observed in the Wilson loop spectra of states $|u_{1}^{v}\rangle+i |u_{2}^{v}\rangle$ and $|u_{1}^{c}\rangle+i |u_{2}^{c}\rangle$ along the surfaces $\Sigma_{\mathrm{X\Gamma}}$ and $\Sigma_{\mathrm{YM}}$ embedded in BZ, where $\Sigma_{\mathrm{AB}}$ is a surface whose boundary is $\partial\Sigma_{\mathrm{AB}}=\mathrm{B}^{+}\cup\mathrm{A}^{-}$. This is because their Berry connection (Berry curvature) coincides with the Euler connection (Euler form) of the real wave functions $\ket{u_{1,2}^{v,c}}$ [see Appendix~\ref{Sec.Chern_basis}]. 
From Fig.~\ref{fig:RTP}~(a)-(b), we can read off the Euler number of the valence band on $\Sigma=\Sigma_{\mathrm{X\Gamma}}\cup\Sigma_{\mathrm{YM}}$, or the Chern number of $|u_{1}^{v}\rangle+i |u_{2}^{v}\rangle$ on $\Sigma$, which is $1+1=2$. From Fig.~\ref{fig:RTP}~(c)-(d), the Euler number of the conduction band on $\Sigma$ (the Chern number of $|u_{1}^{c}\rangle+i |u_{2}^{c}\rangle$) is $1+(-1)=0$.
This confirms the Whitehead formulae in Eq.~(\ref{eq.Whitehead +}),~(\ref{eq.Whitehead -}).
On the other hand, the data in Fig.~\ref{fig:RTP}~(e)-(h) indicates that the second model Eq.~(\ref{RTP2}) satisfies the other set of Whitehead formulae Eqs.~(\ref{eq.Whitehead_swap +})-(\ref{eq.Whitehead_swap -}).

In Sec.~\ref{sec.bulk-boundary}, the bulk-boundary correspondence of RHI related $\chi_{z}+\chi_{w}$ to the surface Chern number of a finite slab (Eq.~(\ref{RHIBBC})). However, the bulk-boundary correspondence alone did not reveal the physical significance of the other bulk invariant $\chi_{z}-\chi_{w}$.
Intriguingly, both $\chi_{z}+\chi_{w}$ and $\chi_{z}-\chi_{w}$ take part in the RTP physics regulating the Wannier charge centers. For example, two RHIs with different set of invariants $(\chi_{z},\chi_{w})=(1,1)$ and $(\chi_{z},\chi_{w})=(2,0)$ share the same value of $\chi_{z}+\chi_{w}=2$. As a result, two models look identical on the surface.
Nonetheless, they show different behavior regarding the flow of Wannier charge centers between rotation invariant momenta [see Fig.~\ref{fig:chiz-chiw}].

As a final remark, we point out that the (real) Hopf-RTP relation is an incarnation of multicellularity, which was proposed as the counterpart of Wannier obstruction for delicate topological insulators~\cite{bzdusek2021multicellularity}. Each of the equations (\ref{HopfRTP_mod4}), (\ref{eq.Whitehead +})-(\ref{eq.Whitehead -}), and (\ref{eq.Whitehead_swap +})-(\ref{eq.Whitehead_swap -}) says if the bulk invariant in the left-hand side is not zero mod 4, the Wannier center flow corresponding to the right-hand side is nontrivial; as a result, the Wannier center in $z$ direction makes contacts with at least two unit cells.

\section{Conclusion}
In this paper, we have established the band topology of a spinless $\mathcal{PT}$-symmetric insulator in three-dimensions with two occupied and two unoccupied bands, dubbed ``Real Hopf Insulator (RHI)", whose 1D and 2D (weak) topological invariants are trivial by definition. 

The discovery of 3D RHI is in continuation of the previous studies on $\mathcal{PT}$ symmetric topological insulators (TIs) in 1D and 2D, and thus filling the last box in Fig.5. Namely, together with the 1D TI with obstructive atomic band topology characterized by $\pi$-Berry phase and the 2D Euler TI with fragile band topology characterized by an Euler invariant, the 3D RHI with delicate band topology characterized by two Hopf invariants completes the list of $\mathcal{PT}$ symmetric TIs in spinless real fermion systems.

\begin{figure}[t]
\includegraphics[width=\linewidth]{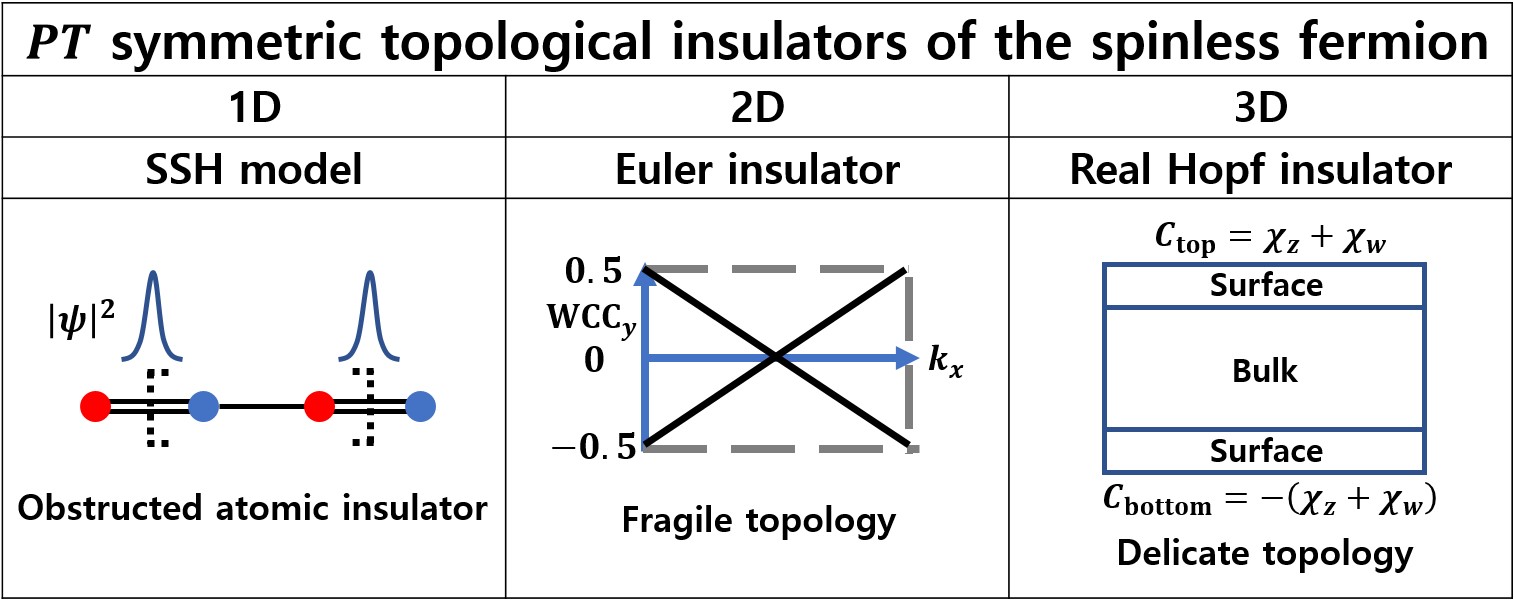}
\caption{
Various topological insulators appearing in $\mathcal{PT}$ symmetric spinless fermion systems. Topological invariants of 1D,2D,and 3D are the Berry phase, the Euler class, and the real Hopf invariants, respectively.}
\label{fig:PTTI}
\end{figure}

With the use of ``oriented" classifying space, which is mathematically the double cover $\bb{S}^2\times\bb{S}^2$ of the original classifying space $\mathrm{Gr}(2,4)$, RHI can be defined using a pair of Hopf maps and their associated shrinking maps. With a fixed pair of unit index Hopf maps, one can enumerate all topological classes of RHI by varying the pair of shrinking maps $(z,w):BZ=\bb{T}^3\to\bb{S}^3\times\bb{S}^3$. Named as the ``enveloping" classifying map, this pair plays a central role in classifying RHIs and computing important physical quantities thereof. 

Enabled by the theory of oriented and enveloping classifying spaces, we identified two novel bulk invariants that are functionals of the Euler connection of both occupied and unoccupied bands [see Eqs.~(\ref{eq.forms & invariants 1})-(\ref{eq.forms & invariants 2})]. The peculiar appearance of Euler connection of both bands indicates the delicate nature of the topology of RHI. 

The delicate character of RHI reveals itself in the bulk-boundary correspondence observed in certain models, which involves the surface Chern number as defined in~\cite{alexandradinata2021HIsurface}. While the surface Chern number of either one of occupied and unoccupied band is not topological, the total surface Chern number is topologically stable and directly related to the bulk RHI invariant. 

Being a $\mathcal{PT}$-symmetric double of Hopf insulator in the sense described in the main text, RTP inherits its physics of returning Thouless pump (RTP)~\cite{bzdusek2021multicellularity,bzdusek2022rotationdelicateTI,
hughes2022spinHopf}.
Two bulk invariants of RHI rules how the polarization of occupied (unoccupied) states jumps between high symmetry lines, provided the system enjoys rotational symmetry. The relevant $\mathcal{PT}$-symmetric version of Hopf-RTP relation~\cite{bzdusek2022rotationdelicateTI} controls the spread of the hybrid Wannier functions, demonstrating the multicellular~\cite{bzdusek2021multicellularity} nature of RHI. 

This paper thus explores a new avenue of delicate topological insulator, rediscovering and building upon the previously reported characteristics of delicate topology. In addition to its own interesting properties, RHI adds to our understanding an abundant set of new examples of TIs. For instance, to the author's knowledge, it provides the first example of an inversion symmetric delicate TI. RHI suggests there is still a rich body of physics of delicate TIs that awaits discovery. 

\begin{acknowledgements}
We thank Yoonseok Hwang for fruitful discussions.
H.L., S.K. and B.-J.Y. were supported by the Institute for Basic Science in Korea (Grant No. IBS-R009-D1), Samsung Science and Technology Foundation under Project
Number SSTF-BA2002-06, the National Research Foundation of Korea (NRF) grant funded by the Korea government (MSIT) (No.2021R1A2C4002773, and No. NRF2021R1A5A1032996).
\end{acknowledgements}

\appendix

\section{Invariance of RHI invariants under the basis transformation}
\label{Sec.invariance}
Here we give detailed proof for the invariance of RHI invariants under the basis change when the invariant is computed by using the occupied wave functions in (\ref{valence states qform_1}).
For convenience, denote
\begin{gather}
	u^{v}_1=A,\quad u^{v}_2=B.
\end{gather}
Under the rotation, $\mathbf{v}_z$ transforms to
\begin{align}
	\nonumber
	\mathbf{v}_z'
	& = (\sin\theta\;A+\cos\theta\;B)(\cos\theta\;\overline{A}-\sin\theta\;\overline{B})
	\n
	& = \sin\theta\cos\theta\;(A\overline{A}-B\overline{B})
	\n
	& \qquad + \cos^2\theta\; B\overline{A} - \sin^2\theta\; A\overline{B}.
\end{align}
From orthonormality of $A,B$ and Eq.~(\ref{R4 inner product quaternion form}), the following facts follow:
\begin{gather}
	A\overline{A}=B\overline{B}=1,
	\\
	\label{RHI eigenstates orthogonality}
	\text{Re}[\;B\overline{A}\;]=0.
\end{gather}
Eq.~(\ref{RHI eigenstates orthogonality}) implies that $B\overline{A}$ is purely imaginary, i.e.
\begin{gather}
	\label{purely imaginary}
	B\overline{A} = - \overline{B\overline{A}} = - A\overline{B}.
\end{gather}
Hence
\begin{gather}
	\mathbf{v}_z'
	= B\overline{A} = \mathbf{v}_z.
\end{gather}
Repeating the above computation for $A=\overline{u^{v}_2}, B=-\overline{u^{v}_1}$ reveals that $\mathbf{v}_w$ is also invariant:
\begin{gather}
	\mathbf{v}_w' = \mathbf{v}_w.
\end{gather}
Therefore, $\mathbf{F}_{z,w}, \mathbf{A}_{z,w}$ hence $\chi_{z,w}$ are all invariant under the rotation Eq.~(\ref{RHI eigenstates rotation}).
\\
We now turn to reflection. As in Eq.~(\ref{purely imaginary}),
\begin{gather}
	\mathbf{v}_z'
	= A\overline{B}
	= - B\overline{A}
	= - \mathbf{v}_z,
	\\
	\mathbf{v}_w'
	= -\overline{B}A
	= \overline{A}B
	= - \mathbf{v}_w.
\end{gather}
Hence $\mathbf{F}_{z,w}, \mathbf{A}_{z,w}$ change their signs under the reflection. However, Eq.~(\ref{def.Hopf Inv vec}) is invariant since it only contains the expression $\mathbf{F}\cdot\mathbf{A}$.

\section{Boundary conditions and weak invariants}
\label{Sec.BC}
In this section, we discuss the lifting problems raised in Sec.~\ref{Sec.modified} and its relation to the weak invariants. As usual, every function appearing in this section is a continuous map between connected CW-complexes. One can avoid this technical nomenclature and regard all spaces as connected manifolds such as spheres, tori, $\mathrm{SO}(n)$, etc. The only purpose of referring to CW-complexes here is to conform to the standard texts in algebraic topology.
\subsection{Lifting problem and Fibration}
\label{Subsec.lifting}
Given a triple of topological spaces $X,Y,\widetilde{Y}$, a continuous map $f:X\rightarrow Y$, and a continuous surjection $p:\widetilde{Y}\rightarrow Y$, the lifting problem is to find another map $\widetilde{f}:X\rightarrow\widetilde{Y}$ such that $p\circ\widetilde{f} = f$ [see \ref{diagram.lift}].
\begin{gather}
\label{diagram.lift}
\begin{tikzcd}[ampersand replacement=\&] 
\& \widetilde{Y} \arrow[d,"p"]\\
X \arrow[ru,dotted,"\widetilde{f}"] \arrow[r,"f"] \& Y
\end{tikzcd}
\end{gather}
In case such $\widetilde{f}$ exists, we say $\widetilde{f}$ is a lift of $f$, or $\widetilde{f}$ covers $f$.

There are special surjective maps $p:\widetilde{Y} \rightarrow Y$, called Serre fibrations, for which the following type of lifting problems can always be solved: Suppose $X=A\times I$ is a product of $A$ with the interval and the restriction $f_0:A\times\{0\}\rightarrow Y$ of the map $f_t:A\times I\rightarrow Y$ has a lift onto $\widetilde{Y}$. Then there exist a lift $\widetilde{f}_t$ that covers $f_t$ and extends $\widetilde{f}_0$. That is, the lifting problem for $(A\times I,Y,\widetilde{Y})$ can be solved so that the resulting lift coincide with the given restricted lift [see~\ref{diagram.Serre}]. This property is commonly referred to as the homotopy lifting property~\cite{hatcher2005algebraic}.
\begin{gather}
\label{diagram.Serre}
\begin{tikzcd}[ampersand replacement=\&] 
\&\& \widetilde{Y} \arrow[d,"p"]\\
A\times\{0\} \arrow[r,hook,"\iota"] \arrow[rru,"\widetilde{f}_0"]\& A\times \mathrm{I} \arrow[ru,dotted,"\widetilde{f}_t" description] \arrow[r,"f_t"] \& Y
\end{tikzcd}
\end{gather}
Covering spaces and the Hopf fibration are typical examples of Serre fibration. Arbitrary compositions and products of Serre fibrations are again Serre fibrations. The latter statement means that whenever $p_i:\widetilde{Y}_i\to Y_i(i=1,2)$ are Serre fibrations, so is $p_1\times p_2:\widetilde{Y}_1\times\widetilde{Y}_2\to Y_1\times Y_2$. For Serre fibration in place of $p$, the lifting problem (\ref{diagram.lift}) is always solvable for $X=\mathrm{I}^n$. Indeed, starting from the point $\mathbf{0}\in \mathrm{I}^n$ and its arbitrary lifting in $p^{-1}(f(\mathbf{0}))$, one can successively apply the homotopy lifting property to get a lift on the entire $\mathrm{I}^n$. Let us apply this fact to the problems relevant to this paper.

Given a general projector function
\begin{gather}
    {P} : \bb{T}^3 \rightarrow \mathrm{Gr}(2,4) \n 
    \mathbf{k} \mapsto [R_{z,w}](\mathbf{k}),
\end{gather}
there is no guarantee that ${P}$ should be lifted to an oriented projector ${P}^{+}:\bb{T}^3\rightarrow\mathrm{Gr}^{+}(2,4)$. However, since the covering map $\mathrm{Gr}^{+}(2,4)\rightarrow\mathrm{Gr}(2,4)$ and the Hopf fibration $\bb{S}^3\rightarrow\bb{S}^2$ are Serre fibrations, the oriented and even enveloping projectors can be defined as a function on $ \mathrm{I}^3$ with the understanding that the original map ${P}$ is a periodic function on $ \mathrm{I}^3$. There are certain boundary conditions on these projectors, however, since ${P}$ is a periodic function on $ \mathrm{I}^3$ [see Sec.~\ref{subsec.construction}]. These boundary conditions are more general than periodic boundary conditions (PBC). In case they can be reduced to the PBC, the lifted projectors can be defined on $\bb{T}^3$ so that the lifting problem on $X=\bb{T}^3$ is solved without compromise. The definability of lifted projectors on $\bb{T}^3$ turns out to be related with the Berry phase and the Euler class, which is the subject of the next subsection.

\subsection{Weak invariants as the obstruction to lifting}
\label{Subsec.obstruction}
Now we prove the aforementioned claim that the oriented projector can be defined on $\bb{T}^3$ if and only if the Berry phases vanish (mod $2\pi$), and the enveloping projector can be defined on $\bb{T}^3$ if and only if the Euler numbers vanish [See Sec.~\ref{subsec.construction}]. The `only if' part of the proposition is easy to show, which goes as follows.

\begin{gather}
\label{diagram.RHI}
\begin{tikzcd}[ampersand replacement=\&] 
\&\& \bb{S}^3\times\bb{S}^3 \arrow[d,"p_2"] \\
\&\& \bb{S}^2\times\bb{S}^2 \arrow[d,"p_1"] \\
\bb{T}^3 \arrow[rru,dotted,"{P}^{+}" description,end anchor={south west}] \arrow[rruu,dotted,"\widetilde{P}",end anchor={south west}] \arrow[rr,"P"] \&\& \mathrm{Gr}(2,4) \&\& \mathrm{I}^3 \arrow[ll,"\underline{{P}}",swap] \arrow[llu,"\underline{{P}}^{+}" description,swap,end anchor={south east}] \arrow[lluu,"\widetilde{\underline{P}}",swap,end anchor={south east}]
\end{tikzcd}
\end{gather}
Definability of lifted projectors is equivalent to the solvability of the corresponding lifting problem, as depicted in the left half of (\ref{diagram.RHI}). Suppose the lifting was achieved for ${P}^{+}$. Since $\pi_1[\bb{S}^2\times\bb{S}^2]=0$, the images of the edges in $\bb{T}^3$ by ${P}^{+}$ are nullhomotopic. Composing the homotopy with $p_1$, we get a nullhomotopy of the edge-images by ${P}$. But an edge is contractible in the classifying space $\mathrm{Gr}(2,4)$ if and only if the Berry phase is trivial. Next, suppose the lifting problem for the enveloping classifying space was solved. Since $\pi_2[\bb{S}^3\times\bb{S}^3]=0$, the images of the faces of $\bb{T}^3$ by $\widetilde{{P}}$ are nullhomotopic. Composing with $p_2$, we get a nullhomotopy of the face-images in $\mathrm{Gr}(2,4)$. This guarantees the well-definedness of the Euler class because the triviality of face maps implies triviality of edge maps, hence the vanishing Berry phases. Moreover, the face maps are trivial if and only if the corresponding Euler class vanishes. This completes the `only if' side of the proof.

The next step is the `if' part, for which the modified classifying BZ $(=\mathrm{I}^3)$ is useful. Recall that the modified oriented projector $\underline{{P}}^{+}$ can be defined on the cube $\mathrm{I}^3$, thanks to the homotopy lifting property of Serre fibrations. Suppose the Berry phases along three edges of BZ vanish, i.e., the edge-images are contractible loops in $\mathrm{Gr}(2,4)$. This implies that $\underline{{P}}^{+}$ sends each edge onto a closed loop in $\bb{S}^2\times\bb{S}^2$, or an element of $\pi_1[\bb{S}^2\times\bb{S}^2]$, which is trivial. Collapsing these edges, the homotopy classes of the face-images of $\underline{{P}}^{+}$ can be regarded as the images of 2-spheres. There are three pairs of opposite faces, each pair projecting onto the same element in $\pi_2[\mathrm{Gr}(2,4)]$ by $p_1$. Since $p_1$ induces the isomorphism $p_1{}_{*}:\pi_2[\bb{S}^2\times\bb{S}^2]\to\pi_2[\mathrm{Gr}(2,4)]$, they were actually identical as an element in $\pi_2[\bb{S}^2\times\bb{S}^2]$. The upshot is that $\underline{{P}}^{+}$ can be continuously deformed so that the opposite faces of $\mathrm{I}^3$ have identical images; after the deformation, $\underline{{P}}^{+}$ becomes a periodic function thus inducing the oriented projector ${P}^{+}:\bb{T}^3\to \bb{S}^2\times\bb{S}^2$.

Finally, we conclude by proving the `if' part for the enveloping projector. To this end, we need a supplementary fact about the Hopf fibration: It is the pullback of the universal $S^1$-bundle, $E\bb{S}^1\to B\bb{S}^1$, by a degree-1 map $h:\bb{S}^2\to B\bb{S}^1$, in the sense that the induced homomorphism of cohomology
\begin{gather}
    h^{*} : H^2(B\bb{S}^1,\bb{Z}) \longrightarrow H^2(\bb{S}^2,\bb{Z}) 
\end{gather}
is an isomorphism.
\begin{gather}
\label{diagram.Hopf}
\begin{tikzcd}[ampersand replacement=\&] 
f^{*}\bb{S}^3 \ar[d,swap] \ar[r] \& \bb{S}^3 \ar[d,"p"] \ar[r] \& E\bb{S}^1 \ar[d] \\ 
X \ar[u,"\sigma",bend left,start anchor={north west},end anchor={[xshift=1.5ex]south west}] \ar[r,"f"] \ar[ru,"\widetilde{f}" description,dotted] \& \bb{S}^2 \ar[r,"h"] \& B\bb{S}^1
\end{tikzcd}
\end{gather} 
Suppose one is trying to solve the lifting problem depicted in (\ref{diagram.Hopf}). What one really needs is a section $\sigma$ of the pullback bundle $f^{*}\bb{S}^3 \longrightarrow X$, so that the composition
\begin{gather}
    \widetilde{f} : X \xrightarrow[]{\;\sigma\;} f^{*}\bb{S}^3 \longrightarrow \bb{S}^3
\end{gather} 
gives the desired lift. The bundle $f^{*}\bb{S}^3\longrightarrow X$ can also be thought as the pullback of the universal bundle by $h\circ f$ and it admits a section if and only if $h\circ f$ is nullhomotopic. Moreover, by the property of classifying space $B\bb{S}^1$, this is equivalent to saying that the cohomology map induced by $h\circ f$
\begin{gather}
    f^{*}\circ h^{*} : H^2(B\bb{S}^1,\bb{Z}) \longrightarrow H^2(\bb{S}^2,\bb{Z}) \longrightarrow H^2(X,\bb{Z}) 
\end{gather}
is zero. Since $h^{*}$ is an isomorphism, it means $f^{*}=0$. This cohomological statement hints at the relevance to the Euler class since characteristic classes represent elements in the de Rham cohomology.

Now we are ready to handle the existence problem of the RHI enveloping projector, which is nothing but a slight modification of (\ref{diagram.Hopf}).
\begin{gather}
\label{diagram.double Hopf}
\begin{tikzcd}[ampersand replacement=\&] 
    (P^{+})^{*}(\bb{S}^3\times\bb{S}^3) \ar[d,swap] \ar[r] \& \bb{S}^3\times\bb{S}^3 \ar[d,"p_2=p\times p"] \ar[r] \& E(\bb{S}^1\times\bb{S}^1) \ar[d] \\ 
    \bb{T}^3 \ar[u,"\sigma",bend left,start anchor={north west},end anchor={[xshift=-2ex]south}] \ar[r,"P^{+}"] \ar[ru,"\widetilde{P}" description,dotted] \& \bb{S}^2\times\bb{S}^2 \ar[r,"h\times h"] \& B(\bb{S}^1\times\bb{S}^1)
\end{tikzcd}
\end{gather} 
Since the universal bundle of $\bb{S}^1\times\bb{S}^1$ is given by the direct product $E(\bb{S}^1\times\bb{S}^1)=E\bb{S}^1\times E\bb{S}^1$, one now has the diagram (\ref{diagram.double Hopf}). Given the oriented projector on $\bb{T}^3$, the sufficient condition for finding its lift to the enveloping level is a section of the leftmost bundle in (\ref{diagram.double Hopf}). The section is available if and only if the bundle is trivial, i.e., the cohomology map induced by $(h\times h)\circ P^{+}$ is nullhomotopic. Since $(h\times h)^{*}$ is an isomorphism, this is to say that the top arrow of the following diagram vanishes. The bottom arrow also vanishes due to the commutativity of the diagram.
\begin{gather}
\label{diagram.cohomology}
\begin{tikzcd}[ampersand replacement=\&]
    H^2(\bb{S}^2\times\bb{S}^2,\bb{Z}) \ar[d,"\approx"] \ar[rr,"(P^{+})^{*}","=0" swap] \&\& H^2(\bb{T}^3,\bb{Z}) \ar[d,"\approx"] \\
    \bb{Z}\times\bb{Z} \ar[rr,"\mathfrak{Ch}","=0" swap] \&\& \bb{Z}\times\bb{Z}\times\bb{Z}
\end{tikzcd}
\end{gather} 
Let us conclude by explaining the bottom arrow of (\ref{diagram.cohomology}), which we name the ``Chern map." 
$\mathfrak{Ch}$ sends the generator $(1,0)$ ($(0,1)$) to the integral (on each face of BZ) of the ``curvature tensor" $f_z$ ($f_w$) associated to the $z$ ($w$) component of the oriented projector. This curvature tensor can be computed by solving the magneto-static equation as in Sec.~\ref{Subsec.RHI inv eig}. Since $BZ=\bb{T}^3$ has three independent faces, this produces a triplet of integers. Accordingly, $\mathfrak{Ch}$ maps into $\bb{Z}^3$. Finally, the element $\mathfrak{Ch}(1,-1)$ ($\mathfrak{Ch}(1,1)$) corresponds to the unoccupied (occupied) band Euler class of three faces [see Sec.~\ref{Subsec.RHI inv Eu}]. Therefore, the condition $\mathfrak{Ch}=0$, being equivalent to $\mathfrak{Ch}(1,-1)=\mathfrak{Ch}(1,1)=0$, means that the Euler classes of both occupied and unoccupied bands all vanish on every face of BZ. 

\section{Free homotopy vs. Based homotopy}
\label{Sec.Free vs Based}

Suppose we have a classifying space $Y$, each point of which representing an occupied subspace of electron states. Any continuous map
\begin{gather}
    f: BZ \longrightarrow Y
    \label{def.classifying map}
\end{gather}
fixes the band structure of the system. As long as we are interested in the topological classification of such systems, any pair of maps $f_0, f_1$ of the form Eq.~(\ref{def.classifying map}) are considered equivalent if they are homotopic (for generality, we replace BZ by an arbitrary space $X$):
\newline\mbox{}\newline
\textbf{(Def. Free Homotopy)} There exists a continuous interpolation
\begin{gather}
    F : X \times [0,1] \longrightarrow Y
    \n 
    F(-,0)=f_0
    \n 
    F(-,1)=f_1.
    \label{def.free homotopy}
\end{gather}
Informally speaking, $f_0$ can be continuously deformed into $f_1$. Eq.~(\ref{def.free homotopy}) is an equivalence relation whose equivalence classes are called \emph{homotopy classes}. Since each equivalence class of band structures corresponds to a homotopy class of Eq.~(\ref{def.classifying map}), it is important to compute the set of homotopy classes of Eq.~(\ref{def.classifying map}). We denote the set as $[X,Y]$.

It turns out that for practical examples of $X$, $[X,Y]$ is related with the homotopy groups $\pi_n[Y]$. Homotopy groups are also defined through homotopy relations but with a variation: We choose a point from each space, $x_0\in X$ and $y_0\in Y$, and only consider the ``based" maps
\begin{gather}
    f: X \longrightarrow Y
    \n 
    \text{bound to the condition } f(x_0)=y_0.
    \label{def.based map}
\end{gather}
Two maps of the form Eq.~(\ref{def.based map}) are homotopic in the based sense if
\newline\mbox{}\newline
\textbf{(Def. Based Homotopy)} There exists a continuous interpolation
\begin{gather}
    F : X \times [0,1] \longrightarrow Y \n 
    F(-,0)=f_0 \n 
    F(-,1)=f_1, \nonumber
\end{gather}
bound to the condition
\begin{gather}
    F(x_0,t)=y_0,\; \forall t\in[0,1].
    \label{def.based homotopy}
\end{gather}
Colloquially, if $f_0$ can be continuously deformed into $f_1$ while obeying the condition $x_0\mapsto y_0$ throughout the deformation process. The set of based homotopy classes, which we will denote by $[X,Y]^{\bullet}$, does not depend on the choice of $x_0, y_0$. The homotopy groups are precisely
\begin{gather}
    \pi_n[Y] \equiv [\bb{S}^n, Y]^{\bullet}.
    \label{def.homotopy grp}
\end{gather}

For the moment, we shall denote the free homotopy class of $f$ as $[f]$ and its based homotopy class as $[f]^{\bullet}$.

Since Eq.~(\ref{def.free homotopy}) allows richer set of interpolations than Eq.~(\ref{def.based homotopy}), $[X,Y]$ is generally ``coarser" than $[X,Y]^{\bullet}$: There are two based maps $f_0, f_1$ that are distinct in the latter while identical in the former, i.e. $[f_0]^{\bullet} \neq [f_1]^{\bullet}$ but $[f_0]=[f_1]$. This difference can be measured by introducing a group action on $[X,Y]^{\bullet}$:
\begin{gather}
    [X,Y]^{\bullet} \times \pi_1[X] \longrightarrow [X,Y]^{\bullet}
    \n 
    ([f]^{\bullet}, [\gamma]^{\bullet}) \mapsto [f\cdot \gamma]^{\bullet}.
    \label{def.pi_1 action}
\end{gather}
This operation is denoted as $\rhd_{\gamma}$ in Ref.~\cite{slager2020geometric}. The definition of the map $f\cdot \gamma: X \rightarrow Y$ will be given in the following paragraphs.

Let $X$ be based at $x_0$ and $Y$ at $y_0$. Then the loop $\gamma: \bb{S}^1\rightarrow Y$ can be interpreted as a free homotopy of the point map
\begin{gather}
    \ast : \{x_0\} \longrightarrow Y
    \n 
    x_0 \mapsto y_0
    \label{ftn.pt}
\end{gather}
to itself. That is,
\begin{gather}
    \gamma : \{x_0\} \times [0,1] \longrightarrow Y
    \n 
    \text{with } \gamma(x_0,0)=\gamma(x_0,1)=y_0
    \label{ftn.point_homotopy}
\end{gather}
continuously interpolates two identical point maps $\gamma(-,0)=\gamma(-,1)=\ast$. Note that we defined $\gamma$ on $\{x_0\} \times [0,1]$ instead of $\bb{S}^1$ and imposed the periodic boundary condition.

Given a based map $f:(X,x_0)\to (Y,y_0)$ and a loop $\gamma$ as in Eq.~(\ref{ftn.point_homotopy}), $\gamma$ specifies a free homotopy of the point map in Eq.~(\ref{ftn.pt}). But $(X,x_0)$, being a CW pair, enjoys the homotopy extension property: There is a free homotopy starting from $f$, which extends the free homotopy $\gamma$ on $\{x_0\}$ onto whole $X$.
\begin{gather}
    \text{There exists a continuous map } \n 
    F : X \times [0,1] \longrightarrow Y \n
    \text{such that } \n 
    F(-,0) = f,\; F(x_0,-)=\gamma(x_0,-).
\end{gather}

\begin{figure}[t!]
\includegraphics[width=8.5cm]{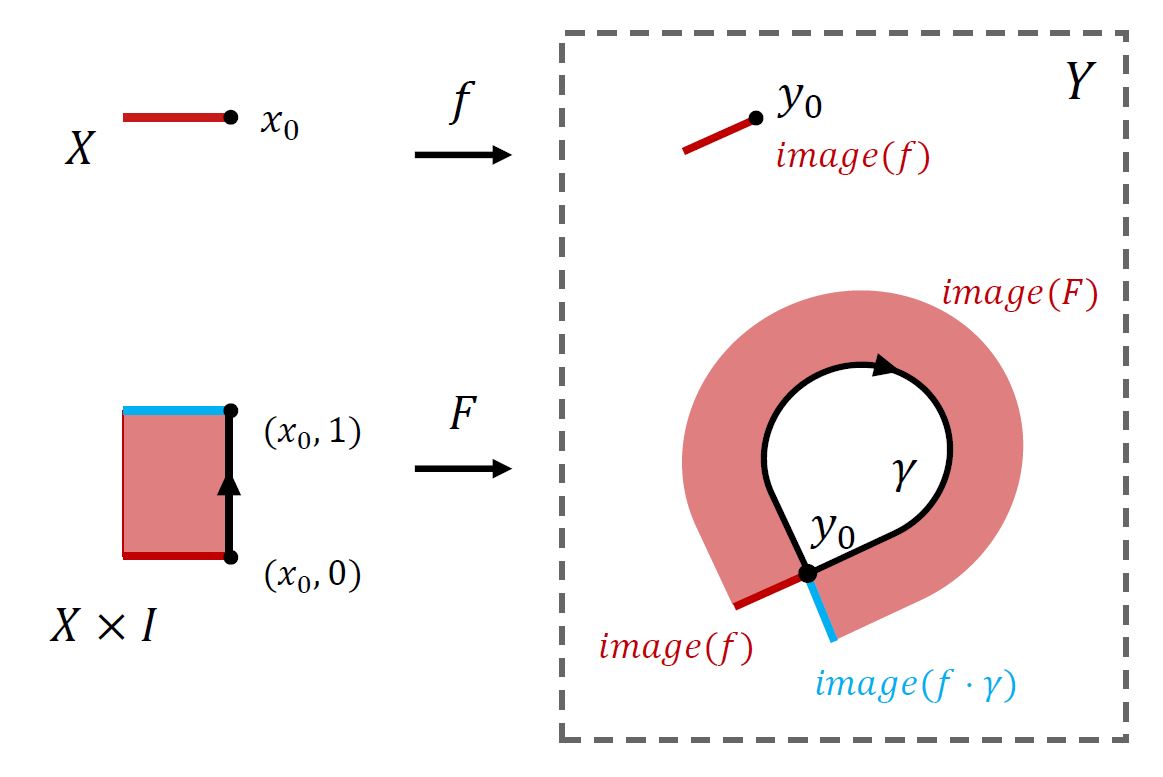}
\caption{
loop $\gamma$ acting on $f$.
} 

\label{fig.loop action}
\end{figure}

\begin{figure}[t!]
\includegraphics[width=8.5cm]{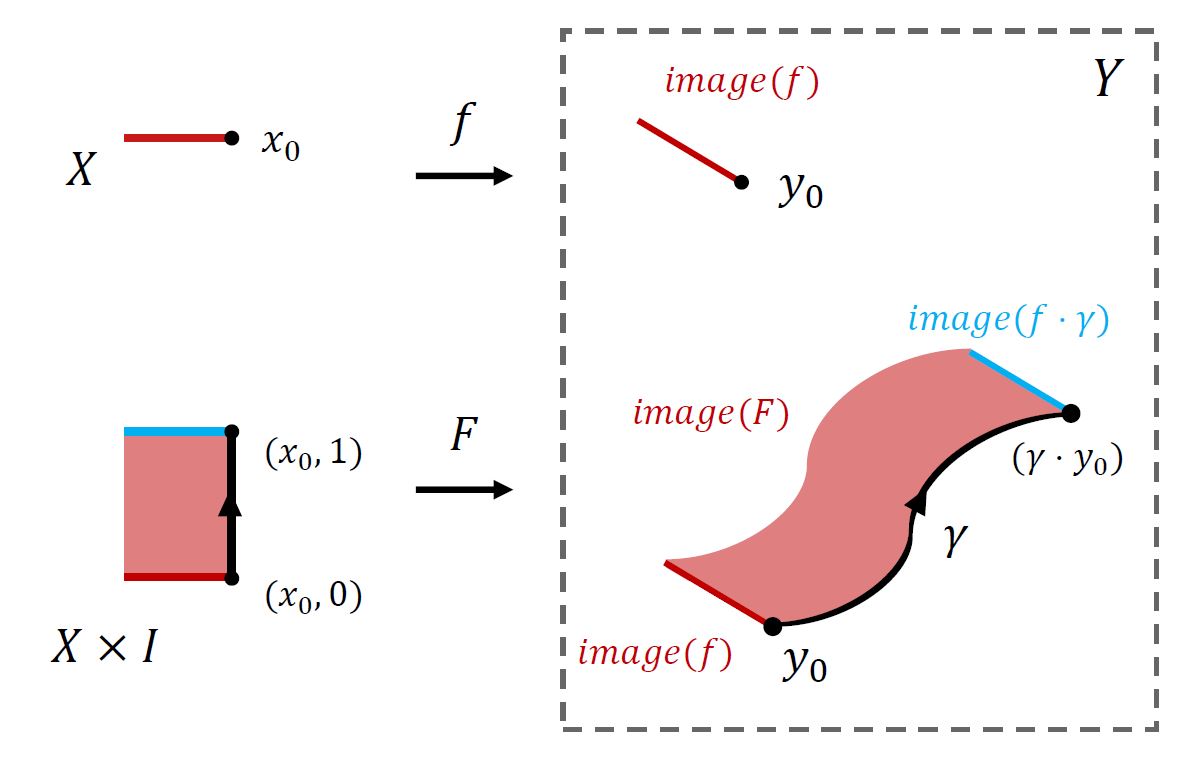}
\caption{
path $\gamma$ acting on $f$.
}
\label{fig.path action}
\end{figure}

\begin{figure}[t!]
\includegraphics[width=8.5cm]{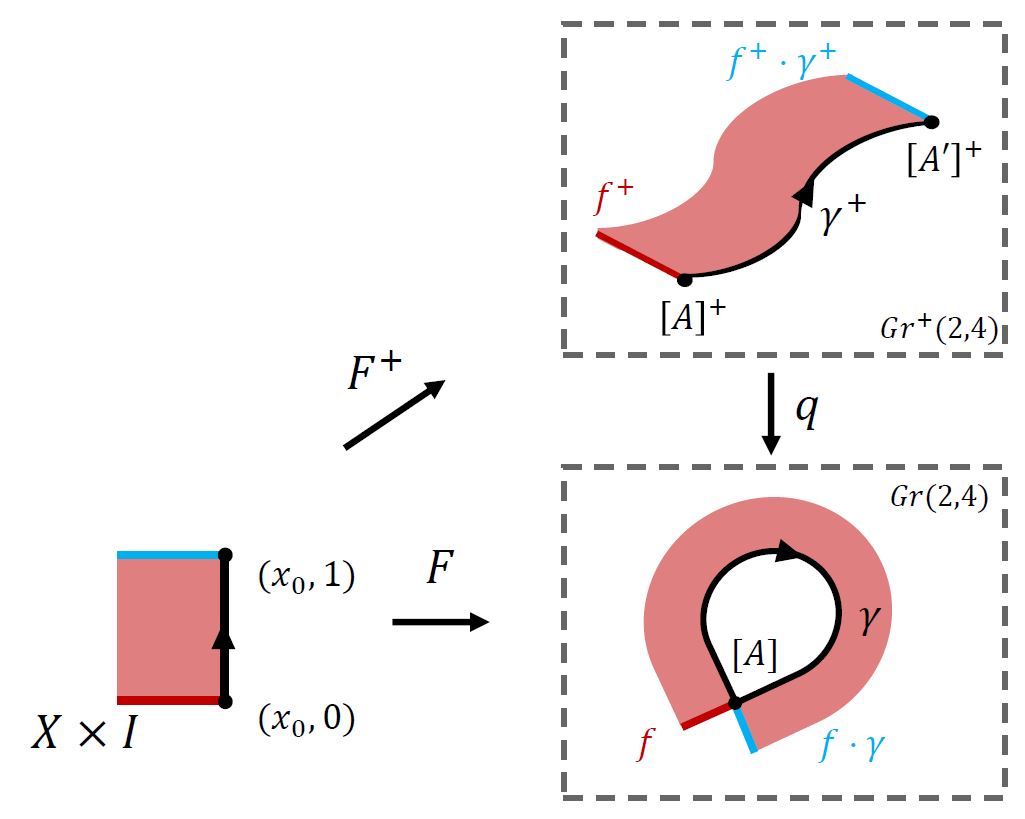}
\caption{
Finding $f\cdot\gamma$: Lift $f$ to $f^{+}$, push along $\gamma^{+}$, and project by $q$.
}
\label{fig.lift}
\end{figure}

To put it another way, $F$ is a (basepoint-free) deformation of $f$ that equals $\gamma$ when restricted to the subset $\{x_0\}\times[0,1] \subseteq X\times[0,1]$. We define $f\cdot\gamma \equiv F(-,1)$ [see Fig.~\ref{fig.loop action}]. Although there are many choices of $F$, the based homotopy class $[f\cdot\gamma]^{\bullet}$ is independent of this choice. Furthermore, $[f\cdot\gamma]^{\bullet}$ only depends on the based homotopy class of $f, \gamma$ so that Eq.~(\ref{def.pi_1 action}) is well-defined. To sum up, $[f\cdot\gamma]^{\bullet}$ is defined by the recipe
\begin{gather}
    \text{Continuously deform $f$ so that } \n 
    \text{the image of $x_0$ moves along $\gamma$.} \n 
    \text{Take $f\cdot\gamma$ to be the end result of the deformation.} \n 
    \text{Then $[f\cdot\gamma]^{\bullet}$ is uniquely determined.}
\label{recipe}
\end{gather}
(\ref{recipe}) determines a group action: Since the identity element $e\in\pi_1[Y]$ is the constant loop,
\begin{gather}
    [f\cdot e]^{\bullet}=[f]^{\bullet}.
\end{gather}
Furthermore,
\begin{gather}
    [(f\cdot\gamma_{1}) \cdot\gamma_{2}]^{\bullet} = [f\cdot(\gamma_{1}\gamma_{2})]^{\bullet}.
\end{gather}
The recipe Eq.~(\ref{recipe}) can be extended to continuous paths whose starting and ending points do not match [see Fig.~\ref{fig.path action}]. In this case, $f\mapsto f\cdot\gamma$ is not a group action; the set $[ I,Y]$ of continuous paths is not a group. Even worse, $f\cdot\gamma$ does not have the base-point condition $x_0\mapsto y_0$ in general. However, $f\mapsto f\cdot\gamma$ for non-loop $\gamma$ is a useful concept for computing Eq.~(\ref{def.pi_1 action}) for $Y$ having a covering space $q:\widetilde{Y}\rightarrow Y$ [see Fig.~\ref{fig.cover}]. In such cases, there exist the ``lifts"
\begin{gather}
    f^{+} : X \longrightarrow \widetilde{Y}, \n 
    \gamma^{+} : [0,1] \longrightarrow \widetilde{Y},
\end{gather}
of $f,\gamma$ such that $q\circ f^{+}=f$, $q\circ\gamma^{+}=\gamma$. $\gamma^{+}$ is a continuous path, but generally not a loop.
The \textit{homotopy lifting property} of covering space guarantees the following:
\begin{gather}
    \text{To find $f\cdot\gamma$ for a loop $\gamma$ in $Y$,} \n 
    \text{Lift $f,\gamma$ to $f^{+},\gamma^{+}$.} \n 
    \text{Find $f^{+}\cdot\gamma^{+}$ and project it down to $Y$.} \n 
    \text{Then the result $[q\circ(f^{+}\cdot\gamma^{+})]^{\bullet}$ equals $[f\cdot\gamma]^{\bullet}$.}
    \label{recipe+}
\end{gather}
For a graphical description, see Fig.~\ref{fig.lift}.

Now we state the main theorem of this subsection.

\mbox{}\newline 
\textbf{Theorem.}
Given two based maps $f$ and $g$, $[f]=[g]$ if and only if $[g]^{\bullet}=[f\cdot\gamma]^{\bullet}$ for some $\gamma$.
\begin{proof}
    The ``if" part is straightforward. Since based homotopies are just special cases of free homotopy, $[g]^{\bullet}=[f\cdot\gamma]^{\bullet}$ automatically implies $[g]=[f\cdot\gamma]$. Moreover, since $f\cdot\gamma$ is defined by a free homotopy from $f$, $[f\cdot\gamma]=[f]$. Then clearly $[g]=[f]$. To show the ``only if" part, suppose $[f]=[g]$, i.e. there is a free homotopy $F$ that interpolates from $f$ to $g$. Since $F(x_0,0)=f(x_0)=y_0=g(x_0)=F(x_0,1)$, $F$ restricted to $\{x_0\}\times[0,1]$ defines a loop with the same basedness condition $x_0\mapsto y_0$. Call this loop $\gamma$. Then $[g]^{\bullet}=[f\cdot\gamma]^{\bullet}$ by the definition of $[f\cdot\gamma]^{\bullet}$.
\end{proof}

Now we can precisely describe how ``coarser" than $[X,Y]^{\bullet}$ is $[X,Y]$. Consider the map
\begin{gather}
    \Phi : [X,Y]^{\bullet} \longrightarrow [X,Y] \n 
    [f]^{\bullet} \mapsto [f].
\end{gather}
Intuitively, as $[f]^{\bullet}$ passes through the function $\Phi$, it ``forgets" the basedness condition $x_0\mapsto y_0$ and identify itself with another map $g$ whenever $f$ can be deformed into $g$ along a free homotopy, even if it is impossible through based homotopy. The theorem says that the class $\Phi^{-1}([f])$ of base-homotopy classes that project onto the free-homotopy class $[f]$ is precisely the orbit of $[f]^{\bullet}$ by the action Eq.~(\ref{def.pi_1 action}). Therefore, we can think of the set $[X,Y]$ as the orbit space of Eq.~(\ref{def.pi_1 action}). Symbolically,
\begin{gather}
    [X,Y]^{\bullet}/\pi_1[Y] \approx [X,Y].
    \label{eq.free vs based}
\end{gather}

\section{Free homotopy and classification of 4 by 4 real Hamiltonian on 1, 2, 3 spatial dimensions}
\label{Sec.Class}
Topological phases of RHI have a one to one correspondence with the set of free homotopy classes of occupied space projectors
\begin{gather}
    {P} : BZ=\bb{T}^3 \longrightarrow \mathrm{Gr}(2,4) \n
    \mathbf{k} \mapsto [R_{z,w}](\mathbf{k}),
\label{bulk map}
\end{gather}
called classifying maps, which assigns to each point $\mathbf{k}$ on the momentum space the projection operator ${P}(\mathbf{k})$ onto the occupied subspace at $\mathbf{k}$. We denote the set of such homotopy classes as $[\bb{T}^3,\mathrm{Gr}(2,4)]$.

$[\bb{T}^3,\mathrm{Gr}(2,4)]$ can be decomposed into different sectors according to the weak invariants, which characterize the band structure along 2D faces of the BZ. We express this fact as
\begin{gather}
    [\bb{T}^3,\mathrm{Gr}(2,4)]
    = \bigcup_{\nu_{xy},\nu_{yz},\nu_{zx}} [\bb{T}^3,\mathrm{Gr}(2,4)]^{(\nu_{xy},\nu_{yz},\nu_{zx})},
\end{gather}
where the weak invariant $\nu_{xy}\in [\bb{T}^2,\mathrm{Gr}(2,4)]$ is the free homotopy class of a given classifying map restricted to the $(k_x,k_y)$-surface in $\bb{T}^3$ (similar for $\nu_{yz}, \nu_{zx}$). Each subset $[\bb{T}^3,\mathrm{Gr}(2,4)]^{(\nu_{xy},\nu_{yz},\nu_{zx})}$ consists of the homotopy classes of continuous maps $\bb{T}^3$ (the bulk of BZ) to $\mathrm{Gr}(2,4)$ whose restriction to each face $\bb{T}^2$ of $\bb{T}^3$ corresponds to $\nu_{xy},\nu_{yz},\nu_{zx}$, respectively. 

The set of free homotopy classes of 2D Hamiltonians $[\bb{T}^2,\mathrm{Gr}(2,4)]$ can be decomposed further as
\begin{gather}
    [\bb{T}^2,\mathrm{Gr}(2,4)]
    = \bigcup_{\alpha_x,\alpha_y} [\bb{T}^2,\mathrm{Gr}(2,4)]^{(\alpha_x,\alpha_y)},
    \label{decomposition of face maps}
\end{gather}
where $\alpha_x\in [\bb{T}^1,\mathrm{Gr}(2,4)]$ is the homotopy class of $\nu_{x,y}$ restricted to the loop along $k_x$ (similarly for $k_y)$. 

Hence, there is a hierarchy of homotopy classes where the lower dimensional data (weak invariants) determine the classification result in one higher dimension. This paper explores the relatively simple sector
\begin{gather}
\label{eq.sector}
    [\bb{T}^3,\mathrm{Gr}(2,4)]^{(0,0,0)} = [\bb{S}^3,\mathrm{Gr}(2,4)]
\end{gather}
where the superscript $(0,0,0)$ means that the restrictions to 2D reduced Brillouin Zones (rBZ) $\nu_{xy}=\nu_{yz}=\nu_{zx}=0 \in[\bb{T}^2,\mathrm{Gr}(2,4)]^{(0,0)} \subset[\bb{T}^2,\mathrm{Gr}(2,4)]$ are the trivial class. Note that 2D (surface) classes being trivial requires every 1D (edge) restrictions to be the trivial element $0\in[\bb{T}^1,\mathrm{Gr}(2,4)]$, hence the superscript (0,0). We will see that triviality of surface and edge maps are equivalent to vanishing Euler numbers and Berry phases, respectively.

The next task is to compute the right-hand side of (\ref{eq.sector}) and its low-dimensional counterparts, which are generally different from the homotopy groups
\begin{gather}
    [\bb{S}^1,\mathrm{Gr}(2,4)]^{\bullet} = \pi_1[\text{Gr(2,4)}], \n 
    [\bb{S}^2,\mathrm{Gr}(2,4)]^{\bullet} = \pi_2[\text{Gr(2,4)}], \n 
    [\bb{S}^3,\mathrm{Gr}(2,4)]^{\bullet} = \pi_3[\text{Gr(2,4)}]. 
\end{gather}
This amounts to computing the $\pi_1[\mathrm{Gr}(2,4)]$-action on the pointed homotopy groups $[\bb{S}^n,\mathrm{Gr}(2,4)]^{\bullet}$ [see Appendix~\ref{Sec.Free vs Based}]. To this end, we consider the transformation
\begin{gather}
\label{transform}
    (z,w) \longrightarrow (z',w')=(\text{j}z,\text{j}w).
\end{gather}
We will show that (\ref{transform}) describes the $\pi_1[\mathrm{Gr}(2,4)]$-action on RHI's associated to the unit quaternion functions $(z,w)$. 

 First, we observe that (\ref{transform}) alters $R_{z,w}$ to $R'_{z,w}$: Directly computing how the columns of $R_{z,w}$ transform, we see
\begin{gather}
    \ket{\overline{z}w} \longrightarrow \ket{\overline{z}w},\quad 
    \ket{\overline{z}\text{i}w} \longrightarrow  -\ket{\overline{z}\text{i}w},\n
    \ket{\overline{z}\text{j}w} \longrightarrow  \ket{\overline{z}\text{j}w},\quad
    \ket{\overline{z}\text{k}w} \longrightarrow  -\ket{\overline{z}\text{k}w}.
\end{gather}
The result is precisely the columns of $R'_{z,w}$. Furthermore, (\ref{transform}) can be interpolated continuously, i.e. it is a result of homotopy
\begin{gather}
    \widetilde{F} : \bb{S}^3 \times \bb{S}^3 \times [0,1] \longrightarrow \bb{S}^3 \times \bb{S}^3 \n 
    (z,w;t) \mapsto (e^{\text{j}\frac{\pi}{2}t} z,e^{\text{j}\frac{\pi}{2}t} w).
    \label{ftn.homotopy}
\end{gather}
We are now ready to describe $f\cdot\gamma$ for a RHI Hamiltonian $f$ and a nontrivial loop $\gamma$ in $\mathrm{Gr}(2,4)$ [see Appendix~\ref{Sec.Free vs Based}].

Since the unique nontrivial loop $[\gamma]^{\bullet}\in\pi_1[\mathrm{Gr}(2,4)]$ can be lifted to a path $\gamma^{+}$ connecting $R_{z_0,w_0}$ to $R'_{z_0,w_0}$ for fixed $(z_0,w_0)$, $f\cdot\gamma$ can be obtained by lifting the Hamiltonian into $f^{+}$ in $\mathrm{Gr}^{+}(2,4)$, applying the path action of $\gamma^{+}$, and projecting $f^{+}\cdot\gamma^{+}$ back onto $\mathrm{Gr}(2,4)$. This is due to the homotopy lifting property of covering spaces.
\subsection{Free homotopy computation in 1D}
\label{Subsec.1d}
In the following sections, dependence on momentum variables $\mathbf{k}$ of an expression is denoted only once for each case, for example, we write $(z,w)(\mathbf{k})$ in place of $(z(\mathbf{k}),w(\mathbf{k}))$. Appropriate use of parentheses will preclude ambiguity.
Suppose we are given a 1D enveloping projector [see Sec.~\ref{subsec.construction}],
\begin{gather}
    \widetilde{\underline{P}}_{1D} :  \mathrm{I}^1 \longrightarrow \bb{S}^3\times\bb{S}^3
    \n 
    k \mapsto (z,w)(k),
\end{gather}
satisfying the boundary condition Eq.~(\ref{eq.BC}) but not necessarily Eq.~(\ref{eq.BC+}). It can alternatively be described by the map
\begin{gather}
    {P}_{1D} : \bb{T}^1 \longrightarrow \mathrm{Gr}(2,4)
    \n 
    k \mapsto [R_{z,w}(k)]
\end{gather}
or the lifted map
\begin{gather}
    \underline{{P}}^{+}_{1D} :  \mathrm{I}^1 \longrightarrow \mathrm{Gr}^{+}(2,4) \approx \bb{S}^2\times\bb{S}^2
    \n 
    k \mapsto  [R_{z,w}(k)]^{+}  \leftrightarrow  (\overline{z}\text{i}z,~\overline{w}\text{i}w)(k) 
\end{gather}
subject to the boundary condition
\begin{gather}
    [R_{z,w}(2\pi)]^{+} =  [R_{z,w}(0)'~]^{+}  \n 
    \leftrightarrow \n  (\overline{z}\text{i}z,~\overline{w}\text{i}w)(2\pi) = -(\overline{z}\text{i}z,~\overline{w}\text{i}w)(0). 
    \label{eq.1D lifted BC}
\end{gather}
The set of topological equivalence classes of 1D systems is
\begin{gather}
    [\bb{T}^1,\mathrm{Gr}(2,4)] = [\bb{S}^1,\mathrm{Gr}(2,4)]
\end{gather}
which we will compute employing the relation Eq.~(\ref{eq.free vs based}). We already know
\begin{gather}
    [\bb{S}^1,\mathrm{Gr}(2,4)]^{\bullet} = \pi_1[\mathrm{Gr}(2,4)] = \{[e]^{\bullet}, [\gamma]^{\bullet}\} \approx \bb{Z}_2,
\end{gather}
and
\begin{gather}
    [\bb{S}^1,\mathrm{Gr}(2,4)] = [\bb{S}^1,\mathrm{Gr}(2,4)]^{\bullet}/\pi_1[\mathrm{Gr}(2,4)],
\end{gather}
where $e\;(\gamma)$ is the contractible (non-contractible) loop in $\mathrm{Gr}(2,4)$. Note that $([\gamma]^{\bullet})^{-1}=[\gamma]^{\bullet}$. Our goal here is to understand how $\pi_1[\mathrm{Gr}(2,4)]$ acts on itself. 

It suffices to take one such projector in the homotopy class and deform it according to (\ref{ftn.homotopy}). As will be shown through Eqs.~(\ref{def.1D model})-(\ref{eq.1D BC}), we can take
\begin{gather}
    \widetilde{\underline{P}}_{1D} :  \mathrm{I}^1 \longrightarrow \bb{S}^3\times\bb{S}^3 \n
    k \mapsto (e^{\text{j}\frac{k}{4}},e^{\text{j}\frac{k}{4}})
\end{gather}
as an initial projector function. The result of homotopy is
\begin{gather}
    \widetilde{\underline{P}}_{1D} \cdot \widetilde{\gamma} :  \mathrm{I}^1 \longrightarrow \bb{S}^3\times\bb{S}^3 \n
    k \mapsto (\text{j}e^{\text{j}\frac{k}{4}},\text{j}e^{\text{j}\frac{k}{4}}).
\end{gather}
Now $\widetilde{{P}}_{1D} \cdot \widetilde{\gamma}$ satisfies the same boundary condition Eq.~(\ref{eq.1D BC-S3}) as $\widetilde{{P}}_{1D}$. This indicates $[\widetilde{{P}}_{1D}\cdot\widetilde{\gamma}]^{\bullet} = [\widetilde{{P}}_{1D}]^{\bullet}$. Projecting the resulting map $\widetilde{\gamma}$ down to $\mathrm{Gr}(2,4)$, we get $[{P}_{1D}\cdot\gamma]^{\bullet} = [{P}_{1D}]^{\bullet}$. That is, the non-contractible loop acts trivially on 1D models. 

The homotopy class of 1D model can also be determined by computing the Berry phase. To this end, one takes a continuous, complex gauge for occupied states as follows:
\begin{gather}
     \ket{u_1} = u_1 \cdot \ket{\overline{z}\text{j}w} + u_2 \cdot \ket{\overline{z}\text{k}w}, 
    \\
     \ket{u_2} = u_3 \cdot \ket{\overline{z}\text{j}w} + u_4 \cdot \ket{\overline{z}\text{k}w}, 
\end{gather}
where
\begin{gather}
    \begin{pmatrix}
    u_1(k) & u_2(k) \\ u_3(k) & u_4(k)
    \end{pmatrix}
    \in U(2).
\end{gather}
Then one computes
\begin{align}
\label{def.Berry phase}
    \int_{0}^{2\pi} dk\;\Tr A 
    & = \int_{0}^{2\pi} dk\;\sum_{j=1}^{2} \bra{u_j} \text{i} \nabla\ket{u_j}
    \n
    & =
    \begin{cases}
    0 \text{ mod } 2\pi \quad (\text{trivial}), \\
    \pi \text{ mod } 2\pi \quad (\text{nontrivial}).
    \end{cases}
\end{align}
We will illustrate one example of a nontrivial 1D model. Define
\begin{gather}
    z(k) = w(k) = e^{\text{j}\frac{k}{4}}
    \label{def.1D model}
\end{gather}
so that $R_{z,w}$ satisfies the boundary condition
\begin{gather}
    R_{z,w}(2\pi) = R_{z,w}(0)'.
    \label{eq.1D BC-mat}
\end{gather}
For the definition of matrix operation $R\to R'$, see Eq.~(\ref{def.A'}). 
Indeed, from
\begin{gather}
    z(2\pi) =  \text{j}z(0),\quad 
    w(2\pi) = \text{j}w(0),
    \label{eq.1D BC-S3}
\end{gather}
it follows that
\begin{alignat}{2}
    &  (\overline{z}w)(2\pi) = (\overline{z}w)(0),  \quad 
    &&  (\overline{z}\text{i}w)(2\pi) = -(\overline{z}\text{i}w)(0), 
    \n 
    &  (\overline{z}\text{j}w)(2\pi) = (\overline{z}\text{j}w)(0),  \quad 
    &&  (\overline{z}\text{k}w)(2\pi) = -(\overline{z}\text{k}w)(0). 
    \label{eq.1D BC}
\end{alignat}
In fact,
\begin{alignat}{2}
    &  (\overline{z}w)(k) = 1,  \quad 
    &&  (\overline{z}\text{i}w)(k) = \text{i} e^{\text{j}(k/2)}, 
    \n 
    &  (\overline{z}\text{j}w)(k) = \text{j},  \quad 
    &&  (\overline{z}\text{k}w)(k) = \text{k} e^{\text{j}(k/2)}. 
    \label{eq.1D eigenstates}
\end{alignat}
Hence, the loop $k\mapsto [R_{z,w}]$ is non-contractible; it lifts to the curve $k\mapsto [R_{z,w}]^{+}$ which connects $[R_{z,w}(0)]^{+}$ to $[R_{z,w}(0)']^{+}$. This can also be seen in the space $\bb{S}^2\times\bb{S}^2$: The lifted classifying map
\begin{gather}
     \mathrm{I}^1 \longrightarrow \bb{S}^2\times\bb{S}^2 \n 
    k \mapsto  (\overline{z}\text{i}z,\overline{w}\text{i}w)(k) 
\end{gather}
satisfies
\begin{gather}
    (\overline{z}\text{i}z,\overline{w}\text{i}w)(2\pi) = -(\overline{z}\text{i}z,\overline{w}\text{i}w)(0), 
    \label{eq.1D BC-S2}
\end{gather}
the boundary condition of Eq.~(\ref{eq.1D lifted BC}).

Using Eq.~(\ref{eq.1D eigenstates}), we form continuous occupied eigenstates
\begin{align}
    \ket{u_1}
    & =  \frac{1}{\sqrt{2}} \bigg[\; 1\cdot \ket{\overline{z}\text{j}w} + e^{\text{i}(k/2)}\cdot \ket{\overline{z}\text{k}w} \;\bigg], 
    \n 
    \ket{u_2}
    & =  \frac{1}{\sqrt{2}} \bigg[\; 1\cdot \ket{\overline{z}\text{j}w} - e^{\text{i}(k/2)}\cdot \ket{\overline{z}\text{k}w} \;\bigg]. 
\end{align}
 Direct computation shows that the Berry phase is $\pi$ (mod $2\pi$) for this model. Furthermore, the action of the non-contractible loop given by (\ref{transform}) amounts to interchanging $\ket{u_1}$ and $\ket{u_2}$, hence leaving the trace in (\ref{def.Berry phase}) unchanged. This again confirms that the $\pi_1$-action on the $1D$ model is trivial.

\subsection{Free homotopy computation in 2D}
\label{Subsec.2d}
 
Henceforth, we assume the boundary condition Eq.~(\ref{eq.BC+}) so that all Berry phases vanish. In this case, each face map belongs to the sector with trivial edges in Eq.~(\ref{decomposition of face maps}), namely
\begin{align}
    [ \mathrm{I}^2,\mathrm{Gr}(2,4)]^{(0,0)}
    & = [\bb{S}^2,\mathrm{Gr}(2,4)] \n
    & =  [\bb{S}^2,\mathrm{Gr}(2,4)]^\bullet/\pi_1[\mathrm{Gr}(2,4)],
    \label{eq.S2 ht}
\end{align}
which we will now compute. As suggested in the Berry phase calculation in the previous section, one can associate a topological invariant to each homotopy class in $[\bb{S}^2,\mathrm{Gr}(2,4)]^\bullet$ and investigate how it transforms under (\ref{transform}). First, we observe
\begin{gather}
    [\bb{S}^2,\mathrm{Gr}(2,4)]^\bullet = 
    \pi_2[\mathrm{Gr}(2,4)] = \pi_2[\mathrm{Gr}^{+}(2,4)] \n 
    = \pi_2[\bb{S}^2\times\bb{S}^2] \approx \bb{Z}\oplus\bb{Z}.
\end{gather}
Here each of the two $\bb{Z}$ components contains the homotopy classes of the map  $\mathbf{k}\mapsto (\overline{z}\text{i}z)(\mathbf{k})$ ($(\overline{w}\text{i}w)(\mathbf{k})$).  However, these integers can be directly computed from  $z$ ($w$) $\in\bb{S}^3$  instead of $\overline{z}\text{i}z$~($ \overline{w}\text{i}w)$, which reveals the relation between these integer invariants and Euler class. 

 Given a 2D model with the enveloping projector
\begin{gather}
    \widetilde{\underline{P}}_{2D} :  \mathrm{I}^2 \longrightarrow \bb{S}^3\times\bb{S}^3 \n
    \mathbf{k} \mapsto (z,w)(\mathbf{k}),
    \label{ftn.2D_model}
\end{gather}
we extract two differential forms as in Sec.~\ref{Subsec.bulk invariants}:
\begin{gather}
     a_z = \Re[ -\text{i} \cdot z~\dd \overline{z} ], \quad
    a_w = \Re[ -\text{i} \cdot w~\dd \overline{w} ], 
    \\
    f_z = \dd a_z, \quad
    f_w = \dd a_w.
\end{gather}
Then the integers in $\pi_2[\mathrm{Gr}(2,4)]=\bb{Z}\oplus\bb{Z}$ are the first Chern class
\begin{gather}
    \nu_z = \frac{1}{2\pi} \int_{ \mathrm{I}^2} f_z, \quad
    \nu_w = \frac{1}{2\pi} \int_{ \mathrm{I}^2} f_w.
\end{gather} 
Now we will see how $\pi_1[\mathrm{Gr}(2,4)]$ acts on these numbers. The result of homotopy in Eq.~(\ref{ftn.homotopy}) acted on the function (\ref{ftn.2D_model}) is represented by the $(z,w)$ function
\begin{gather}
    \widetilde{\underline{P}}_{2D}\cdot\widetilde{\gamma} :  \mathrm{I}^2 \longrightarrow \bb{S}^3\times\bb{S}^3 \n 
    \mathbf{k} \mapsto  (\text{j}z,\text{j}w)(\mathbf{k}). 
\end{gather}
It can be shown that the transformation
\begin{gather}
    z(\mathbf{k}) \rightarrow \text{j}z(\mathbf{k})
\end{gather}
flips the sign of $a_z$:
\begin{align}
    a'_{z} 
    &  = \Re[ -\text{i} \cdot z'~\dd \overline{z}'~]  \n
    &  = \Re[ -\text{i} \cdot \text{j}z \dd\overline{z}(-\text{j})~]  \n
    &  = - \Re[\; \text{j} ( -\text{i} \cdot z \dd \overline{z} ) (-\text{j})~]  \n 
    &  = - \Re[ -\text{i} \cdot z \dd \overline{z}~] \quad\text{ (use Eq.~(\ref{eq.Re conj}).)}  \n 
    & = - a_z.
\end{align}
Therefore, the invariants corresponding to $\widetilde{\underline{P}}_{2D}\cdot\widetilde{\gamma}$ is $(-m, -n)$, the orbit space of $(m,n)$ being $\{\pm(m,n)\}$. As a result,
\begin{gather}
    [ \mathrm{I}^2,\mathrm{Gr}(2,4)]^{(0,0)} \approx \bb{Z}\oplus\bb{Z}/\{\pm(1,1)\}.
\end{gather}

A (3D) real Hopf insulator, which is the main subject of this paper, is characterized by a 3D enveloping projector
\begin{gather}
    \widetilde{\underline{P}}_{3D}\cdot\widetilde{\gamma} :  \mathrm{I}^3 \longrightarrow \bb{S}^3\times\bb{S}^3\n 
    \mathbf{k} \mapsto (z,w)(\mathbf{k})
\end{gather}
with Euler classes $\nu_z,\nu_w=0$ on every 2D face of the 3D BZ. This vanishing of Euler numbers can be expressed in terms of the Euler classes of the occupied (unoccupied) eigenstates, which is physically more relevant than abstract quantities $\nu_z,\nu_w$. Namely,
\begin{gather}
    \nu^v = \frac{1}{2\pi} \int_{\bb{T}^2} \text{Eu}^v, \quad
    \nu^c = \frac{1}{2\pi} \int_{\bb{T}^2} \text{Eu}^c,
\end{gather}
where the Euler forms $\text{Eu}^{v, c}$ are constructed from the real-valued eigenstates $u^{v,c}_1, u^{v,c}_2$ of the respective band following the recipe
\begin{gather}
\label{def.euler}
    a^{v,c} = \bra{u^{v,c}_2} \dd \ket{u^{v,c}_1},\quad
    \text{Eu}^{v,c} = \dd a^{v,c}.
\end{gather}
A simple computation [see~\ref{Subsec.RHI inv Eu}] shows that
\begin{gather}
    \text{a}^v = a_z + a_w,\quad
    \text{a}^c = a_z - a_w,
\end{gather}
which in turn implies
\begin{gather}
    \text{Eu}^v = f_z + f_w,\quad
    \text{Eu}^c = f_z - f_w,
\end{gather}
hence the equality
\begin{gather}
    \nu^v = \nu_z + \nu_w,\quad
    \nu^c = \nu_z - \nu_w.
\end{gather}
We can now state that the vanishing of $\nu_z,\nu_w$ is equivalent to the vanishing of physical Euler classes $\nu^v,\nu^c$.

\subsection{Free homotopy computation in 3D}
\label{Subsec.3D}
A (3D) RHI, being trivial when restricted to every 2D section of BZ, is represented by the homotopy classes in
\begin{align}
    [ \mathrm{I}^3,\mathrm{Gr}(2,4)]^{(0,0,0)}
    & = [\bb{S}^3,\mathrm{Gr}(2,4)] \n 
    & = [\bb{S}^3,\mathrm{Gr}(2,4)]^\bullet/\pi_1[\mathrm{Gr}(2,4)] \n 
    & = \pi_3[\mathrm{Gr}(2,4)]/\pi_1[\mathrm{Gr}(2,4)].
	\label{eq.S3 ht}
\end{align}
The machinery to compute this has already been prepared. The third homotopy group of the (honest) classifying space is
\begin{gather}
    \pi_3[\mathrm{Gr}(2,4)] \approx \bb{Z}\oplus\bb{Z}
\end{gather}
where the two integer invariants in the right-hand side is given by
\begin{gather}
    (\chi_z, \chi_w) = \left( \frac{-1}{4\pi^2}\int_{BZ} a_z \wedge f_z, \frac{-1}{4\pi^2}\int_{BZ} a_w \wedge f_w \right).
\end{gather}
From the previous section we know that the transformation $(z,w)\rightarrow(\text{j}z,\text{j}w)$ flips the sign of $(a_z,a_w)$ and $(f_z,f_w)$. As a result, this transformation has no effects on 3D invariants and the $\pi_1[\mathrm{Gr}(2,4)]$-action on these invariants is trivial. Therefore
\begin{gather}
    [ \mathrm{I}^3,\mathrm{Gr}(2,4)]^{(0,0,0)} = \pi_3[\mathrm{Gr}(2,4)] \approx \bb{Z}\oplus\bb{Z}.
\end{gather}

\section{Uniqueness of the integral formulae for bulk RHI invariants} 
\label{Sec.uniqueness} 

For convenience, we will use the notation
\begin{gather}
     x_0=1,~x_1=\text{i},~x_2=\text{j},~x_3=\text{k}. 
\end{gather}
The non-abelian Berry connection of RHI is
\begin{gather}
    A_{\alpha\beta} \nonumber
     = \bra{\overline{z} x_{\beta}w} \dd \ket{\overline{z} x_{\alpha}w}  \n
     = \Re[x_\alpha\overline{x}_\beta\; z \dd \overline{z}] + \Re[\overline{x}_\alpha x_\beta\; w \dd \overline{w}] 
     \n
    = A^{1}_{\alpha\beta}\;dk_x + A^{2}_{\alpha\beta}\;dk_y + A^{3}_{\alpha\beta}\;dk_z
     \n
    (\alpha,\beta=0, 1, 2, 3).
\end{gather}
Since $A_{\beta\alpha}=-A_{\alpha\beta}$, there are $6\times 3=18$ independent components, namely
\begin{gather}
    A_{01}^{i}, A_{02}^{i}, A_{03}^{i}, A_{12}^{i}, A_{23}^{i}, A_{31}^{i}
    \quad (i=1,2,3).
\end{gather}
The Berry curvature is given by
\begin{gather}
    F_{\alpha\beta}=\dd A_{\alpha\beta}.
\end{gather}
We expect that the topological invariants of RHI can be expressed as the form
\begin{align}
    \chi[p]
    = \int_{BZ}\; p(A_{\alpha\beta},F_{\alpha\beta}),
\end{align}
where $p$ is a polynomial of $A_{\alpha\beta}, F_{\alpha\beta}$'s $(\alpha,\beta=1,i,j,k)$. For dimensional reason, $p$ only consists of the following two types of terms:
\begin{gather}
    A_{\mu\nu} \wedge F_{\rho\sigma},
    \\
    A_{\mu\nu} \wedge A_{\rho\sigma} \wedge A_{\tau\lambda},
\end{gather}
where each of $\mu, \nu, \rho, \sigma, \tau, \lambda$ individually runs over $\{1, \text{i} , \text{j}, \text{k}\}$ while $i,j,k$ are dummy indices.
Gauge invariance strongly restricts the form of $p$.
There are two independent gauge transformations,
\begin{gather}
    (z, w) \rightarrow  (e^{\text{i}\theta_1}z,~e^{\text{i}\theta_1}w ), 
    \label{eq.GT1}
    \\
    (z, w) \rightarrow  (e^{\text{i}\theta_2}z,~e^{-i\theta_2}w ). 
    \label{eq.GT2}
\end{gather}
By Eq.~(\ref{eq.GT1}), $A_{\alpha\beta}$ transforms as
\begin{gather}
    \begin{matrix}
    A_{01}\\A_{02}\\A_{03}\\A_{23}\\A_{31}\\A_{12}\\
    \end{matrix}
    \quad\longrightarrow\quad
    \begin{matrix}
    A_{01} \\
     \cos2\theta~A_{02}-\sin2\theta~A_{03}  \\
     \sin2\theta~A_{02}+\cos2\theta~A_{03}  \\
     A_{23}-2\dd\theta  \\
     \cos2\theta~A_{31}-\sin2\theta~A_{12}  \\
     \sin2\theta~A_{31}+\cos2\theta~A_{12}  \\
    \end{matrix}
\end{gather}
while Eq.~(\ref{eq.GT2}) takes action as
\begin{gather}
    \begin{matrix}
    A_{01}\\A_{02}\\A_{03}\\A_{23}\\A_{31}\\A_{12}\\
    \end{matrix}
    \quad\longrightarrow\quad
    \begin{matrix}
     A_{01}-2 \dd \theta  \\
     \cos2\theta~A_{02}-\sin2\theta~A_{12}  \\
     \sin2\theta~A_{31}+\cos2\theta~A_{03}  \\
    A_{23}\\
     \cos2\theta~A_{31}-\sin2\theta~A_{03}  \\
     \sin2\theta~A_{02}+\cos2\theta~A_{12}.  \\
    \end{matrix}
\end{gather}
it can be observed that gauge invariance disagrees with the dimensional consideration. To see this, observe that the gauge invariance enforces some of $A_{\alpha\beta}$'s to pair up; for example, the first transformation, along with the dimensional analysis, only permits
\begin{gather}
     A_{02}\wedge F_{03},~
    A_{31}\wedge F_{12},~
    F_{02}\wedge A_{03},~
    F_{31}\wedge A_{12},  \n
     A_{02}\wedge F_{31} + A_{03}\wedge F_{12},~
    F_{02}\wedge A_{31} + F_{03}\wedge A_{12},  \n
     A_{02}\wedge F_{12} - A_{03}\wedge F_{31},~
    F_{02}\wedge A_{12} - F_{03}\wedge A_{31}. 
\end{gather}
Nonetheless, none of these terms are invariant under the second transformation. The only polynomials that are gauge invariant when integrated over BZ are
\begin{gather}
    A_{01}\wedge F_{01},~
    A_{23}\wedge F_{23},~
    A_{01}\wedge F_{23},~
    A_{23}\wedge F_{01}. 
\end{gather}
Among these, only the first two are constructed purely from the occupied (or unoccupied) bands. However, it is easy to show that (e.g. by testing for numerical models with tunable parameters) they have non-quantized values.

\section{Proof of Eq.~(\ref{tslabtot})}
\label{sec:prooftslabtot}
In Ref.~\cite{vanderbilt2017BBCofCSA} a quantity $\theta_{0}^{v}$ is defined for an insulator with the open boundary condition in $x,y,z$,
\begin{gather}
\theta_{0}^{v}=-8\pi^{2}\mathrm{ImTr}\left[PxPyPz\right],
\label{t0occ}
\end{gather}
where $P$ is the projection operator onto the occupied states and the trace is over not only the occupied states but also the unoccupied states.
Suppose that there are $N$ unit cells in $x,y$-directions and $N_{z}$ unit cells in $z$-direction.
When there are lots of unit cells in $x,y$-directions, the geometry of this insulator can be regarded as a slab geometry.
Under this limit, we can compare $\theta_{\mathrm{slab}}^{v}$ in Eq.~(\ref{th2occ}) with $\theta_{0}^{v}$.
In Ref.~\cite{vanderbilt2017BBCofCSA}, it is shown that
\begin{gather}
\frac{\theta_{0}^{v}}{N^{2}A_{\mathrm{cell}}}\rightarrow N_{z}c~\theta_{\mathrm{slab}}^{v}~~\mathrm{for}~N\rightarrow \infty ,
\label{t0t2thm}
\end{gather}
where $A_{\mathrm{cell}}$ is the area of the unit cell in $xy$ plane.
Eq.~(\ref{t0t2thm}) is satisfied when the Chern number of the occupied states of the entire slab is trivial;
this is satisfied for the Hopf insulator.
For the unoccupied states, $\theta_{0}^{c}$ can be defined similarly as Eq.~(\ref{t0occ}), and $\theta_{0}^{c},~\theta_{\mathrm{slab}}^{c}$ satisfy the equation Eq.~(\ref{t0t2thm}) after replacing occupied states by unoccupied states.
If we show that $\theta_{0}^{v}+\theta_{0}^{c}=0$, then Eq.~(\ref{tslabtot}) is straightforward.
After substituting $Q=\mathbb{1}-P$ in $\theta_{0}^{c}$, $\theta_{0}^{v}+\theta_{0}^{c}$ is given by
\begin{gather}
\theta_{0}^{v}+\theta_{0}^{c}=-8\pi^{2}\mathrm{ImTr}(xyz-Pxyz-xPyz\n
-xyPz+PxPyz+PxyPz+xPyPz).
\label{t0occ+unocc}
\end{gather}
Obviously, $\mathrm{ImTr}[xyz]=0$ since $xyz$ is a Hermitian operator.
$\mathrm{Tr}[Pxyz]$ is given by
\begin{gather}
\mathrm{Tr}[Pxyz]=\sum_{n\in occ}\langle \psi_{n}|xyz|\psi_{n} \rangle ,
\end{gather}
where $|\psi_{n}\rangle$ is an eigenstate and $n$ is an index for the occupied eigenstates.
Since $\mathrm{Im}\left[\bra{\psi_{n}}xyz\ket{\psi_{n}}\right]=0$ for all $n\in occ$, the second term in Eq.~(\ref{t0occ+unocc}) is zero.
Also, $\mathrm{Tr}[xPyz]=\mathrm{Tr}[Pyzx]$ and $\mathrm{Tr}[xyPz]=\mathrm{Tr}[Pzxy]$.
Therefore, the third and fourth terms in Eq.~(\ref{t0occ+unocc}) are zero.
$\mathrm{Tr}[PxPyz]$ is given by
\begin{gather}
\mathrm{Tr}[PxPyz]=\sum_{n,m\in occ}\langle \psi_{n}|x|\psi_{m}\rangle\langle\psi_{m}|yz|\psi_{n}\rangle ,
\end{gather}
The complex conjugated $\mathrm{Tr}[PxPyz]$ is given by
\begin{gather}
\mathrm{Tr}[PxPyz]^{*}=\sum_{n,m\in occ}\langle \psi_{m}|x|\psi_{n}\rangle\langle\psi_{n}|yz|\psi_{m}\rangle\n
=\mathrm{Tr}[PxPyz].
\end{gather}
Therefore, $\mathrm{ImTr}[PxPyz]=0$.
Note that $\mathrm{Tr}[PxyPz]=\mathrm{Tr}[PzPxy]$ and $\mathrm{Tr}[xPyPz]=\mathrm{Tr}[PyPzx]$.
Therefore, $\mathrm{ImTr}[PxyPz]=\mathrm{ImTr}[xPyPz]=0$ due to the same reason.

 \section{Interpretation of Euler connection as the Berry connection of the Chern basis wave functions} 
 \label{Sec.Chern_basis} 
Here we describe the method of Chern basis as a convenient way to interpret the Euler connection of a pair of real wave functions as the Berry connection of a complex wave function, which is used several times in this paper.

Given a two-band real wave functions,
\begin{gather}
    \ket{u_1(\mathbf{k})}, \ket{u_2(\mathbf{k})},
    \label{real_wf}
\end{gather}
one often computes the Euler connection $\text{a}=\bra{u_2}\dd\ket{u_1}$ and other quantities derived from it. For example, Eqs.~(\ref{eq.Whitehead +})-(\ref{eq.Whitehead -}) relates $\text{Eu}=\dd\text{a}$ with the bulk invariants of RHI. It is in turn interpreted in Sec.~\ref{subsec.RTP_numerical} as the Wilson loop spectral flow
\begin{gather}
    \mathsf{p}_{+}(k_x,k_y) = \frac{1}{2\pi} \int_{0}^{2\pi} \bra{u_2}\partial_{k_z}\ket{u_1} dk_z \mod1, \n
    \mathsf{p}_{-}(k_x,k_y) = \frac{-1}{2\pi} \int_{0}^{2\pi} \bra{u_2}\partial_{k_z}\ket{u_1} dk_z \mod1,
    \label{ftn.Wilson_spectrum}
\end{gather}
as a function along a curve in the $(k_x,k_y)$-plane. But passing to the complex gauge defined by the ``Chern basis"
\begin{gather}
    \ket{\psi_{+}(\mathbf{k})} = \frac{1}{\sqrt{2}} \bigg( \ket{u_1(\mathbf{k})} + i \ket{u_2(\mathbf{k})}\bigg), \n 
    \ket{\psi_{-}(\mathbf{k})} = \frac{1}{\sqrt{2}} \bigg( \ket{u_1(\mathbf{k})} - i \ket{u_2(\mathbf{k})}\bigg), \n
    \label{def.Chern_basis}
\end{gather}
it is easy to see that the Berry connection of the complex wave function is given by
\begin{gather}
    i\bra{\psi_{\pm}}\dd\ket{\psi_{\pm}} = \pm \bra{u_2}\dd\ket{u_1}.
    \label{eq.Chern_connection}
\end{gather}
When the reality of wave functions in Eq.~(\ref{real_wf}) arises from an anti-unitary symmetry $\mathcal{I}$ realized by the complex conjugation $\mathcal{K}$, the complex wave functions in Eq.~(\ref{def.Chern_basis}) are a $\mathcal{I}$-symmetric pair, hence conveniently exposing the symmetry of the given problem. In our case, $\mathcal{I}=\mathcal{PT}$.

We also point out that the surface Chern number described in Sec.~\ref{sec.bulk-boundary}, which takes the opposite sign at the top and bottom surfaces of a slab, is another manifestation of Chern basis. If $\ket{\psi_{+}}$ in Eq.~(\ref{def.Chern_basis}) describes the surface state localized at the top surface, $\ket{\psi_{-}}$ corresponds to the bottom side of the slab. They necessarily carry opposite Chern numbers due to $\mathcal{PT}$ symmetry.

\section{Wannier sheet crossings of RHI with approximate rotational symmetry}
\label{sec.crossing} 
 We demonstrate that those RHIs with ``large polarization differences" necessarily have crossing points, robust against small perturbations, interconnecting the bulk Wannier sheets of adjacent unit cells. To be precise, the robust Wannier crossings are unavoidable as soon $|\Delta\mathsf{p}_{\mathrm{AB}}|\geq2$ for either occupied or unoccupied band and for any pair of points $\mathrm{A,B}$ on the 2D rBZ. As a result, the Wannier sheets cannot be detached and packaged into completely decoupled groups, resulting in a breakdown of the surface theorem [see Sec.~\ref{sec.bulk-boundary}].

\begin{figure}[t]
\includegraphics[width=\linewidth]{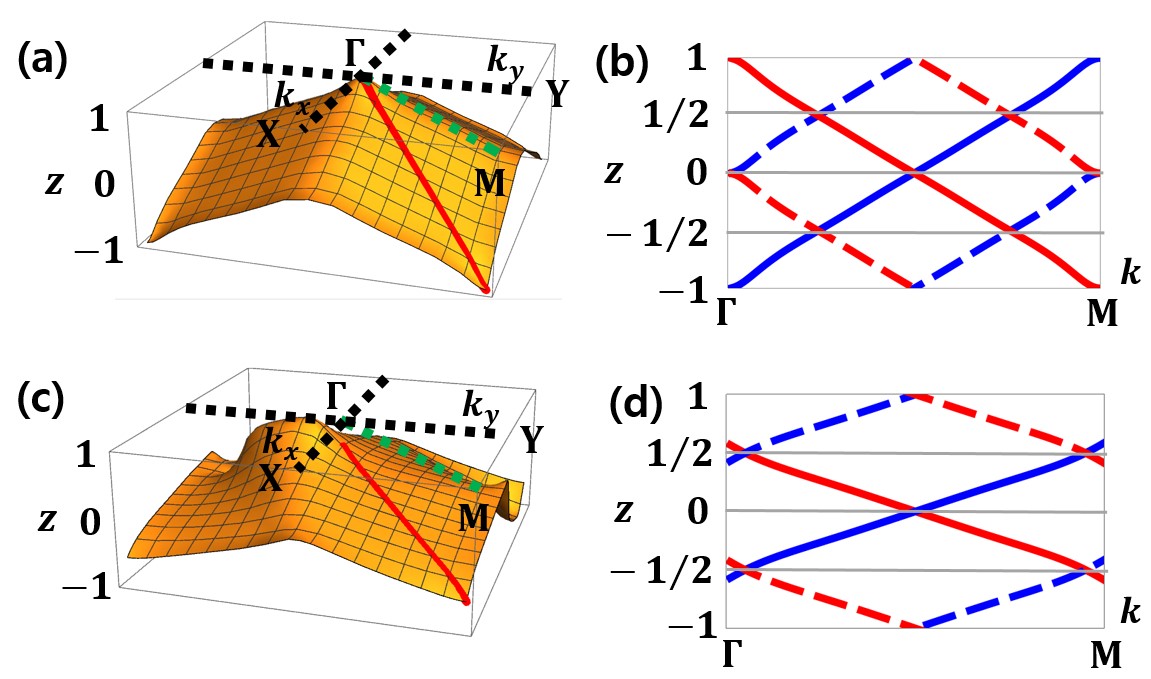}
\caption{
Wannier sheet crossing of a real Hopf insulator model in Eq.~(\ref{RTP3}) with $(\chi_z,\chi_w)=(1,1)$.
(a) One of the bulk Wannier sheets of $|u_{1}^{c}\rangle+i |u_{2}^{c}\rangle$ of Eq.~(\ref{RTP1}), which is the $\lambda_1=\lambda_2=0$ limit of Eq.~(\ref{RTP3}). 
In the $(k_x,k_y)$ plane, the green dotted line connects $\Gamma$ and $\mathrm{M}$ without passing through the $C_{2z}$-invariant momenta along the way, thus being a path $\gamma$ as described in the main text. Along $\gamma$, the Wannier center travels two unit layers (red solid line).
(b) Cross-section of the Wannier sheet in (a) along the $\Gamma \mathrm{M}$ line. 
The red solid line is the cross-section of the Wannier sheet in (a) and the blue solid line is its $\mathcal{PT}$ symmetric partner.
The red (blue) dashed lines are given by the parallel translation of the red (blue) solid line and depict Wannier sheets of adjacent unit cells. 
(c) One of the bulk Wannier sheets of $|u_{1}^{c}\rangle+i |u_{2}^{c}\rangle$ of Eq.~(\ref{RTP3}) with $\lambda_{1}=\lambda_{2}=0.7$.
(d) Cross-section of the Wannier sheets in (c) and their partners along $\Gamma \mathrm{M}$.}

\label{fig:RTP2}
\end{figure}

In this section, we illustrate this principle with the $C_{4z}$-symmetric RHI in Eq.~(\ref{RTP3}). The model describes a family of RHIs with small $C_{4z}$ breaking perturbation, free of gap closing while $|\lambda_{1}|,|\lambda_{2}|<\frac{\sqrt{3}}{2}$. In the limit $\lambda_1=\lambda_2=0$, the model reduces to the model in Eq.~(\ref{RTP1}) and recovers $C_{4z}$ symmetry. However, the point remains valid for the other rotation symmetric models, inasmuch as Sec.~\ref{sec.RTP} provides a general scheme applicable to any kind of rotation symmetry.
\begin{gather}
    z_{0}=\sin k_{x},~z_{1}=\sin k_{y}+\lambda_{1},~z_{2}=\sin k_{z},\n
    z_{3}=\cos k_{x}+\cos k_{y}+\cos k_{z} -\frac{3}{2},\n
    w_{0}=\sin k_{x}+\lambda_{2},~w_{1}=\sin k_{y},~w_{2}=\sin k_{z},\n
    w_{3}=\cos k_{x}+\cos k_{y}+\cos k_{z} +\frac{3}{2},
    \label{RTP3}
\end{gather}
When $C_{4z}$-symmetry is present, one can find many $C_{4z}$-symmetric RHIs that harbor $|\Delta\mathsf{p}_{\mathrm{AB}}|\geq2$.
As will be seen, this enforces a series of crossing points interconnecting the bulk Wannier sheets. This phenomenon is often observed when $\chi_z,\,\chi_w$ are both nonzero; a case of $(\chi_z,\chi_w)=(1,1)$ is illustrated in Fig.~\ref{fig:RTP2}.

In the following, let us suppose without loss of generality that $|\mathsf{p}^{c}_{\Gamma\mathrm{M}}|=m\geq2$. This implies that the bulk Wannier sheets constructed from the Chern basis state $\ket{u_{1}^{c}}+i\ket{u_{2}^{c}}$ [see Appendix~\ref{Sec.Chern_basis}] range across at least two unit cells. Fig.~\ref{fig:RTP2}~(a) depicts such a Wannier sheet.

Under the condition $|\mathsf{p}^{c}_{\Gamma\mathrm{M}}|=m\geq2$, we can choose a continuous path $\gamma$ in the $(k_x,k_y)$-plane in the BZ that begins at $\Gamma$, ends at $M$, and otherwise never passes through any one of $\Gamma,\mathrm{M,X,Y}$.

Along such $\gamma$, the Wannier centers traverse $m$ times the unit cell length. Different Wannier sheets must intersect each other due to the large value of $m\geq2$ (Fig.~\ref{fig:RTP2}~(b)). These crossing points between conduction band Wannier sheets can be operationally divided into two groups of integer-spaced points. Setting the positions of the crossing points to be the union of integers and half integers, we name the group of crossing points sitting on the integers as the even crossings, while those sitting on the half odd integers will be called odd. In Fig.~\ref{fig:RTP2}~(b), the odd crossings arise as the touching points between two Wannier sheets from adjacent unit cells. On the other hand, the even crossings are made between the $\mathcal{PT}$-symmetric pairs of Wannier sheets inside a single unit cell. Since the Wannier sheets do not merely osculate but completely pass over each other on both groups of crossing points, these points do not disappear after perturbing the Hamiltonian unless the perturbation severely destroys the Wannier sheet configuration (Fig.~\ref{fig:RTP2}~(c)-(d)). The spacing between the crossing points remains smaller than a unit cell length, preventing the $\mathcal{PT}$-pairs of Wannier sheets from being decoupled from the other pairs and lying in a single unit cell. This nullifies the scheme of the surface theorem to define surface Chern numbers.

\section{Directions of the Berry curvature of the $C_{4z}$ symmetric Hopf insulator along the $C_{4z}$ invariant lines}
\label{sec.orientation}
To apply the Whitehead formula~\cite{Whitehead1947Hopf} to a surface $\Sigma$, we need to specify the orientation on every component of $\partial\Sigma$. The three rules completely determine the orientations $\partial\Sigma$~\cite{bzdusek2021multicellularity}:

First, due to the trivial first Chern class, the number of lines with $+\hat{k}_z$ direction and that of $-\hat{k}_z$ direction are the same. 

Second, due to the rotational symmetry, the line passing through a momentum $\mathbf{k}$ has the same orientation with that passing through $C_{nz}\mathbf{k}$. 

Finally, the direction of any line component $\gamma$ coincides with the sign of the $z$-component of the Berry curvature $F_z = -2\Im\braket{\partial_{k_x}\psi^v|\partial_{k_y}\psi^v}$. Note that the last rule is a master principle that determines the orientation for an arbitrary line component. The first two rules serve to minimize the labor. 

Let us apply these rules to the model in Sec.~\ref{sec.RTP}, Eq.~(\ref{RTP1}).
The Bloch Hamiltonian $H(\mathbf{k})$ satisfies 
\begin{gather}
    R_{C_{4z}} H(\mathbf{k}) R_{C_{4z}}^{-1} = H(C_{4z}\mathbf{k}), \quad 
    R_{C_{4z}} = \begin{pmatrix}
    i && 0 \\
    0 && 1
    \end{pmatrix}, 
    \label{eq.K1}
\end{gather}
where $C_{4z}\mathbf{k}=(-k_{y},k_{x},k_{z})$. For any invariant momentum $\Pi=C_{4z}\Pi=(\Pi_x,\Pi_y,\Pi_z)$, the Hamiltonian can be expanded as
\begin{gather}
    H(\Pi+\bm{\kappa}) = a_{x}(\bm{\kappa}) \sigma_x + a_{y}(\bm{\kappa}) \sigma_y + a_{z}(\bm{\kappa}) \sigma_z, \\
    a_{x}(\bm{\kappa}) = A_{0}(\Pi) + A_{x}(\Pi)\kappa_{x} + A_{y}(\Pi)\kappa_{y},\n 
    a_{y}(\bm{\kappa}) = B_{0}(\Pi) + A_{x}(\Pi)\kappa_{x} + B_{y}(\Pi)\kappa_{y},\n 
    a_{z} (\bm{\kappa}) = C_{0}(\Pi) + A_{x}(\Pi)\kappa_{x} + C_{y}(\Pi)\kappa_{y} 
\label{eq.K2}
\end{gather}
for small $\bm{\kappa}=(\kappa_x,\kappa_y)$. The term proportional to the identity matrix is ignored because it does not affect the eigenstates. The constraint~(\ref{eq.K1}) implies 
\begin{gather}
    A_{0}=B_{0}=C_{x}=C_{y}=0, \; B_{x}=-A_{y}, \; B_{y}=A_{x}.
\end{gather}
Then $H(\mathbf{k})$ near $\Pi$ is given by
\begin{gather}
H(\mathbf{k})=C_{0}\sigma_{z}+(A_{x}\kappa_{x}+A_{y}\kappa_{y})\sigma_{x}+(-A_{y}\kappa_{x}+A_{x}\kappa_{y})\sigma_{y}.
\end{gather}
The $z$-directional Berry curvature can now be computed:
\begin{gather}
    F_{z} = -2\mathrm{Im}\langle \partial_{\kappa_{x}} \psi^v| \partial_{\kappa_{y}} \psi^v\rangle \n 
    = -2\mathrm{Im}\frac{\langle \psi^v|\partial_{\kappa_{x}}H|\psi^c\rangle\langle \psi^c|\partial_{\kappa_{y}}H|\psi^v\rangle}{(E_{c}-E_{v})^{2}},
\label{Berrycurvatureinz}
\end{gather}
where $\ket{\psi^v}$ ($\ket{\psi^c}$) is the occupied (unoccupied) state with energy $E_{v}$ ($E_{c}$). 

To determine the orientation of the invariant line anchored at $\Pi$, Eq.~(\ref{Berrycurvatureinz}) is evaluated on $\Pi$. 
Along the $C_{4z}$ invariant line, $H(\mathbf{k})\equiv C_{0}\sigma_{3}$.
Since $H(\mathbf{k})$ is insulating, $C_{0}\neq 0$. The answer depends on the sign of $C_0$.
First, let $C_{0}>0$.
In this case, $|\psi^v\rangle=(0,1)^{\top}$, $|\psi^c\rangle=(1,0)^{\top}$, and
\begin{gather}
    F_{z} = \frac{1}{2|C_{0}|^{2}}(A_{x}^{2}+A_{y}^{2}) > 0.
\end{gather}
That is, the orientation of the $\Pi$-line is upward. 

Now suppose the opposite $C_{0}<0$.
In this case, $|\psi^v\rangle=(1,0)^{\top}$ and $|\psi^c\rangle=(0,1)^{\top}$, which leads to
\begin{gather}
    F_{z} = -\frac{1}{2|C_{0}|^{2}}(A_{x}^{2}+A_{y}^{2}) < 0.
\end{gather}
The corresponding orientation is downward.  

Using the fact that $\psi^v, \psi^v$ are always the eigenvectors of $R_{C_{4z}}$, one can deduce the orientation from the corresponding eigenvalues (angular momenta), even without the functional form of the Hamiltonian. For the present case, where Eq.~(\ref{eq.K1}) is known, the angular momentum $l^{v,c}$ is determined (modulo 4) by the relation $R_{C_{4z}}\ket{\psi^{v,c}}=\exp(2\pi i l^{v,c}/4)$. Then previous analysis yields
\begin{gather}
    \Delta l = l^v - l^c \equiv -1 \mod 4 \Rightarrow \text{upward orientation}, \n 
    \Delta l = l^v - l^c \equiv 1 \mod 4 \Rightarrow \text{downward orientation}.
\end{gather}

\section{Symmetries of Moore-Ran-Wen RHI}
\label{Sec.Symmetries}
The real Hopf insulators we discussed so far have $\mathcal{PT}$ symmetry by construction (i.e. they have real Hamiltonians). Each homotopy class of RHIs can be realized using two Moore-Ran-Wen models for Hopf insulator, and such RHI models have additional symmetries.

The fact that a general Hamiltonian $H$ has a symmetry $S$ with its operator representation $R_S$ can be stated as
\begin{gather}
\label{symmetry_ham}
    R_S H(\mathbf{k}) R_S^{-1} = H(S\mathbf{k}).
\end{gather}
Without changing the homotopy class of the Hamiltonian, the energy spectrum can be flattened so that the restriction of the Hamiltonian onto the occupied subspace is a constant multiple of the occupied state projector,
\begin{gather}
    \overline{H}(\mathbf{k}) = \bb{1} - 2\sum_{n\in occ} \ket{\psi_n(\mathbf{k})}\bra{\psi_n(\mathbf{k})} = \bb{1} - 2 {P}_{occ}(\mathbf{k}).
\end{gather}
The conduction (valence) energy level is fixed at constant $+1$ ($-1$), placing the Fermi level at $E_F=0$. Then (\ref{symmetry_ham}) can be recast as
\begin{gather}
\label{symmetry_proj}
     R_S {P}_{occ}(\mathbf{k}) R_S^{-1} = {P}_{occ}(S\mathbf{k}).
\end{gather}
(\ref{symmetry_proj}) states that $\ket{\psi(\mathbf{k})}$ is an occupied (unoccupied) state with momentum $\mathbf{k}$ if and only if $R_S \ket{\psi(\mathbf{k})}$ is an occupied(unoccupied) state with momentum $S\mathbf{k}$; in other words,
\begin{gather}
\label{symmetry_state}
    R_S \ket{\psi^\alpha(\mathbf{k})} = U^\alpha{}_\beta(\mathbf{k}) \ket{\psi^\beta(\mathbf{Sk})}
\end{gather}
where $\psi^{\alpha=1,\ldots,N}$ run over all (independent) occupied eigenstates and $U(\mathbf{k})$ is a 2 by 2  unitary matrix. For the symmetries not involving time reversal, $R_S$ is just an orthogonal matrix whereas for those involving time reversal, $R_S=\mathcal{O}\cdot \mathcal{K}$ for some $\mathcal{O}\in \mathrm{O}(4)$ and the complex conjugation $\mathcal{K}$. 

 In the remaining part of this section, we fix 
\begin{gather}
    z
    = z_0 + z_1 \text{i} + z_2 \text{j} + z_3 \text{k} ,
    \\
    w
    = w_0 + w_1 \text{i} + w_2 \text{j} + w_3 \text{k} ,
\end{gather}
where
\begin{alignat}{1}
\nonumber
    & z_0 = \sin k_x,
    \quad
    z_1 = \sin k_y,
    \quad
    z_2 = \sin [m k_z],
    \\
\nonumber
    & z_3 = \cos k_x+\cos k_y+\cos[m k_z] - \frac{3}{2}, 
    \\
\nonumber
    & w_0 = \sin k_x,
    \quad
    w_1 = \sin k_y,
    \quad
    w_2 = \sin [n k_z],
    \\
\nonumber
    & w_3 = \cos k_x+\cos k_y+\cos[n k_z] - \frac{3}{2}, 
    \\
\label{Moore model zw vectors}
    & m, n \in \bb{Z}, 
\end{alignat}
so that
\begin{gather}
    \chi_z = m, \quad \chi_w = n.
\end{gather}
The resulting Hamiltonian $H_{m,n}$ has the conduction band spanned by
\begin{alignat}{2}
\nonumber
    &  \overline{z}w 
    = && (z_0w_0+z_1w_1+z_2w_2+z_3w_3)
    \\
\nonumber
    & && +  (z_0w_1 - z_1w_0 - z_2w_3 + z_3w_2 )~\text{i} 
    \\
\nonumber
    & && +  (z_0w_2 + z_1w_3 - z_2w_0 - z_3w_1 )~\text{j} 
    \\
\nonumber
    & && +  (z_0w_3 - z_1w_2 + z_2w_1 - z_3w_0)~\text{k}, 
    \\
    \mbox{}\nonumber
    \\
\nonumber
    &  \overline{z}\text{i}w 
    = &&  (-z_0w_1 + z_1w_0 - z_2w_3 + z_3w_2) 
    \\
\nonumber
    & && +  (z_0w_0 + z_1w_1 - z_2w_2 - z_3w_3)~\text{i} 
    \\
\nonumber
    & && +  (z_0w_3 + z_1w_2 + z_2w_1 + z_3w_0)~\text{j} 
    \\
\label{Moore model zwbar ziwbar}
    & && +  (z_0w_2 + z_1w_3 + z_2w_0 + z_3w_1)~\text{k}. 
\end{alignat}

\subsection{$C_{4z}$ symmetry of Moore-Ran-Wen RHI} 
\label{subsec.C4}
For $S=C_{4z}$ symmetry, the crystal momentum $\mathbf{k}$ transforms as
\begin{gather}
\label{sym.C4z}
    S(k_x,k_y,k_z) = (-k_y,k_x,k_z).
\end{gather}
First we observe how $S$ acts on the unoccupied eigenstates. Substituting (\ref{sym.C4z}) to Eq.~(\ref{Moore model zw vectors}), we get
\begin{gather}
\nonumber
    z(S\mathbf{k})
    = e^{\text{i}\frac{\pi}{4}} z(\mathbf{k}) e^{\text{i}\frac{\pi}{4}},
    \quad
    w(S\mathbf{k})
    = e^{\text{i}\frac{\pi}{4}} w(\mathbf{k}) e^{\text{i}\frac{\pi}{4}}.
\end{gather}
Colloquially, the coefficients of $1, \text{i}$ in $z,w$ rotates by $\pi/2$ while the coefficients of $\text{j}, \text{k}$ stay fixed.
This in turn implies that the quaternions corresponding to unoccupied states satisfy
\begin{alignat}{2}
    & (\overline{z}w)(S\mathbf{k}) 
    && = e^{-\text{i}\frac{\pi}{4}} \overline{z}(\mathbf{k}) e^{-\text{i}\frac{\pi}{4}} e^{\text{i}\frac{\pi}{4}} w(\mathbf{k}) e^{\text{i}\frac{\pi}{4}}
    \n
    & && =  e^{-\text{i}\frac{\pi}{4}} (\overline{z}w)(\mathbf{k}) e^{\text{i}\frac{\pi}{4}},
    \\
    & (\overline{z}\text{i}w)(S\mathbf{k}) 
    && = e^{-\text{i}\frac{\pi}{4}} \overline{z}(\mathbf{k}) e^{-\text{i}\frac{\pi}{4}}e^{\text{i}\frac{\pi}{4}} w(\mathbf{k}) e^{\text{i}\frac{\pi}{4}}
    \n
    & && = e^{-\text{i}\frac{\pi}{4}} (\overline{z}\text{i}w)(\mathbf{k}) e^{\text{i}\frac{\pi}{4}}.
\end{alignat}
In terms of eigenstates $\ket{\psi^1_{occ}} = \ket{\overline{z}\text{j}w}, \ket{\psi^2_{occ}} = \ket{\overline{z}\text{k}w}$, this means
\begin{gather}
    R_S \ket{\psi^\alpha(\mathbf{k})} = U^\alpha{}_\beta\ket{\psi^\beta(S\mathbf{k})}
\end{gather}
with
\begin{gather}
    R_S =
    \begin{pmatrix}
    1&0&&\\0&1&&\\&&0&1\\&&-1&0
    \end{pmatrix},~
    U =
    \begin{pmatrix}
    0&1\\-1&0
    \end{pmatrix},
\end{gather}
so that Eq.~(\ref{symmetry_state}) is fulfilled. 

\subsection{$\mathcal{P}$ symmetry under the condition $\chi_w=-\chi_z$}
\label{subsec.P}
We will see that $S=\mathcal{P}$ symmetry for any nontrivial RHI with $\mathcal{PT}$ symmetry (thus having both $\mathcal{P}$ and $\mathcal{T}$) is allowed only when
\begin{gather}
    \chi_w = -\chi_z
\end{gather}
and must be represented by a matrix $R_S$ with negative determinant
\begin{gather}
    \det R_S=-1,
    \quad S=\mathcal{P},
\end{gather}
while the matrix representation with positive determinant forces the trivial RHI $\chi_z=\chi_w=0$. Before proving this fact, we shall observe how $\mathcal{P}$ is embodied in the MRW-RHI Hamiltonian $H_{m,-m}$. Note that in this case
\begin{gather}
    (w_0,w_1,w_2,w_3)(\mathbf{k}) = (z_0,z_1,-z_2,z_3)(\mathbf{k}).
\end{gather}
Direct computation yields
\begin{gather}
    \overline{z}w = (z_0^2+z_1^2-z_2^2+z_3^2)-2z_2z_3\text{i}-2z_0z_2\text{j}+2z_1z_2\text{k},
    \n 
    \overline{z}\text{i}w = -2z_2z_3+(z_0^2+z_1^2+z_2^2-z_3^2)\text{i}+2z_0z_3\text{j}+2z_1z_3\text{k}.
\end{gather}
Again, we take the unoccupied states $\ket{\psi^1}=\ket{\overline{z}w}, \ket{\psi^2}=\ket{\overline{z}\text{i}w}$. Noting $z_i(S\mathbf{k}) = \eta_iz_i(\mathbf{k})$ with $\eta_{3}=1,~\eta_{0,1,2}=-1$, we get
\begin{gather}
    \ket{\psi^1(S\mathbf{k})} = R_S \ket{\psi^1(\mathbf{k})},
    \n
    \ket{\psi^2(S\mathbf{k})} = -R_S \ket{\psi^2(\mathbf{k})},
\end{gather}
where
\begin{gather}
    R_S =
    \begin{pmatrix}
    1&&&\\&-1&&\\&&1&\\&&&1
    \end{pmatrix},~
    U =
    \begin{pmatrix}
    1&0\\0&-1
    \end{pmatrix}
\end{gather}
fulfills Eq.~(\ref{symmetry_state}). 

By the discussion in the previous subsection, $C_{2z}$ (even $C_{4z}$) symmetry is still present in $H_{m,-m}$. Hence, $H_{m,-m}$ has the symmetries
\begin{gather}
    \mathcal{P}, \mathcal{T}, C_{4z}, M_z
\end{gather}
separately. 

Now we prove the claim stated at the beginning of this subsection. Since $S=\mathcal{P}$ does not involve time reversal, it is represented by a unitary operator. In our context,
\begin{gather}
    R_S \in \mathrm{O}(4).
\end{gather}
To begin with, we assume
\begin{gather}
    \det R_S = 1,
    \;\text{i.e. }
    R_S \in \mathrm{SO}(4),
\end{gather}
to arrive at the conclusion $\chi_z=\chi_w=0$. Recall that surjectivity of Eq.~(\ref{SO(4) qform}) implies that there exist $\alpha,\beta\in \bb{S}^3\subset\bb{H}$ such that
\begin{gather}
    R_S \mathbf{x}
    =  \overline{\alpha} \mathbf{x} \beta, 
    \quad
    \forall \mathbf{x}\in\bb{R}^4=\bb{H}.
\end{gather}
Using $R_S$, we can compute $\chi_{z,w}$ in two different ways; first, we can apply the procedure Eq.~(\ref{eq.Gauss curvature})-(\ref{def.Hopf Inv vec}) to the functions Eq.~(\ref{ftn.Gauss map 1})-(\ref{ftn.Gauss map 2}). Alternatively, we can apply the same process to
\begin{gather}
\label{alternative basis at k}
    R_S~u^v_1(S\mathbf{k})~\text{and}~
    R_S~u^v_2(S\mathbf{k}), 
\end{gather}
which is just another basis of the occupied space at $\mathbf{k}$. The two results must coincide. 

Temporarily using the notation
\begin{gather}
    \mathbf{x}
    = u^v_1(S\mathbf{k}),
    \quad
    \mathbf{y}
    = u^v_2(S\mathbf{k}),
\end{gather}
we shall keep track of how the functions Eq.~(\ref{ftn.Gauss map 1})-(\ref{ftn.Gauss map 2}) changes as we pass from $u^v_{1,2}(\mathbf{k})$ to Eq.~(\ref{alternative basis at k}). For simplicity, we will only discuss Eq.~(\ref{ftn.Gauss map 1}) and arrive at the conclusion $\chi_z = -\chi_z$. The same reasoning about Eq.~(\ref{ftn.Gauss map 2}) leads to $\chi_w = -\chi_w$.
\begin{align}
    \mathbf{v}_z(\mathbf{k}) \rightarrow 
    \mathbf{v}_z'(\mathbf{k})
    & = \overline{\alpha}\mathbf{y}\beta \cdot \overline{\overline{\alpha}\mathbf{x}\beta} \n
    & = \overline{\alpha}\mathbf{y}\beta \cdot
    \overline{\beta} \overline{\mathbf{x}} \alpha \n 
    & =  \overline{\alpha} \cdot
    \mathbf{y} \overline{\mathbf{x}} \cdot \alpha \n 
    & =  \overline{\alpha} \cdot
    \mathbf{v}_z(S\mathbf{k}) \cdot \alpha \n 
    & =  \overline{\alpha} \cdot \mathbf{v}_z(S\mathbf{k}) \cdot \alpha. 
\label{change of Gauss map}
\end{align}
Since $x\mapsto\overline{\alpha} x\alpha$ is a rotation of
\begin{gather}
    \bb{R}^3
    = \{x_0 + x_1 \text{i} + x_2 \text{j} + x_3 \text{k} \in\bb{H}: x_0 = 0\},
\end{gather}
and Eq.~(\ref{eq.Gauss curvature}) is invariant under the rotation of $\mathbf{v}$,
\begin{align}
    \mathbf{F}_z'(\mathbf{k})
    & = (F_z'^1,F_z'^2,F_z'^3)(k_x,k_y,k_z) \n
    & = (F_z^1,F_z^2,F_z^3)(-k_x,-k_y,-k_z) \n 
    & = \mathbf{F}_z(S\mathbf{k}).
\end{align}
If $\mathbf{A}=\mathbf{A}_z(\mathbf{k})$ solves the magneto-static equations Eq.~(\ref{eq.magnetostat 1})-(\ref{eq.magnetostat 2}) for $\mathbf{F}=\mathbf{F}_z(\mathbf{k})$, then
\begin{align}
    \mathbf{A}_z'(\mathbf{k})
    & = (A_z'^1,A_z'^2,A'^3)(k_x,k_y,k_z) \n
    & = (-A_z^1,-A_z^2,-A_z^3)(-k_x,-k_y,-k_z) \n 
    & = -\mathbf{A}_z(S\mathbf{k})
\end{align}
solves the primed equations
\begin{gather}
    \nabla \times \mathbf{A}_z' = \mathbf{F}_z',
    \\
    \nabla \cdot \mathbf{A}_z'=0.
\end{gather}
Therefore,
\begin{align}
    \chi_z
    & = -\frac{1}{4\pi^2}\int_{BZ}d^3k\;(\mathbf{F}_z'\cdot\mathbf{A}_z')(\mathbf{k}) \n 
    & = -\frac{1}{4\pi^2}\int_{BZ}d^3k\;(-\mathbf{F}_z\cdot\mathbf{A}_z)(S\mathbf{k}) \n
    & = -\frac{1}{4\pi^2}\int_{BZ}d^3k\;(-\mathbf{F}_z\cdot\mathbf{A}_z)(\mathbf{k}) \n 
    & = -\chi_z.
\end{align}
Next, we tackle the case
\begin{gather}
    \det R_S = -1,
    ~\text{i.e.}~R_S \in \mathrm{O}(4)\backslash\mathrm{SO}(4).
\end{gather}
Any such operation $R_S$ acting on $\bb{R}^4=\bb{H}$ is the composition of the inversion
\begin{gather}
    \mathbf{x} \mapsto \overline{\mathbf{x}}
\end{gather}
and a rotation
\begin{gather}
    x\mapsto  \overline{\alpha}\mathbf{x}\beta. 
\end{gather}
Hence Eq.~(\ref{change of Gauss map}) is modified to
\begin{align}
    \mathbf{v}_z(\mathbf{k}) \rightarrow 
    \mathbf{v}_z'(\mathbf{k})
    & =  \overline{\alpha} \overline{\mathbf{y}}\beta\cdot
    \overline{ \overline{\alpha} \overline{\mathbf{x}}\beta } \n 
    & =  \overline{\alpha} \overline{\mathbf{y}}\beta\cdot
    \overline{\beta}\mathbf{x}\alpha \n 
    & = \overline{\alpha}\cdot\overline{\mathbf{y}}\mathbf{x}\cdot\alpha \n 
    & =  \overline{\alpha} \cdot
    ( - \overline{\mathbf{x}}\mathbf{y} )\cdot\alpha \n 
    & =  \overline{\alpha} \cdot
    \mathbf{v}_w(S\mathbf{k})\cdot\alpha \n
    & =  \overline{\alpha} \cdot
    \mathbf{v}_w(S\mathbf{k})\cdot\alpha 
\label{change of Gauss map2}
\end{align}
Repeating the previous discussion, we arrive at
\begin{gather}
    \chi_z = -\chi_w.
\end{gather}


\end{document}